\documentclass[aps,prd,twocolumn,nofootinbib]{revtex4-1}

\usepackage{graphicx}
\usepackage{hyperref}
\usepackage{slashed}
\usepackage{color}

\begin{document}

\title{Heavy Quarkonium Production at LHC through $W$ Boson Decays}

\author{Qi-Li Liao$^{1}$}
\author{Xing-Gang Wu$^{1,2}$}
\email{wuxg@cqu.edu.cn}
\author{Jun Jiang$^{1}$}
\author{Zhi Yang$^{1}$}
\author{Zhen-Yun Fang$^{1}$}

\address{$^{1}$ Department of Physics, Chongqing University, Chongqing 401331, P.R. China\\
$^{2}$ SLAC National Accelerator Laboratory, 2575 Sand Hill Road, Menlo Park, CA 94025, USA}

\date{\today}

\begin{abstract}

The production of the heavy $(c\bar{c})$-quarkonium, $(c\bar{b})$-quarkonium and $(b\bar{b})$-quarkonium states [$(Q\bar{Q'})$ quarkonium for short], via the $W^+$ semi-inclusive decays, has been systematically studied within the framework of the non-relativistic QCD. In addition to the two color-singlet $S$-wave states, we also discuss the production of the four color-singlet $P$-wave states $|(Q\bar{Q'})(^1P_1)_{\bf 1}\rangle$ and $|(Q\bar{Q'})(^3P_J)_{\bf 1}\rangle$ [with $J=(0,1,2)$] together with the two color-octet components $|(Q\bar{Q'})(^1S_0)_{\bf 8}\rangle$ and $|(Q\bar{Q'})(^3S_1)_{\bf 8}\rangle$. Improved trace technology is adopted to derive the simplified analytic expressions at the amplitude level, which shall be useful for dealing with the following cascade decay channels. At the LHC with the luminosity ${\cal L}\propto 10^{34}cm^{-2}s^{-1}$ and the center-of-mass energy $\sqrt{S}=14$ TeV, sizable heavy-quarkonium events can be produced through the $W^+$ boson decays, i.e. $2.57\times10^6$ $\eta_c$, $2.65\times10^6$ $J/\Psi$ and $2.40\times10^6$ $P$-wave charmonium events per year can be obtained; and $1.01\times10^5$ $B_c$, $9.11\times10^4$ $B^*_c$ and $3.16\times10^4$ $P$-wave $(c\bar{b})$-quarkonium events per year can be obtained. Main theoretical uncertainties have also been discussed. By adding the uncertainties caused by the quark masses in quadrature, we obtain $\Gamma_{W^+\to (c\bar{c})+c\bar{s}} =524.8^{+396.3}_{-258.4}$ KeV, $\Gamma_{W^+\to (c\bar{b})+b\bar{s}} =13.5^{+4.73}_{-3.29}$ KeV, $\Gamma_{W^+\to (c\bar{b})+c\bar{c}}= 1.74^{+1.98}_{-0.73}$ KeV and $\Gamma_{W^+\to (b\bar{b})+c\bar{b}}= 38.6^{+13.4}_{-9.69}$ eV.  \\


\end{abstract}

\maketitle

\section{Introduction}

The study of the $W$ boson is helpful for understanding the electroweak interactions and for searching new physics beyond the standard model. With the LHC luminosity rising up to ${\cal L}\propto 10^{34}cm^{-2}s^{-1}$ and running at the center-of-mass energy $\sqrt{S}=14$ TeV, large amount of $W$ bosons about $10^{10}$ events per year will be produced. This makes the LHC a much better $W$ boson factory than the TEVATRON \cite{Teva,Tevb,Tevc}, and more $W$ boson rare decays can be adopted for precise studies. In Refs.\cite{w,w1}, the authors have discussed a class of $W$ boson semi-inclusive decays to the lowest $S$-wave heavy-quarkonium states $(Q\bar{Q'})$ with $Q$ and $Q'$ stands for the $c$-quark or the $b$-quark respectively. Their results show that large number of heavy-quarkonium events through the $W$ boson decays can be found at LHC, so these channels shall be helpful for studying heavy-quarkonium properties.

Intuitively, the heavy-quarkonium production process could be understood in terms of two distinct steps: the production of the $Q\bar{Q'}$ pair and the subsequent evolution of the $Q\bar{Q'}$ pair into the quarkonium. Different treatment of the evolution leads to different theoretical models, among which the non-relativistic QCD (NRQCD) \cite{nrqcd} is widely adopted. In the framework of NRQCD, a doubly heavy meson is considered as an expansion of various Fock states. And in addition to the two color-singlet $S$-wave states $|(Q\bar{Q'})(^1S_0)_{\bf 1}\rangle$ and $|(Q\bar{Q'})(^3S_1)_{\bf 1}\rangle$, the naive NRQCD scaling rule shows that the four color-singlet $P$-wave states $|(Q\bar{Q'})(^1P_1)_{\bf 1}\rangle$ and $(Q\bar{Q'})(^3P_J)_{\bf 1}\rangle$ [with $J=(0,1,2)$] together with the two color-octet components $|(Q\bar{Q'})(^1S_0)_{\bf 8}\rangle$ and $|(Q\bar{Q'})(^3S_1)_{\bf 8}\rangle$ shall also give sizable contributions to the production. Here the thickened subscripts of $(Q\bar{Q'})$ stand for the color indices, ${\bf 1}$ for color singlet and ${\bf 8}$ for color octet; the relevant angular momentum quantum numbers are shown in the parentheses accordingly. These higher excited $(Q\bar{Q'})$-quarkonium states may directly or indirectly decay to their ground state via the electromagnetic or hadronic interactions with high probability. It is interesting to study higher Fock states' contributions to make a sound estimation on the heavy-quarkonium production, and hence to be a more useful reference for experimental studies.

Moreover, the heavy-quarkonium production itself is very useful for testing perturbative QCD \cite{yellow1,yellow2,cms}. For example, since its discovery by the CDF collaboration \cite{cdf}, the $B_c$ meson being the unique `doubly heavy-flavored' meson in the standard model has aroused people's great interest. The `direct' hadronic production of the $B_c$ meson has been studied systematically in Refs.\cite{bc1,bc2,bc3,bcvegpy}. As a compensation, it would be helpful to study its `indirect' production mechanisms. Because sizable top-quark and $W$ boson events shall be produced at the LHC, the production of $B_c$ through their decay shall be helpful for determining the $B_c$-meson properties, since too many directly produced $B_c$ events shall be cut off by the trigging condition at the LHC \cite{yellow1,yellow2,cms}. A systematical study on the $B_c$-meson production through the top-quark and the $Z^0$ boson decay can be found in the literature \cite{tbc1,tbc2,zbc0,zbc1,zbc2,zbc3}. In the present paper, we shall make a systematic study on the $B_c$ meson production through the $W$ boson decays.

To deal with the heavy-quarkonium production through the $W$ boson semi-inclusive decays, one needs to derive the squared amplitude, which is usually done by the conventional trace technique. The analytical expression for the squared amplitude of the $S$-wave case can be found in Ref.\cite{w}, however it is hard to write down the squared amplitudes for the $P$-wave cases, which is much too complex and lengthy. One important way to solve this is to deal with the process directly at the amplitude level. For this purpose, the helicity amplitude approach and the improved trace amplitude approach have been suggested in the literature. As for the helicity amplitude approach \cite{helicity}, all the amplitudes can be expressed by the complex valued helicity amplitudes that can be numerically calculated, an explicit example of which to deal with the case of massive spinors can be found in Ref.\cite{bcvegpy}. While for the improved trace amplitude approach suggested and developed by Refs.\cite{tbc2,zbc0,zbc1,zbc2}, the hard-scattering amplitude can also be expressed by the dot-products of the particle momenta as that of the squared amplitude, it is, however, done at the amplitude level and is much more simpler. In the present paper, we shall adopt the improved trace amplitude approach to derive analytical expressions for all the mentioned Fock states, and to be a useful reference, we simplify its form as compactly as possible when fully applying the symmetries and relations among them.

The paper is organized as follows. In Sec.II, we show our calculation techniques for the mentioned $W^+$ semi-inclusive decays to the heavy-quarkonium. In Sec.III, we present the numerical results and discuss on the properties of the heavy-quarkonium production through $W^+$ decays. The final section is reserved for a summary. To make the paper more compact, we present the detailed formulas for dealing with the process under the improved trace amplitude approach in the Appendix.

\section{Calculation Technology}

We shall deal with some typical $W$ boson semi-inclusive processes for the heavy-quarkonium production, i.e. $W^{+}(k) \to (Q\bar{Q'})[n](q_3) +q(q_2) + \bar{q}'(q_1)$, where $q$ and $\bar{q}'$ stand for the Cabibbo-Kobayashi-Maskawa (CKM) favored quark and anti-quark accordingly, and $k$ and $q_i$ are momenta of the corresponding particles. According to the NRQCD factorization formula \cite{petrelli}, its total decay width $d\Gamma$ can be factorized as
\begin{equation}
d\Gamma=\sum_{n} d\hat\Gamma(W^+ \to (Q\bar{Q'})[n]+q \bar{q}') \langle{\cal O}^H(n) \rangle ,
\end{equation}
where $\langle{\cal O}^{H}(n)\rangle$ describes the hadronization of a $Q\bar{Q'}$ pair into the observable quark state $H$ and is proportional to the transition probability of the perturbative state $(Q\bar{Q'})[n]$ into the bound state. As for the color-singlet components, their matrix elements can be directly related to the wave functions at the origin for the $S$-wave states or the first derivative of the wave functions at the origin for the $P$-wave states \cite{nrqcd}, which can be computed via the potential models \cite{pot1,pot2,pot3,pot4,pot5,pot6} and/or potential NRQCD \cite{pnrqcd1,pnrqcd2} and/or lattice QCD \cite{lat1} respectively. As for the color-octet components, their matrix elements are to be determined experimentally, which are smaller than the color-singlet matrix elements by a certain $v^2$ order. Here $v$ is the typical velocity of the heavy quark or anti-quark in the quarkonium rest frame, $v^2\simeq 0.3$ for $J/\Psi$ and $v^2\simeq0.1$ for $\Upsilon$. More specifically, based on the velocity scaling rule and under the vacuum-saturation approximation, we have \cite{yellow1,nrqcd,bc3,bcdecay}
\begin{eqnarray}
&&\langle (Q\bar{Q'})_{\bf 8}[^1S_0]|{\cal O}_{\bf 8}(^1S_0)| (Q\bar{Q'})_{\bf 8}[^1S_0] \rangle \nonumber\\
&\simeq&\Delta_S(v)^2 \langle (Q\bar{Q'})_{\bf 1}[^1S_0]|{\cal O}_{\bf 1}(^1S_0)| (Q\bar{Q'})_{\bf 1}[^1S_0] \rangle
\end{eqnarray}
and
\begin{eqnarray}
&& \langle (Q\bar{Q'})_{\bf 8}[^3S_1]|{\cal O}_{\bf 8}(^3S_1)| (Q\bar{Q'})_{\bf 8}[^3S_1] \rangle \nonumber\\
&\simeq& \Delta_S(v)^2 \langle (Q\bar{Q'})_{\bf 1}[^3S_1] |{\cal O}_1(^3S_1)| (Q\bar{Q'})_{\bf 1}[^3S_1] \rangle\,,
\end{eqnarray}
where $\Delta_S(v)$ is of order $v^2$.

The short-distance decay width
\begin{equation}
d\hat\Gamma(W^{+}\to (Q\bar{Q}')[n]+q\bar{q}')= \frac{1}{2k^0} \overline{\sum}  |M|^{2} d\Phi_3,
\end{equation}
where $\overline{\sum}$ means that we need to average over the spin states of the initial particles and to sum over the color and spin of all the final particles. In the $W^+$ rest frame, the three-particle phase space can be written as
\begin{equation}
d{\Phi_3}=(2\pi)^4 \delta^{4}\left(k - \sum_f^3 q_{f}\right)\prod_{f=1}^3 \frac{d^3{\vec{q}_f}}{(2\pi)^3 2q_f^0}.
\end{equation}
The $1 \to 3$ phase space with massive quark/antiqark in the final state can be found in Refs.\cite{tbc2,zbc1}. To shorten the paper, we shall not present it here. With the help of the formulas listed in Refs.\cite{tbc2,zbc1}, one can not only derive the whole decay width but also obtain the corresponding differential decay widths that are helpful for experimental studies, such as $d\Gamma/ds_1$, $d\Gamma/ds_2$, $d\Gamma/d\cos\theta_{13}$ and $d\Gamma/d\cos\theta_{23}$, where $s_1=(q_1+q_3)^2$, $s_2=(q_1+q_2)^2$, $\theta_{13}$ is the angle between $\vec{q}_1$ and $\vec{q}_3$, and $\theta_{23}$ is the angle between $\vec{q}_2$ and $\vec{q}_3$ in the $W^+$ rest frame, respectively.

And then our task left to deal with is the hard-scattering amplitude for the specified processes
\begin{eqnarray}
&& W^{+}\rightarrow (c\bar{c})[n] + c \bar{s} ,\;\; W^{+}\rightarrow (b\bar{b})[n] + c \bar{b} ,\nonumber\\
&& W^{+}\rightarrow (c\bar{b})[n] + b \bar{s} ,\;\; W^{+}\rightarrow (c\bar{b})[n] + c \bar{c} . \nonumber
\end{eqnarray}
Their amplitudes can be generally expressed as
\begin{equation} \label{amplitude}
iM = {\cal{C}} {\bar u_{s i}}({q_2}) \sum\limits_{n = 1}^{m} {{\cal A} _n } {v_{s' j}}({q_1}),
\end{equation}
where $m$ stands for the number of the Feynman diagrams, $s$ and $s'$ are spin indices, $i$ and $j$ are color indices for the outgoing quark and antiquark. The overall factor ${\cal C}={\cal C}_s$ or ${\cal C}_o$ stands for the specified quarkonium in the color-singlet and the color-octet states respectively. ${\cal C}_s=\frac{2gg_s^2 V_{CKM}}{3\sqrt{6}}\delta_{ij}$ and ${\cal C}_o=\frac{gg_s^2 V_{CKM} }{2\sqrt{2}}(\sqrt{2}T^aT^bT^a)_{ij}$, where $\sqrt{2}T^b$ stands for the color factor of the color-octet quarkonium state. $V_{CKM}$ stands for the CKM matrix element, $V_{CKM}=V_{cs}$ for $W^{+}\rightarrow (c\bar{c})[n] + c \bar{s}$ and $W^{+}\rightarrow (c\bar{b})[n] + b \bar{s}$; $V_{CKM}=V_{cb}$ for $W^{+}\rightarrow (c\bar{b})[n] +c\bar{c}$ and $W^{+}\rightarrow (b\bar{b}) +c \bar{b}$.

\subsection{${\cal A}_n$ for $W^{+}\rightarrow (c\bar{c})[n] + c \bar{s}$ and $W^{+}\rightarrow (c\bar{b})[n] + b \bar{s}$}

\begin{figure}
\includegraphics[width=0.45\textwidth]{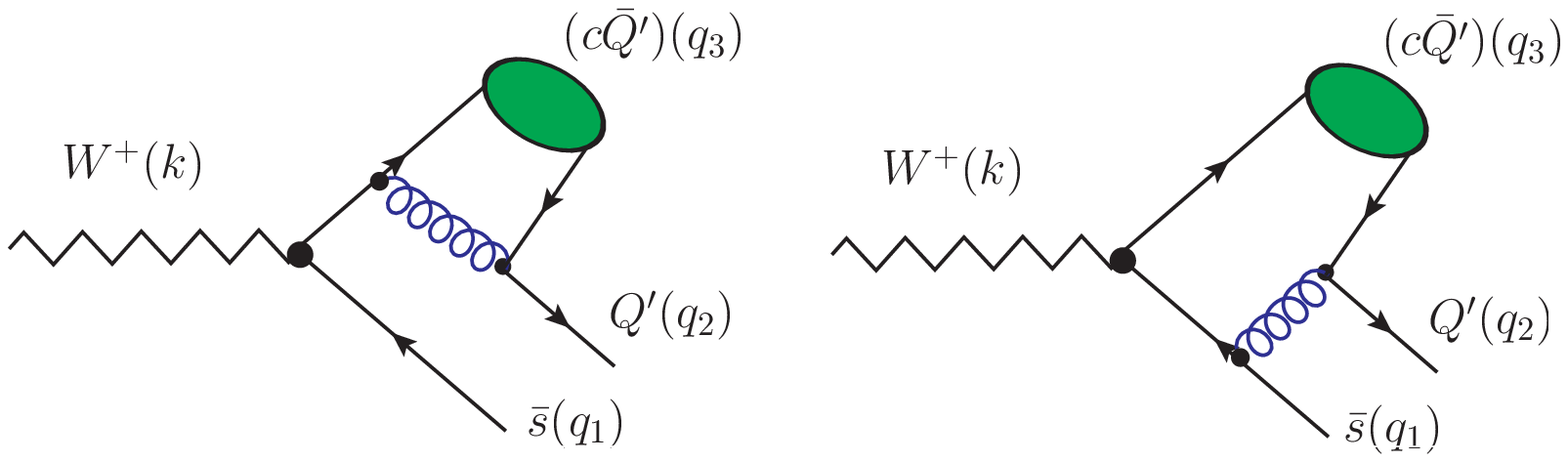}
\caption{Feynman diagrams for the precess $W^+(k)\rightarrow (c\bar{Q'})(q_3) + Q'(q_2) \bar{s}(q_1)$, where $(c\bar{Q'})$ stands for the quarkonium Fock states $|(c\bar Q')_{\bf 1}[^{1}S_{0}]\rangle$, $|(c\bar Q')_{\bf 1}[^{3}S_{1}] \rangle$, $|(c\bar Q')_{\bf 1}[^{1}P_{1}]\rangle$, $|(c\bar Q')_{\bf 8}[^{1}S_{0}] g\rangle$, $|(c\bar Q')_{\bf 1}[^{3}P_{J}]\rangle$ and $|(c\bar Q')_{\bf 8}[^{3}S_{1}] g\rangle$ respectively. } \label{feyn1}
\end{figure}

For convenience, we present these two processes as $W^+(k) \rightarrow (c\bar{Q'})[n](q_3) + Q'(q_2)\bar{s}(q_1)$, where $Q'$ stands for $c$ or $b$ quark accordingly. The Feynman diagrams of the process are presented in Fig.(\ref{feyn1}), where the intermediate gluon should be hard enough to produce a $c\bar{c}$ pair or $b\bar{b}$ pair, so the amplitude is perturbative QCD calculable.

For $(c\bar{Q'})$ quarkonium in $S$-wave states, ${\cal A}_n$ can be written as
\begin{widetext}
\begin{eqnarray}
{\cal A}_1 &=& \left[{\gamma_\alpha}\frac{\Pi^{0(\nu)}_{q_3}(q)}{(q_{32} + {q_2})^2}{\gamma_\alpha}\frac{\slashed{q}_2 + \slashed{q}_3 + {m_c}}{(q_2+q_3 )^2 - m_c^2}\slashed\epsilon(k)({1-\gamma^5})\right]_{q=0}, \\
{\cal A}_2 &=& \left[{\gamma_\alpha}\frac{\Pi^{0(\nu)}_{q_3}(q)}{(q_{32} + {q_2})^2}\slashed\epsilon(k)({1-\gamma^5})\frac{{m_s}-\slashed{q}_1-\slashed{q}_2-\slashed{q}_{32} }{(q_1+q_2+q_{32} )^2 - m_s^2}\gamma_\alpha\right]_{q=0} .
\end{eqnarray}
And for the $P$-wave states, ${\cal A}_n$ can be written as
\begin{eqnarray}
{\cal A}^{S=0,L=1}_1 &=& \varepsilon_l^{\mu}(q_3) \frac{d}{dq_\mu} \left[{\gamma_\alpha}\frac{\Pi^0_{q_3}(q)}{(q_{32} + {q_2})^2}{\gamma_\alpha}\frac{\slashed{q}_2 + \slashed{q}_3 + {m_c}}{(q_2+q_3 )^2 - m_c^2}\slashed{\epsilon}(k)({1-\gamma^5})\right]_{q=0}, \\
{\cal A}^{S=0,L=1}_2 &=& \varepsilon_l^{\mu}(q_3) \frac{d}{dq_\mu} \left[{\gamma_\alpha}\frac{\Pi^0_{q_3}(q)}{(q_{32} + {q_2})^2}\slashed{\epsilon}(k)({1-\gamma^5})\frac{{m_s}-\slashed{q}_1-\slashed{q}_2-\slashed{q}_{32\emph{}} }{(q_1+q_2+q_{32} )^2 - m_s^2}\gamma_\alpha\right]_{q=0}
\end{eqnarray}
and
\begin{eqnarray}
{\cal A}^{S=1,L=1}_1 &=& \varepsilon^{J}_{\mu\nu}(q_3) \frac{d}{dq_\mu} \left[{\gamma_\alpha}\frac{\Pi^\nu_{q_3}(q)}{(q_{32} + {q_2})^2}{\gamma_\alpha}\frac{\slashed{q}_2 + \slashed{q}_3 + {m_c}}{(q_2+q_3 )^2 - m_c^2}\slashed{\epsilon}(k)({1-\gamma^5})\right]_{q=0},\\
{\cal A}^{S=1,L=1}_2 &=& \varepsilon^{J}_{\mu\nu}(q_3) \frac{d}{dq_\mu} \left[ {\gamma_\alpha}\frac{\Pi^\nu_{q_3}(q)}{(q_{32} + {q_2})^2}\slashed{\epsilon}(k)({1-\gamma^5})\frac{{m_s}-\slashed{q}_1-\slashed{q}_2-\slashed{q}_{32} }{(q_1+q_2+q_{32} )^2 - m_s^2}\gamma_\alpha\right]_{q=0}.
\end{eqnarray}
\end{widetext}
Here $q$ stands for the relative momentum between the two constituent quarks in $(c\bar{Q'})$ quarkonium. $q_{31}$ and $q_{32}$ are the momenta of the two constituent quarks, i.e.
\begin{equation}
q_{31} = \frac{m_c}{M}{q_3} + q , q_{32} = \frac{m_{Q'}}{M}{q_3} - q.
\end{equation}
where $M\simeq m_c + m_{Q'}$ are adopted to ensure the gauge invariance of the hard-scattering amplitude. $\varepsilon(k)$ is the polarization vector of $W^+$. $\varepsilon_{s}(q_3)$ and $\varepsilon_{l}(q_3)$ are the polarization vectors relating to the spin and the orbit angular momentum of $(c\bar{Q'})$ quarkonium, and $\varepsilon^{J}_{\mu\nu}(q_3)$ is the polarization tensor for the spin-triplet $P$-wave states with $J=0$, $1$ and $2$ respectively. The projectors $\Pi^0_{q_3}(q)$ and $\Pi^\nu_{q_3}(q)$ are for spin-singlet and spin-triplet quarkonium states respectively, and their covariant form can be conveniently written as
\begin{equation}
\Pi^0_{q_3}(q)=\frac{-\sqrt{M}}{4{m_c}{m_{Q'}}}(\slashed{q}_{32}- m_{Q'})\gamma_5 (\slashed{q}_{31}+ m_c)
\end{equation}
and
\begin{equation}
\Pi^\nu_{q_3}(q)=\frac{-\sqrt{M}}{4{m_c}{m_{Q'}}}(\slashed{q}_{32}-m_{Q'}) \gamma_{\nu}(\slashed{q}_{31}+m_c).
\end{equation}
After substituting these projectors into the amplitudes, the amplitudes then can be squared, summed over the freedoms in the final state and averaged over the ones in the initial state. Selection of the appropriate total angular momentum quantum number is done by performing the proper polarization sum, which for a spin-triplet $S$ state or a spin-singlet $P$ state is given by
\begin{equation}
\sum_{J_z}\varepsilon_\alpha \varepsilon^*_{\alpha'} =\Pi_{\alpha\alpha'} ,\label{3s1}
\end{equation}
where $\Pi_{\alpha\beta}=-g_{\alpha\beta}+\frac{q_{3\alpha} q_{3\beta}}{M^2}$, and $J_z=s_z$ or $l_z$ respectively. And for the case of $^3P_J$ states, the sum over polarization is given by \cite{petrelli,projector}
\begin{eqnarray}
\varepsilon^{(0)}_{\alpha\beta} \varepsilon^{(0)*}_{\alpha'\beta'} &=& \frac{\Pi_{\alpha\beta}\Pi_{\alpha'\beta'}}{3} , \nonumber\\
\sum_{J_z}\varepsilon^{(1)}_{\alpha\beta} \varepsilon^{(1)*}_{\alpha'\beta'} &=& \frac{\Pi_{\alpha\alpha'}\Pi_{\beta\beta'}- \Pi_{\alpha\beta'}\Pi_{\alpha'\beta}}{2} \nonumber
\end{eqnarray}
and
\begin{displaymath}
\sum_{J_z}\varepsilon^{(2)}_{\alpha\beta} \varepsilon^{(2)*}_{\alpha'\beta'} = \frac{\Pi_{\alpha\alpha'}\Pi_{\beta\beta'}+ \Pi_{\alpha\beta'}\Pi_{\alpha'\beta}}{2} -\frac{\Pi_{\alpha\beta}\Pi_{\alpha'\beta'}}{3} .
\end{displaymath}

\subsection{${\cal A}_n$ for $W^{+}\rightarrow (c\bar{b})[n] +c\bar{c}$ and $W^{+}\rightarrow (b\bar{b})[n] +c \bar{b}$}

\begin{figure}
\includegraphics[width=0.45\textwidth]{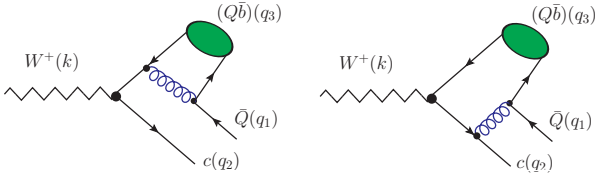}
\caption{Feynman diagrams for the process $W^+(k)\rightarrow (Q\bar{b})(q_3) + c(q_2) \bar{Q}(q_1)$, where $(c\bar{Q'})$ stands for the quarkonium Fock states $|(Q\bar{b})_{\bf 1}[^{1}S_{0}]\rangle$, $|(Q\bar{b})_{\bf 1}[^{3}S_{1}] \rangle$, $|(Q\bar{b})_{\bf 1}[^{1}P_{1}]\rangle$, $|(Q\bar{b})_{\bf 8}[^{1}S_{0}] g\rangle$, $|(Q\bar{b})_{\bf 1}[^{3}P_{J}]\rangle$ and $|(Q\bar{b})_{\bf 8}[^{3}S_{1}] g\rangle$ respectively. } \label{feyn2}
\end{figure}

For convenience, we present these two processes as $W^+(k) \rightarrow (Q\bar{b})[n](q_3) + c(q_2)\bar{Q}(q_1)$, where $Q$ stands for the $c$ or $b$ quark accordingly. The Feynman diagrams of the process are presented in Fig.(\ref{feyn2}). For the $(Q\bar{b})$ quarkonium in $S$-wave states, ${\cal A}_n$ reads
\begin{widetext}
\begin{eqnarray}
{\cal A}_1 &=& \left[\slashed\epsilon(k) ({1-\gamma^5}) \frac{{m_b}-\slashed{q}_1 - \slashed{q}_3} {(q_1+q_3 )^2 - m_b^2} {\gamma_\alpha} \frac{\Pi^{0(\nu)}_{q_3}(q)}  {(q_{31} + {q_1})^2} {\gamma_\alpha} \right]_{q=0}, \\
{\cal A}_2 &=& \left[{\gamma_\alpha} \frac{\slashed{q}_1+\slashed{q}_2+\slashed{q}_{31}+{m_c}} {(q_1+q_2+q_{31} )^2 - m_c^2} \slashed\epsilon(k)({1-\gamma^5}) \frac{\Pi^{0(\nu)}_{q_3}(q)}{(q_{31} + {q_1})^2}\gamma_\alpha\right]_{q=0} .
\end{eqnarray}
And for the $P$-wave states, ${\cal A}_n$ can be written as
\begin{eqnarray}
{\cal A}^{S=0,L=1}_1 &=& \varepsilon_l^{\mu}(q_3) \frac{d}{dq_\mu} \left[\slashed\epsilon(k)({1-\gamma^5})\frac{{m_b}-\slashed{q}_1 - \slashed{q}_3}{(q_1+q_3 )^2 - m_b^2} {\gamma_\alpha}\frac{\Pi^{0}_{q_3}(q)}{(q_{31} + {q_1})^2} {\gamma_\alpha} \right]_{q=0}, \\
{\cal A}^{S=0,L=1}_2 &=& \varepsilon_l^{\mu}(q_3) \frac{d}{dq_\mu} \left[{\gamma_\alpha} \frac{\slashed{q}_1+\slashed{q}_2+\slashed{q}_{31}+{m_c}} {(q_1+q_2+q_{31} )^2 - m_c^2} \slashed\epsilon(k)({1-\gamma^5}) \frac{\Pi^{0}_{q_3}(q)}{(q_{31} + {q_1})^2}\gamma_\alpha\right]_{q=0}
\end{eqnarray}
and
\begin{eqnarray}
{\cal A}^{S=1,L=1}_1 &=& \varepsilon^{J}_{\mu\nu}(q_3) \frac{d}{dq_\mu} \left[\slashed\epsilon(k)({1-\gamma^5})\frac{{m_b}-\slashed{q}_1 - \slashed{q}_3}{(q_1+q_3 )^2 - m_b^2} {\gamma_\alpha}\frac{\Pi^{\nu}_{q_3}(q)}{(q_{31} + {q_1})^2}{\gamma_\alpha}\right]_{q=0}, \\
{\cal A}^{S=1,L=1}_2 &=& \varepsilon^{J}_{\mu\nu}(q_3) \frac{d}{dq_\mu} \left[{\gamma_\alpha} \frac{\slashed{q}_1+\slashed{q}_2+\slashed{q}_{31}+{m_c}} {(q_1+q_2+q_{31} )^2 - m_c^2} \slashed\epsilon(k)({1-\gamma^5}) \frac{\Pi^{\nu}_{q_3}(q)}{(q_{31} + {q_1})^2}\gamma_\alpha\right]_{q=0} .
\end{eqnarray}
\end{widetext}
The quark momenta $q_{31}$ and $q_{32}$, and the projectors $\Pi^{0}_{q_3}(q)$ and $\Pi^{\nu}_{q_3}(q)$ have the same meaning as those listed in the last subsection, and one needs to change the two constituent quarks there ($c$ and $Q'$) to be ($Q$ and $b$).

\subsection{Analytical expressions for the mentioned amplitudes}

As mentioned in the Introduction, we adopt the improved trace technology to deal with the hard-scattering amplitude (\ref{amplitude}). Detailed processes of the approach can be found in Refs.\cite{tbc2,zbc0,zbc1,zbc2}, here for self-consistency, we shall present its main idea and our main results.

For the present considered $W$ boson semi-inclusive decay process, $W^{+}(k) \to (Q\bar{Q'})[n](q_3) + q(q_2) + \bar{q}'(q_1)$, there are one quark and one antiquark in the final state. Under the approach, we first arrange the whole amplitude into four orthogonal sub-amplitudes $M_{\pm{s}\pm{s'}}$ according to the spins of the outgoing quark with spin-index $s$ and antiquark with spin-index $s'$, then transform these sub-amplitudes into the trace form by properly dealing with the massive spinors with the help of an arbitrary light-like momentum $k_0$ and an arbitrary space-like momentum $k_1$, $k_1^2=-1$ and $k_0\cdot k_1 =0$ \footnote{The final results are independent of $k_0$ and $k_1$, which provides another way to check the rightness of the derived results. And one can choose them to be those that can maximumly simply the analytical expressions for the amplitude. }. And then we do the trace of the Dirac $\gamma$ matrix strings at the amplitude level, which shall result in explicit series over some independent Lorentz-structures.

After summing up the spin states of the outgoing quark/antiquark, the squared amplitude can be divided into four orthogonal parts,
\begin{equation}
|M|^2 = |M_{1}|^2 + |M_{2}|^2 + |M_{3}|^2 + |M_{4}|^2,
\end{equation}
where the four amplitudes $M_i$ can be written as
\begin{eqnarray}
M_1 &=& \frac{M_{ss'}+M_{-s-s'}}{\sqrt{2}} = \frac{N {\rm Tr}\left[ {({\slashed{q}_1} - m_{q'}){\slashed{k}_0}({\slashed{q}_2} + {m_{q}})A } \right]}{{\sqrt 2 }}  \nonumber\\
M_2 &=& \frac{M_{ss'}-M_{-s-s'}}{\sqrt{2}} = \frac{N {\rm Tr}\left[ {({\slashed{q}_1} - m_{q'}){\gamma _5}{\slashed{k}_0}({\slashed{q}_2} + {m_{q}})A } \right]}{{\sqrt 2 }}  \nonumber\\
M_3 &=& \frac{M_{s-s'}-M_{-ss'}}{\sqrt{2}} = \frac{N {\rm Tr}\left[ {({\slashed{q}_1} - m_{q'}){\slashed{k}_0}{\slashed{k}_1}({\slashed{q}_2} + {m_q})A } \right]}{{\sqrt 2 }} \nonumber
\end{eqnarray}
and
\begin{equation}
M_4 = \frac{M_{s-s'}+M_{-ss'}}{\sqrt{2}} = \frac{N {\rm Tr}\left[ {({\slashed{q}_1} - m_{q'}){\gamma _5}{\slashed{k}_1}{\slashed{k}_0}({\slashed{q}_2} + {m_q})A } \right]}{{\sqrt 2 }} \nonumber
\end{equation}
Here, $A=\sum\limits_{n = 1}^{2} {{\cal A} _n }$ and $N = 1/\sqrt{4({k_0}\cdot{q_1})({k_0}\cdot{q_2})}$ is the normalization constant. As a viable choice to simplify the amplitude, we take $k_0 = {q_2} - \alpha {q_1}$ with $ \alpha = \frac{{q_1} \cdot {q_2} + \sqrt{({q_1} \cdot {q_2})^2 - q_1^2 q_2^2}}{q_1^2}$ and $k_1^\mu  = i{N_0}{\varepsilon ^{\mu \nu \rho \sigma }}{q_{1\nu }}{k_{\rho }}{q_{2\sigma }}$ ( $N_0$ ensures $k_1\cdot k_1=-1$), which leads to
\begin{displaymath}
\slashed{k}_1 = {N_0}{\gamma _5}\left[ {{q_1} \cdot {k}{\slashed{q}_2} + {\slashed{q}_1}{k} \cdot {q_2} - {q_1} \cdot {q_2}{\slashed{k}} - {\slashed{q}_1}{\slashed{k}}{\slashed{q}_2}} \right]. \\
\end{displaymath}
Then the resultant $M_i$ are,
\begin{eqnarray}
M_1 &=& {L_1} \times Tr [({\slashed{q}_1} - m_{q'})({\slashed{q}_2} + {m_q})A] \\
M_2 &=& {L_2} \times Tr [({\slashed{q}_1} - m_{q'}){\gamma _5}(\slashed{q}_{2}+{m_q})A] \\
M_3 &=& M_{3'} - {N_0}[{m_q} ({q_1} \cdot {k}) + m_{q'}({q_2} \cdot {k})]M_2 \\
M_4 &=& M_{4'} + {N_0}[{m_q} ({q_1} \cdot {k}) - m_{q'}({q_2} \cdot {k})]M_1
\end{eqnarray}
where $L_{1,2}=1/(2\sqrt{q_1\cdot q_2 \mp m_q m_{q'}})$ and
\begin{eqnarray}
M_{3'} &=&\frac{N_0}{4L_2} Tr\left[ {({\slashed{q}_1} - m_{q'}){\gamma _5}{\slashed{k}}({\slashed{q}_2} + {m_q})A } \right] , \\
M_{4'} &=&-\frac{N_0}{4L_1} Tr\left[ {({\slashed{q}_1} -m_{q'}){\slashed{k}}({\slashed{q}_2} + {m_q})A } \right] .
\end{eqnarray}

Furthermore, the amplitudes $M_{i^{(')}}$ can be expanded over some basic Lorentz structures:
\begin{equation}
M_i(n)=\sum^m_{j=1} A^i_j(n) B_j(n) (i=1-4)
\end{equation}
and
\begin{equation}
M_{i'}(n)=\sum^m_{j=1} A^{i'}_j(n) B_j(n) \;\; (i'=3,4)
\label{amat}
\end{equation}
where $m$ stands for the number of basic Lorentz structures $B_j(n)$, whose values depend on the heavy-quarkonium state $[n]$. The independent lorentz structures $B_j(n)$ for all the Fock states and the explicit expressions for the Lorentz-invariant coefficients $A^{1,2}_j(n)$ and $A^{3',4'}_j(n)$ are put in the Appendix.

\section{Numerical Results}

We adopt the following values to do the numerical calculation \cite{wtd,pdg}: $m_W=80.399$GeV, $\Gamma_{W^+}=2.085$GeV, $m_c=1.35$GeV, $m_b=4.90$GeV, $m_s=0.105$GeV, $|V_{cs}|=1.023\pm0.036$ and $|V_{cb}|=0.0406\pm0.0013$. Leading-order $\alpha_s$ running is adopted and we set the renormalization scale to be $2m_c$ for charmonium and $(c\bar{b})$ quarkonium, and $2m_b$ for bottomonium accordingly, which lead to $\alpha_s(2m_c)=0.26$ and $\alpha_s(2m_b)=0.18$. Non-perturbative matrix elements can be related to the wave function at the origin $\Psi_S(0)=\sqrt{{1}/{4\pi}}R_S(0)$ and the first derivative of the wave function at the origin $\Psi'_P(0)=\sqrt{{3}/{4\pi}}R'_P(0)$, where we adopt \cite{pot6}
\begin{eqnarray}
&&|R_S(c\bar{c})(0)|^2=0.810\;{\rm GeV}^3 \;,\; |R'_P(c\bar{c})(0)|^2=0.075 \;{\rm GeV}^5 ,\nonumber\\ &&|R_S(c\bar{b})(0)|^2=1.642\;{\rm GeV}^3 \;,\; |R'_P(c\bar{b})(0)|^2=0.201 \;{\rm GeV}^5 ,\nonumber\\
&&|R_S(b\bar{b})(0)|^2=6.477\;{\rm GeV}^3 \;,\; |R'_P(b\bar{b})(0)|^2=1.417 \;{\rm GeV}^5 . \nonumber
\end{eqnarray}

As a cross-check, in addition to the improved trace technology, we also adopt the traditional trace technology for dealing with the mentioned processes. Numerically, we obtain a nice agreement between these two approaches for all the above mentioned decay channels and heavy-quarkonium states. Moreover, it is found that our numerical results for the color-singlet $S$-wave cases agree with those of Ref.\cite{w} under the same input values.

\subsection{Basic results}

As a reference, we calculate the decay widths for the basic processes $W^+\to c+\bar{s}$ and $W^+\to c+\bar{b}$. Their decay width can be written as
\begin{displaymath}
\Gamma =\frac{G_F |V_{CKM}|^{2}}{\sqrt{2}}\left[3\sqrt{(m_1^2 +|p|^2)(m^2_2 +|p|^2)} +|p|^2\right] ,
\end{displaymath}
where  $m_1 =m_c$, $m_2=m_s$ or $m_b$, and
\begin{displaymath}
|p| = \frac{\sqrt{(m_W^2 -(m_1 -m_2)^2)(m_W^2 -(m_1 + m_2)^2)}}{2 m_W} .
\end{displaymath}
$V_{CKM}=V_{cs}$ for $W^{+}\rightarrow c\bar{s}$ and $V_{CKM}=V_{cb}$ for $W^{+}\rightarrow c\bar{b}$ respectively. Then, we obtain $\Gamma_{W^+\to c+\bar{s}} = 713.6$ MeV and $\Gamma_{W^+\to c+\bar{b}} = 1.118$ MeV, the sum of which is about $34\%$ for the total width $\Gamma_{W^+}$.

\begin{table}
\begin{tabular}{|c||c|c|c|}
\hline
~~$W^+\rightarrow (c\bar{c})[n] +c\bar{s}$~~ & ~~$\Gamma$(KeV)~~ & ~~$\frac{\Gamma_{W^+\rightarrow (c\bar{c})[n]}}{\Gamma_{W^+\rightarrow c\bar{s}}}$~~ \\
\hline\hline
$W^+\rightarrow\eta_c$ & 174.8 & $2.45\times10^{-4}$ \\
\hline
$W^+\rightarrow J/\psi$ & 180.6 & $2.53\times10^{-4}$ \\
\hline
$W^+\rightarrow {|(c\bar{c})_{\bf 1}[^1P_1]\rangle}$ & 37.9 & $5.31\times10^{-5}$ \\
\hline
$W^+\rightarrow{|(c\bar{c})_{\bf 1}[^3P_0]\rangle}$ & 42.5 & $5.95\times10^{-5}$ \\
\hline
$W^+\rightarrow {|(c\bar{c})_{\bf 1}[^3P_1]\rangle}$ & 45.1 & $6.32\times10^{-5}$ \\
\hline
$W^+\rightarrow {|(c\bar{c})_{\bf 1}[^3P_2]\rangle}$& 39.9& $5.59\times10^{-5}$ \\
\hline
$W^+\rightarrow {|(c\bar{c})_{\bf 8}[^1S_0]\rangle}$& $21.9v^4$ & $3.07 v^4\times10^{-5}$ \\
\hline
$W^+\rightarrow {|(c\bar{c})_{\bf 8}[^3S_1]\rangle}$& $22.6 v^4$ & $3.17 v^4\times10^{-5}$ \\
\hline
\end{tabular}
\caption{Decay widths and branching fractions for the charmonium production through $W^+\rightarrow (c\bar{c})[n] +c\bar{s}$.}
\label{tabrpa}
\end{table}

\begin{table}
\begin{tabular}{|c||c|c|c|}
\hline
~~$W^+\rightarrow (c\bar{b})[n] +b\bar{s}$~~ & ~~$\Gamma$(KeV)~~ & ~~$\frac{\Gamma_{W^+\rightarrow (c\bar{b})[n]}}{\Gamma_{W^+\rightarrow c\bar{s}}}$~~ \\
\hline\hline
$W^+\rightarrow B_c $ & 6.32 & $8.86\times10^{-6}$ \\
\hline
$W^+\rightarrow B^*_c$ & 5.38 & $7.54\times10^{-6}$ \\
\hline
$W\rightarrow {|(c\bar{b})_{\bf 1}[^1P_1]\rangle}$ & 0.300 & $4.20\times10^{-7}$ \\
\hline
$W\rightarrow{|(c\bar{b})_{\bf 1}[^3P_0]\rangle}$ & 0.851 & $1.19\times10^{-6}$ \\
\hline
$W\rightarrow {|(c\bar{b})_{\bf 1}[^3P_1]\rangle}$ & 0.583 & $8.17\times10^{-7}$ \\
\hline
$W\rightarrow {|(c\bar{b})_{\bf 1}[^3P_2]\rangle}$& 0.0326 & $4.57\times10^{-8}$ \\
\hline
$W\rightarrow {|(c\bar{b})_{\bf 8}[^1S_0]\rangle}$& $0.789 v^4$ & $1.11 v^4\times10^{-6}$ \\
\hline
$W\rightarrow {|(c\bar{b})_{\bf 8}[^3S_1]\rangle}$& $0.672 v^4$ & $0.94 v^4\times10^{-6}$ \\
\hline
\end{tabular}
\caption{Decay widths and branching fractions for the $(c\bar{b})$-quarkonium production through $W^+\rightarrow (c\bar{b})[n] +b\bar{s}$.}
\label{tabrpb}
\end{table}

\begin{table}
\begin{tabular}{|c||c|c|c|}
\hline
~~$W^+\rightarrow (c\bar{b})[n] +c\bar{c}$~~ & ~~$\Gamma$(KeV)~~ & ~~$\frac{\Gamma_{W\rightarrow (c\bar{b})[n]}} {\Gamma_{W\rightarrow c\bar{b}}}$~~ \\
\hline\hline
$W^+\rightarrow B_c$ & 0.546 & $4.88\times10^{-4}$ \\
\hline
$W^+\rightarrow B^*_c$ & 0.810 & $7.25\times10^{-4}$ \\
\hline
$W\rightarrow {|(c\bar{b})_{\bf 1}[^1P_1]\rangle}$ & 0.170 & $1.52\times10^{-4}$ \\
\hline
$W\rightarrow{|(c\bar{b})_{\bf 1}[^3P_0]\rangle}$ & 0.037 & $3.31\times10^{-5}$ \\
\hline
$W\rightarrow {|(c\bar{b})_{\bf 1}[^3P_1]\rangle}$ & 0.079 & $7.02\times10^{-5}$ \\
\hline
$W\rightarrow {|(c\bar{b})_{\bf 1}[^3P_2]\rangle}$& 0.094 & $8.36\times10^{-5}$ \\
\hline
$W\rightarrow {|(c\bar{b})_{\bf 8}[^1S_0]\rangle}$& $0.068v^4$ & $6.11 v^4\times10^{-5}$ \\
\hline
$W\rightarrow {|(c\bar{b})_{\bf 8}[^3S_1]\rangle}$& $0.101v^4$ & $9.03 v^4\times10^{-5}$ \\
\hline
\end{tabular}
\caption{Decay widths and branching fractions for the $(c\bar{b})$-quarkonium production through $W^+\rightarrow (c\bar{b})[n] +c\bar{c}$.}
\label{tabrpc}
\end{table}

\begin{table}
\begin{tabular}{|c||c|c|c|}
\hline
~~$W^+\rightarrow (b\bar{b})[n] + c\bar{c}$~~ & ~~$\Gamma$(eV)~~ &~~ $\frac{\Gamma_{W\rightarrow (b\bar{b})[n]}} {\Gamma_{W\rightarrow c\bar{b}}}$~~ \\
\hline\hline
$W^+\rightarrow\eta_b$ & 17.6 & $1.57\times10^{-5}$ \\
\hline
$W^+\rightarrow \Upsilon$ & 18.6 & $1.66\times10^{-5}$ \\
\hline
$W\rightarrow {|(b\bar{b})_{\bf 1}[^1P_1]\rangle}$ & 0.469 & $4.19\times10^{-7}$ \\
\hline
$W\rightarrow{|(b\bar{b})_{\bf 1}[^3P_0]\rangle}$ & 0.771 & $6.90\times10^{-7}$ \\
\hline
$W\rightarrow {|(b\bar{b})_{\bf 1}[^3P_1]\rangle}$ & 0.821 & $7.34\times10^{-7}$ \\
\hline
$W\rightarrow {|(b\bar{b})_{\bf 1}[^3P_2]\rangle}$& 0.284 & $2.54\times10^{-7}$ \\
\hline
$W\rightarrow {|(b\bar{b})_{\bf 8}[^1S_0]\rangle}$& $2.20 v^4$ & $1.97 v^4\times10^{-6}$ \\
\hline
$W\rightarrow {|(b\bar{b})_{\bf 8}[^3S_1]\rangle}$& $2.33 v^4$ & $2.08 v^4\times10^{-6}$ \\
\hline
\end{tabular}
\caption{Decay widths and branching fractions for the bottomonium production through $W^+\rightarrow (b\bar{b})[n] + c\bar{c}$.}
\label{tabrpd}
\end{table}

Total decay widths and their branching fractions for the typical channels of $W^{+} \to (Q\bar{Q'})[n] +q + \bar{q}'$ are listed in TABs.(\ref{tabrpa},\ref{tabrpb},\ref{tabrpc},\ref{tabrpd}). It is found that the squared amplitude for the color-octet decay width is suppressed by eight times to that of the color-singlet case. As a combined effect of such color suppression and the relative velocity suppression ($v^4$-suppression), the color-octet channels are quite small in comparison to their corresponding color-singlet production channels as shown by TABs.(\ref{tabrpa},\ref{tabrpb},\ref{tabrpc},\ref{tabrpd}) explicitly. So in the following discussion, if not specially stated, we shall not include the color-octet states' contributions. We should point out that for the processes with a much more complicated color structures, due to the cancellation and enhancement of different color structures of the heavy-quarkonium, those color states other than the color-singlet state may also give sizable contributions. Two such examples for the direct hadronic production of $B_c$ and $\Xi_{cc}$ can be found in Refs.\cite{bc3,bcvegpy,genxicc,xicc}, where the color-octet $(c\bar{b})$-quarkonium and the color-sextuplet $(cc)$-diquark can provide sizable contributions up to $\sim 10\%-20\%$ to the final meson/baryon production cross section.

For the charmonium production channel $W^{+}\rightarrow (c\bar{c})[n] + c \bar{s}$, its total decay width for all the $P$-wave states is $165$ KeV, which is comparable to that of $\eta_c$ or $J/\psi$, i.e. it is about $95\%$ ($92\%$) of that of $\eta_c$ ($J/\psi$). For the $(c\bar{b})$-quarkonium production, the total decay width for all the $P$-wave states is about $28\%$ ($33\%$) of that of $B_c$ ($B^*_c$) for $W^{+}\rightarrow (c\bar{b})[n] + b\bar{s}$; and is about $69\%$ ($47\%$) of that of $B_c$ ($B^*_c$) for $W^{+}\rightarrow (c\bar{b})[n] + c\bar{c}$. Note even though $W^{+}\rightarrow (c\bar{b})[n] + c\bar{c}$ is CKM suppressed to $W^{+}\rightarrow (c\bar{b})[n] + b\bar{s}$ by $|V_{cb}|^2/|V_{cs}|^2 \sim 0.2\%$, it is enhanced by the phase space, since it is easier to generate a $(c\bar{c})$-pair than a $(b\bar{b})$-pair. So as a combined result, the decay width of $W^{+}\rightarrow (c\bar{b})[n] + c\bar{c}$ is smaller than that of $W^{+}\rightarrow (c\bar{b})[n] + b\bar{s}$ by only $1/10$. For the bottomonium production, the total decay width for all the $P$-wave states is about $13\%$ of that of $\eta_b$ or $\Upsilon$. Sizable decay width for the $P$-wave quarkonium states shows that one needs to take the $P$-wave states into consideration for a sound estimation, especially for the channels of the charmonium and the $(c\bar{b})$-quarkonium.

\begin{figure}[!htb]
\includegraphics[width=0.38\textwidth]{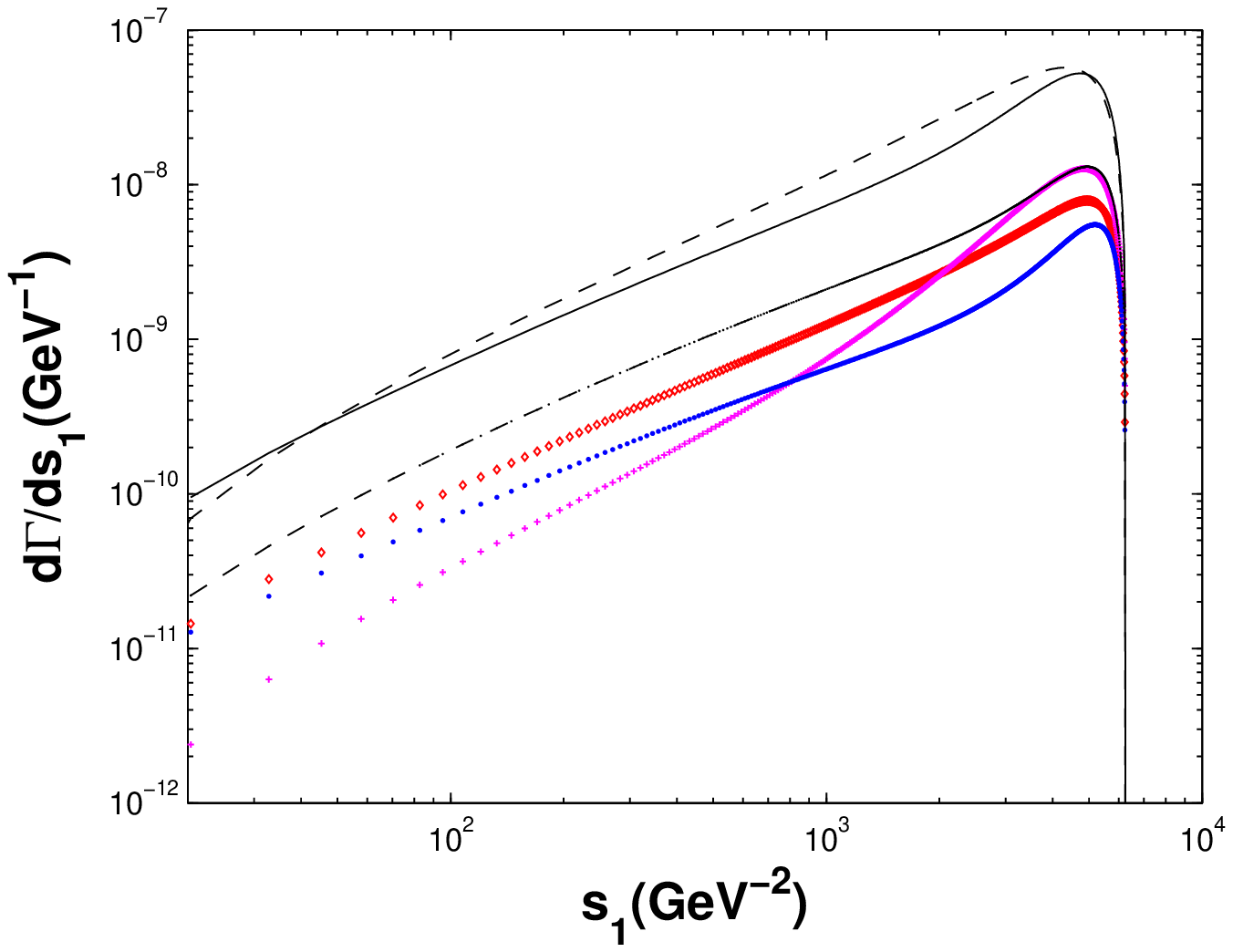}
\includegraphics[width=0.38\textwidth]{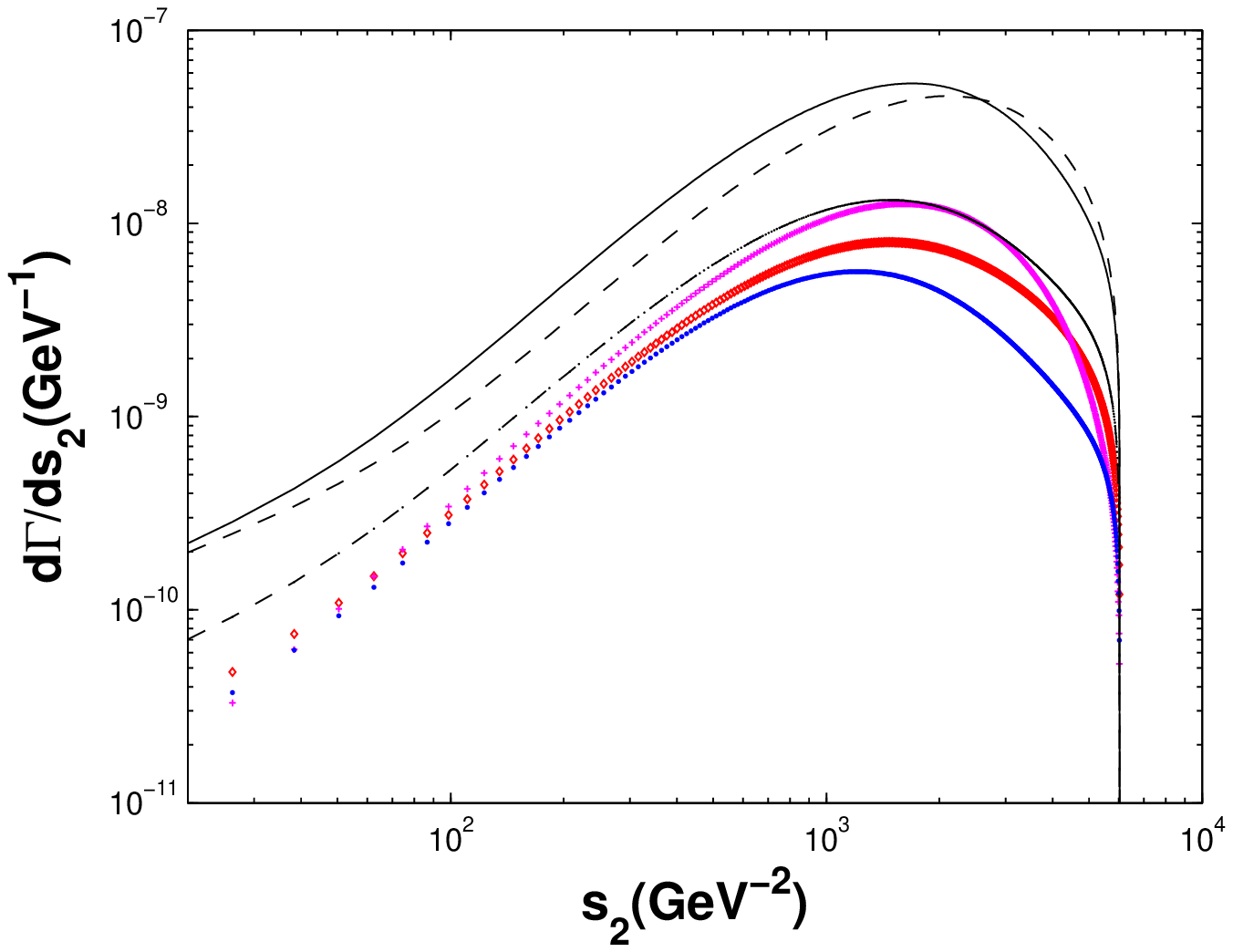}
\caption{Differential decay widths $d\Gamma/ds_1$ (Up) and $d\Gamma/ds_2$ (Down) for $W^+\rightarrow (c\bar{c})[n] +c\bar{s}$, where the dashed line, the solid line, the diamond line, the crossed line, the dash-dot line and the dotted line are for $|(c\bar{c})_{\bf 1}[^1S_0]\rangle$, $|(c\bar{c})_{\bf 1}[^3S_1]\rangle$, $|(c\bar{c})_{\bf 1}[^1P_1]\rangle$, $|(c\bar{c})_{\bf 1}[^3P_0]\rangle$, $|(c\bar{c})_{\bf 1}[^3P_1]\rangle$ and $|(c\bar{c})_{\bf 1}[^3P_2]\rangle$ respectively. } \label{CCcsdiss1s2}
\end{figure}

\begin{figure}[!htb]
\includegraphics[width=0.38\textwidth]{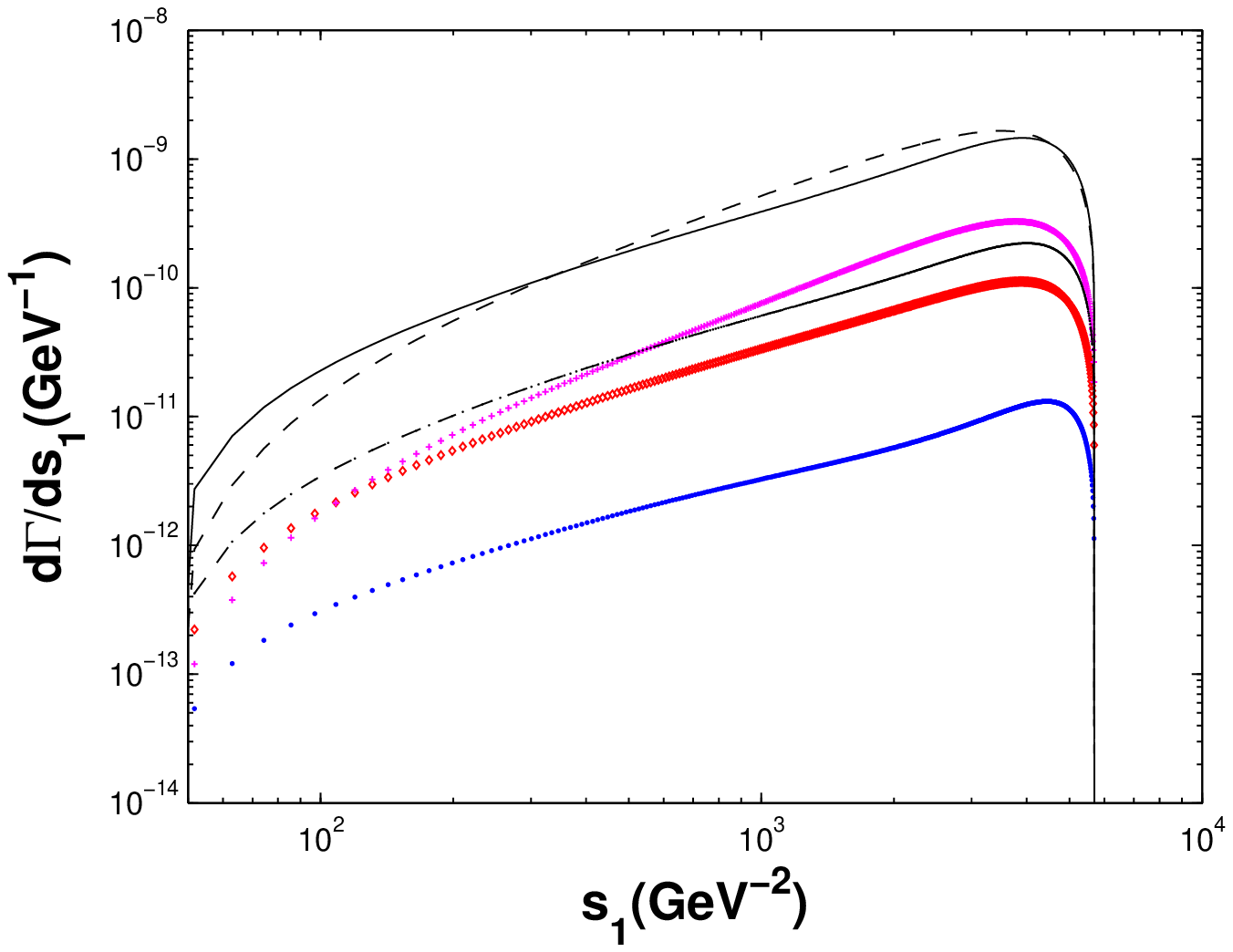}
\includegraphics[width=0.38\textwidth]{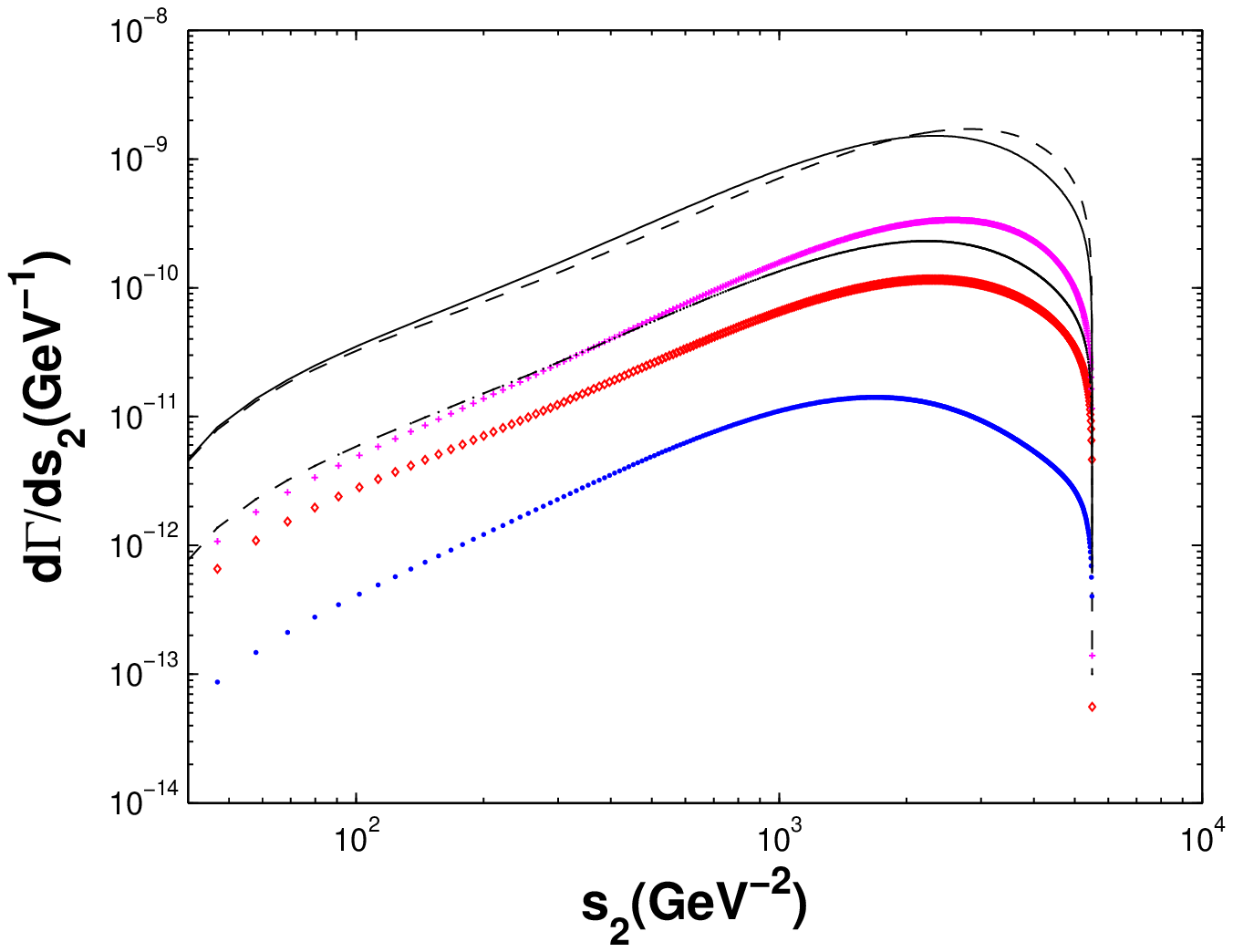}
\caption{Differential decay widths $d\Gamma/ds_1$ (Up) and $d\Gamma/ds_2$ (Down) for $W^+\rightarrow (c\bar{b})[n] +b\bar{s}$, where the dashed line, the solid line, the diamond line, the crossed line, the dash-dot line and the dotted line are for $|(c\bar{b})_{\bf 1}[^1S_0]\rangle$, $|(c\bar{b})_{\bf 1}[^3S_1]\rangle$, $|(c\bar{b})_{\bf 1}[^1P_1]\rangle$, $|(c\bar{b})_{\bf 1}[^3P_0]\rangle$, $|(c\bar{b})_{\bf 1}[^3P_1]\rangle$ and $|(c\bar{b})_{\bf 1}[^3P_2]\rangle$ respectively. } \label{W+Bcbsds}
\end{figure}

\begin{figure}[!htb]
\includegraphics[width=0.38\textwidth]{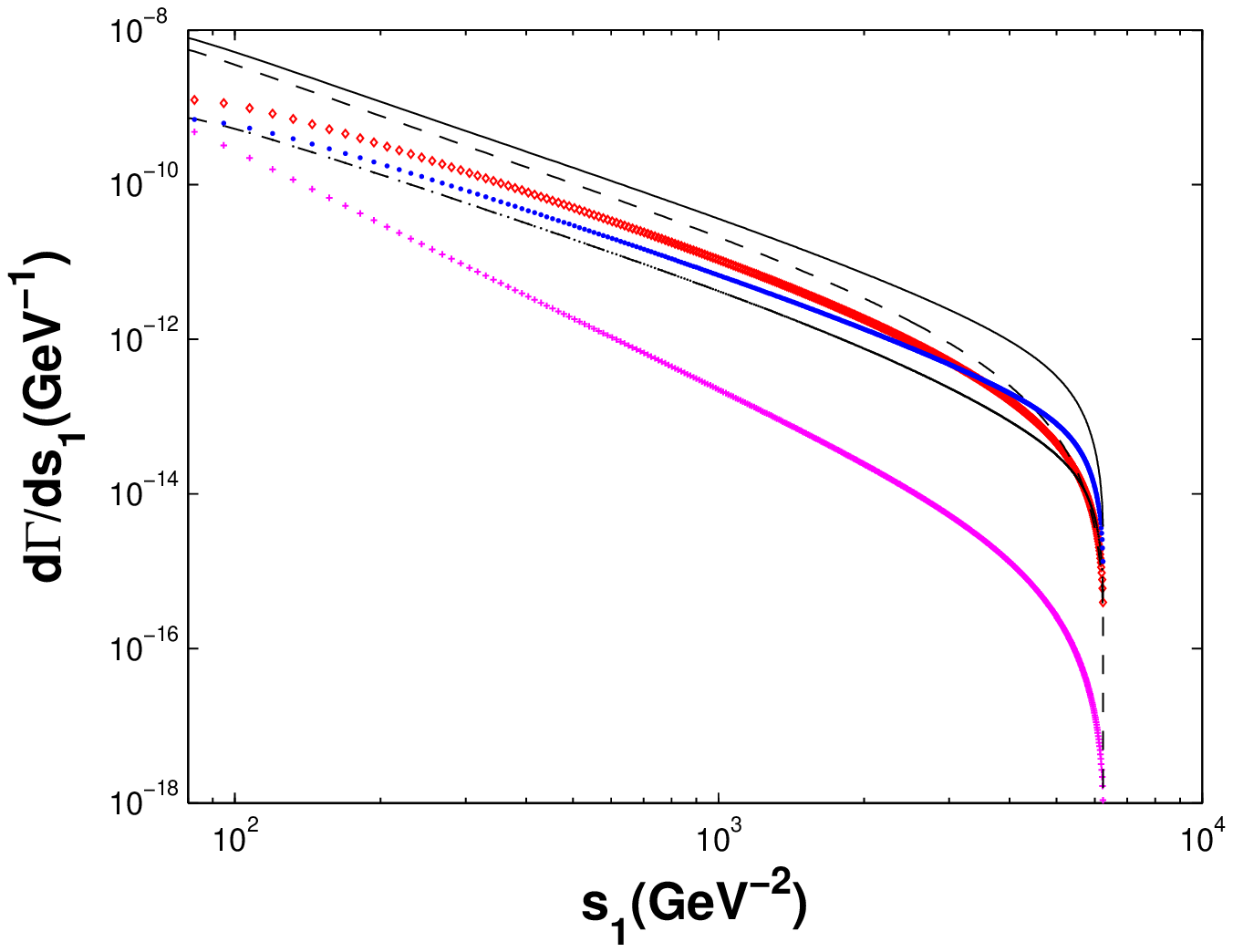}
\includegraphics[width=0.38\textwidth]{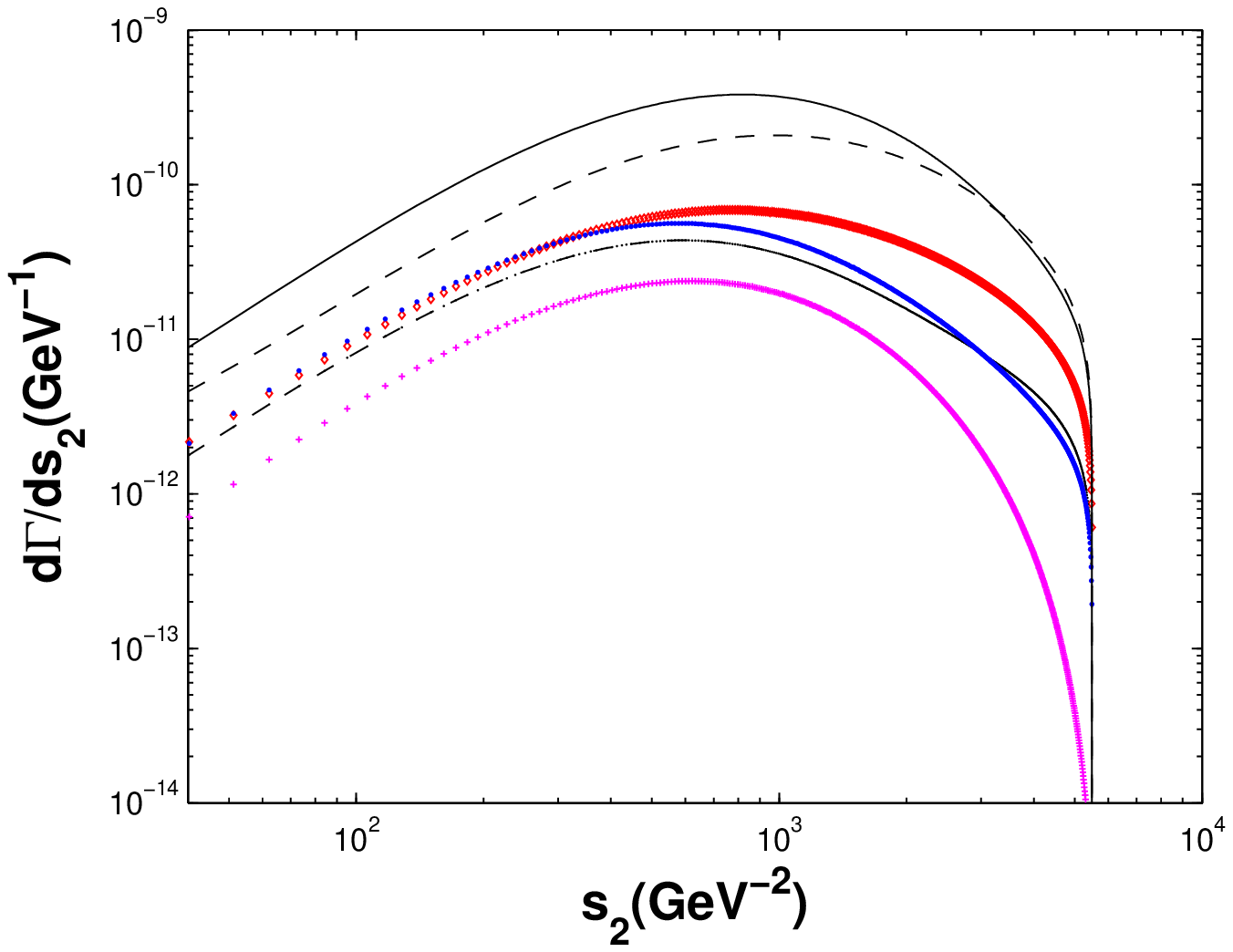}
\caption{Differential decay widths $d\Gamma/ds_1$ (Up) and $d\Gamma/ds_2$ (Down) for $W^+\rightarrow (c\bar{b})[n]+ c\bar{c}$, where the dashed line, the solid line, the diamond line, the crossed line, the dash-dot line and the dotted line are for $|(c\bar{b})_{\bf 1}[^1S_0]\rangle$, $|(c\bar{b})_{\bf 1}[^3S_1]\rangle$, $|(c\bar{b})_{\bf 1}[^1P_1]\rangle$, $|(c\bar{b})_{\bf 1}[^3P_0]\rangle$, $|(c\bar{b})_{\bf 1}[^3P_1]\rangle$ and $|(c\bar{b})_{\bf 1}[^3P_2]\rangle$ respectively. } \label{W+Bcccds}
\end{figure}

\begin{figure}[!htb]
\includegraphics[width=0.38\textwidth]{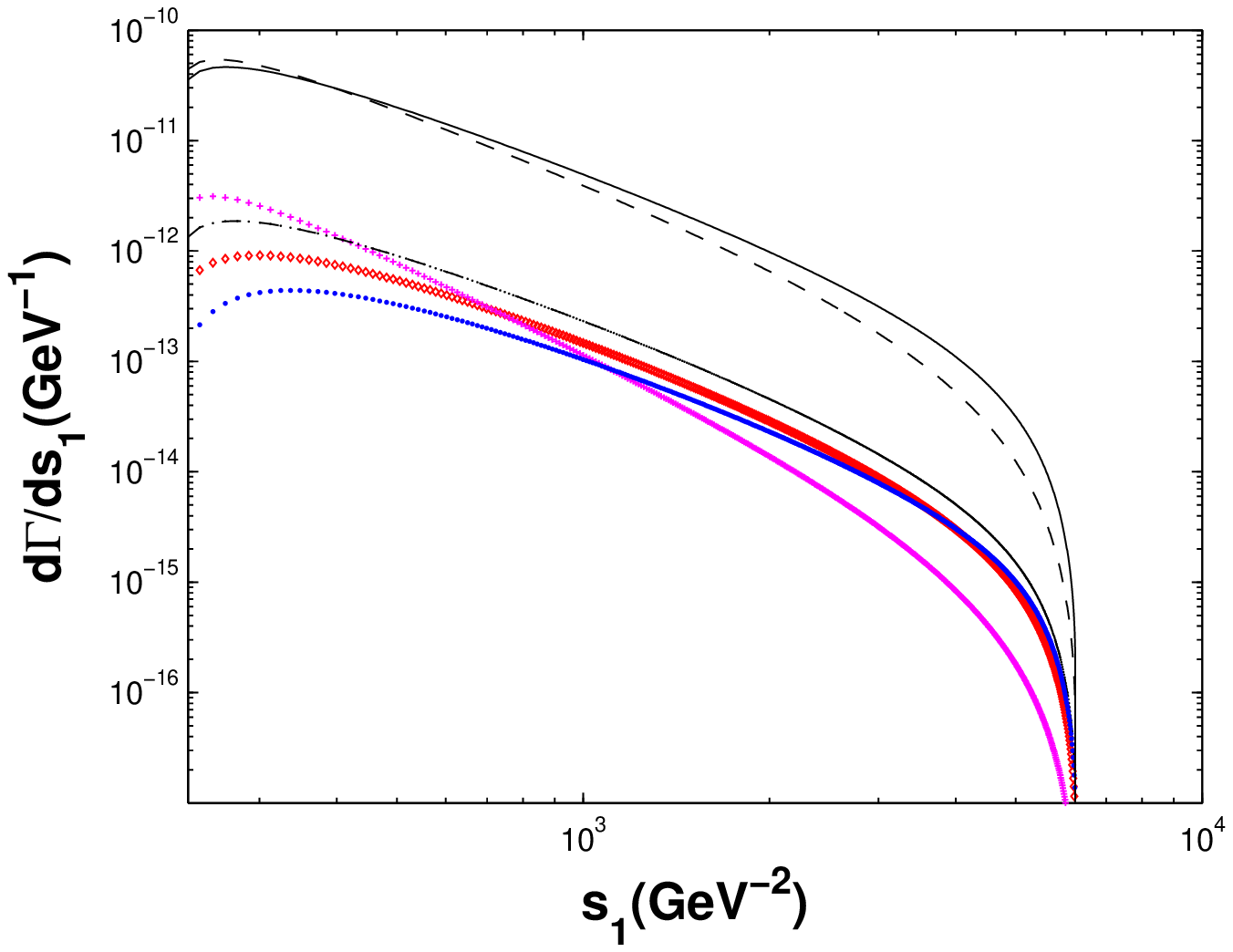}
\includegraphics[width=0.38\textwidth]{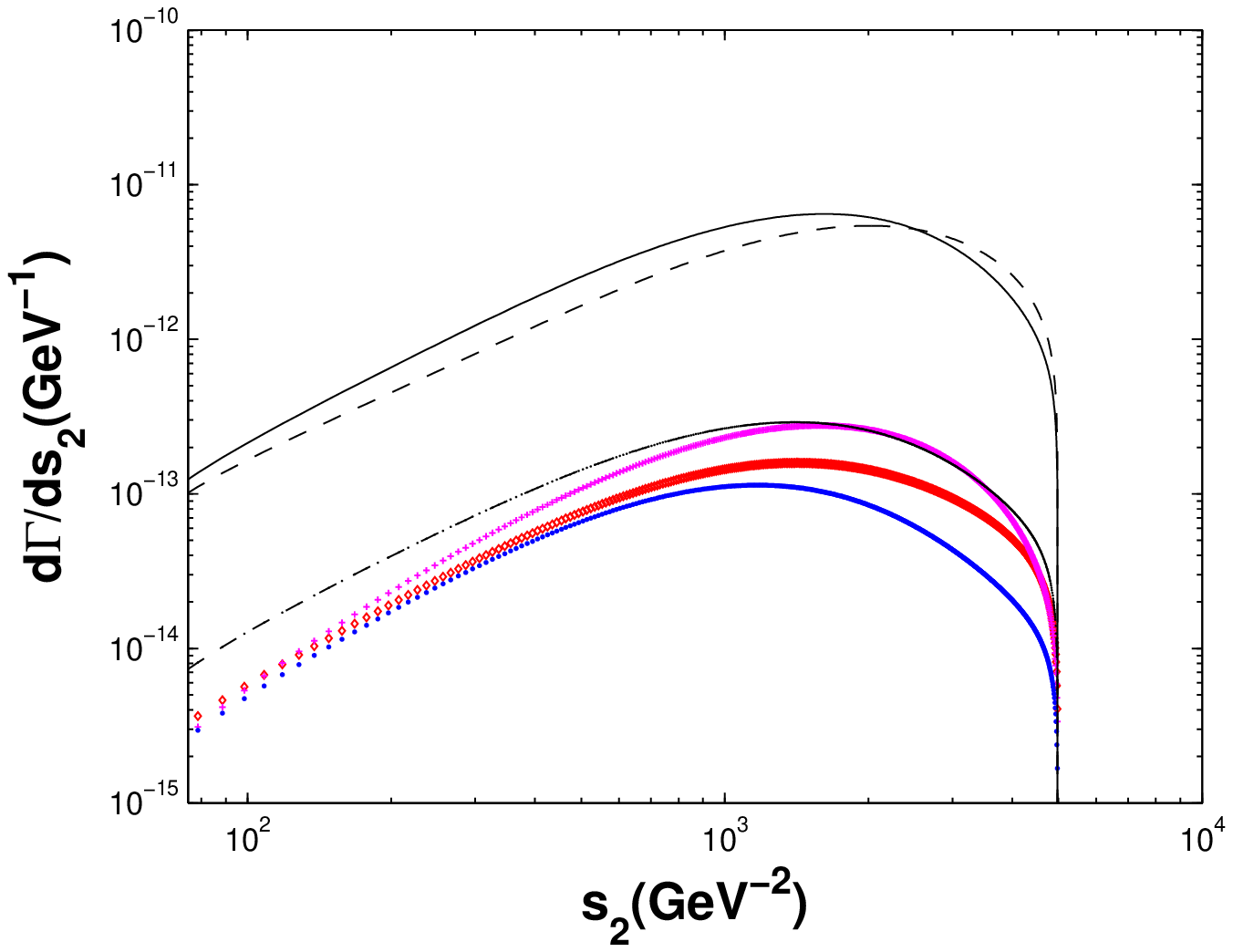}
\caption{Differential decay widths $d\Gamma/ds_1$ (Up) and $d\Gamma/ds_2$ (Down) for $W^+\rightarrow (b\bar{b})[n]+ c\bar{b}$, where the dashed line, the solid line, the diamond line, the crossed line, the dash-dot line and the dotted line are for $|(b\bar{b})_{\bf 1}[^1S_0]\rangle$, $|(b\bar{b})_{\bf 1}[^3S_1]\rangle$, $|(b\bar{b})_{\bf 1}[^1P_1]\rangle$, $|(b\bar{b})_{\bf 1}[^3P_0]\rangle$, $|(b\bar{b})_{\bf 1}[^3P_1]\rangle$ and $|(b\bar{b})_{\bf 1}[^3P_2]\rangle$ respectively. } \label{rbcdiss1s2}
\end{figure}

\begin{figure}[!htb]
\includegraphics[width=0.38\textwidth]{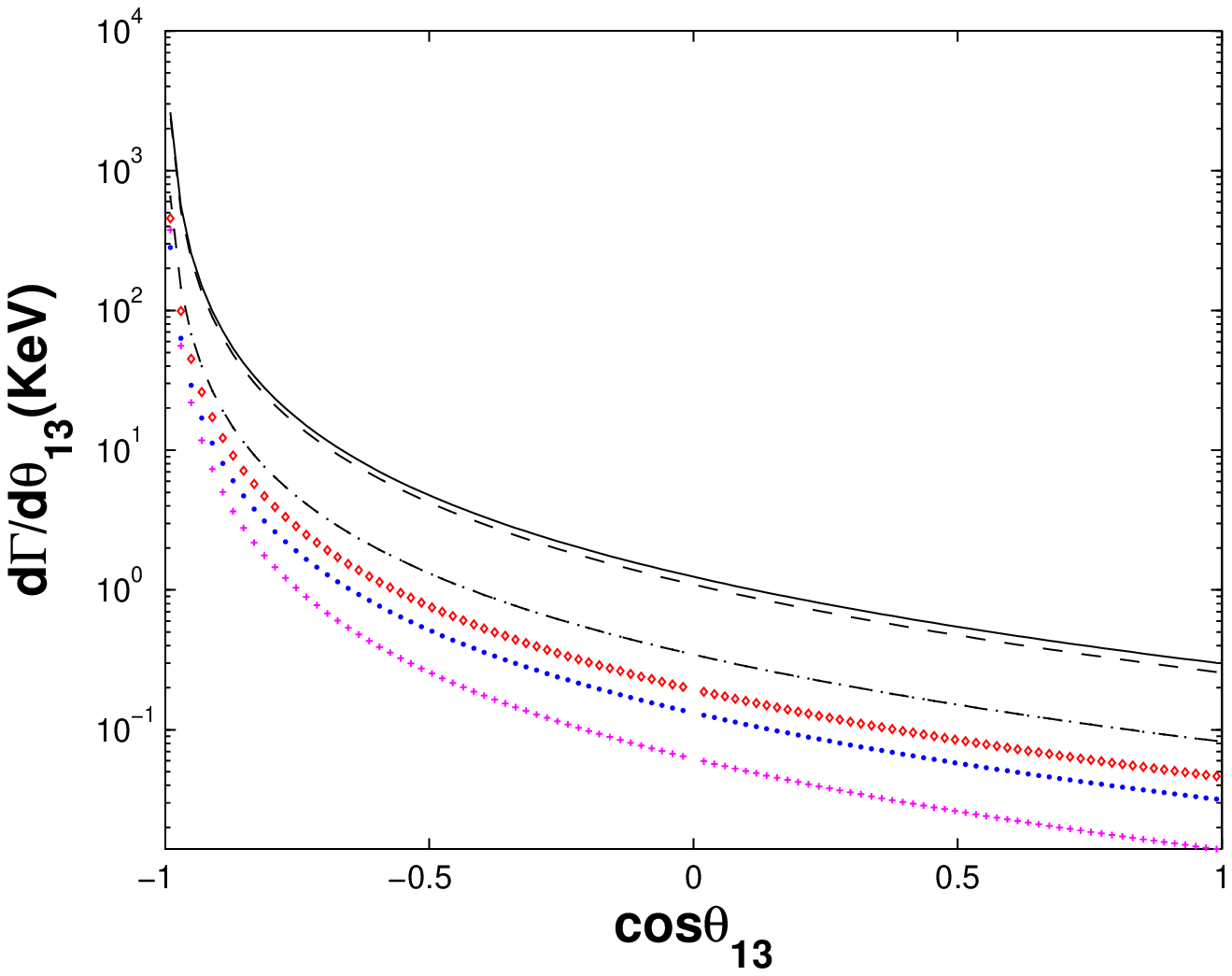}
\hspace{0.2cm}
\includegraphics[width=0.38\textwidth]{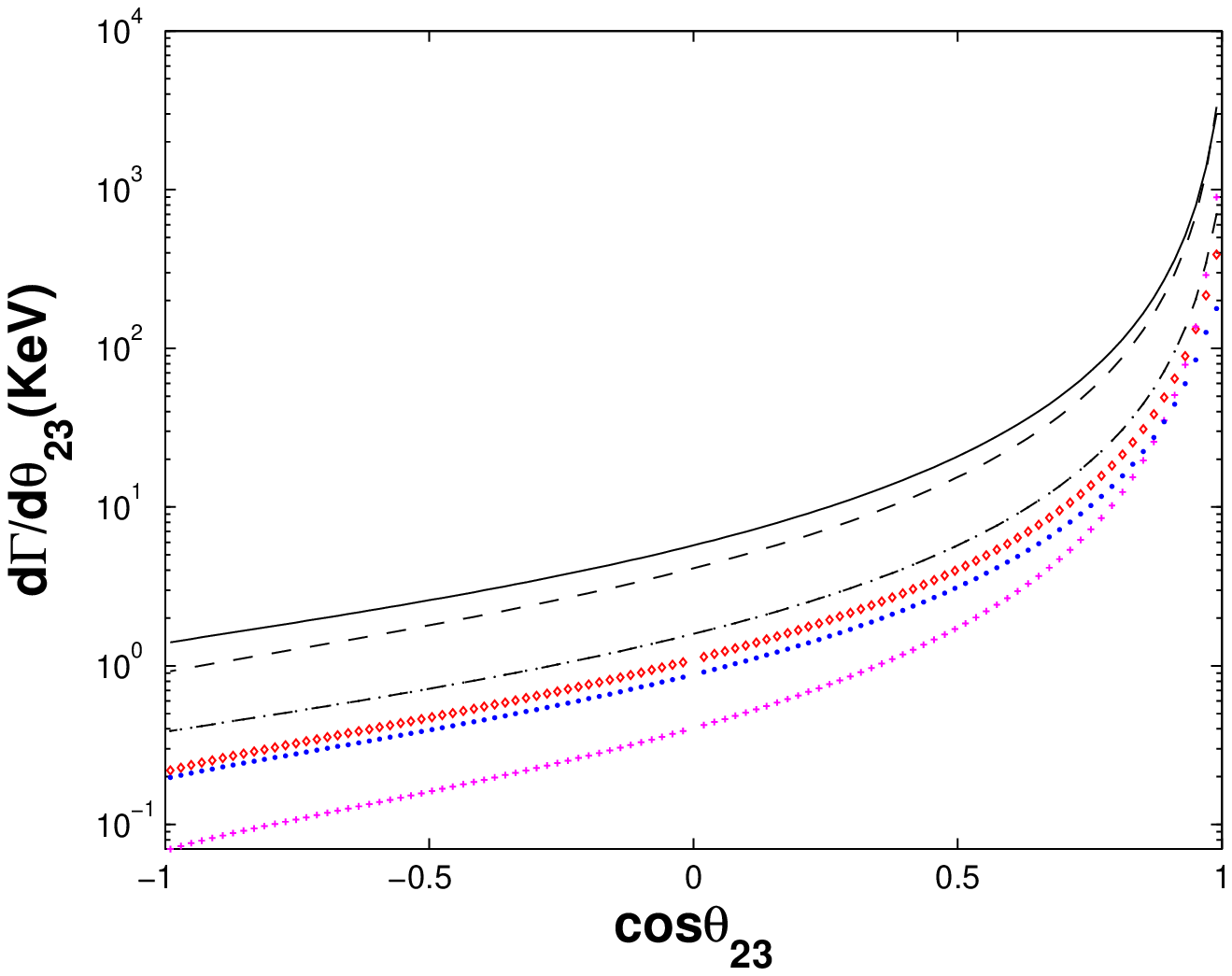}
\caption{Differential decay widths $d\Gamma/d\cos{\theta_{13}}$ (Up) and $d\Gamma/d\cos{\theta_{23}}$ (Down) for $W^+\rightarrow (c\bar{c})[n] +c\bar{s}$, where the dashed line, the solid line, the diamond line, the crossed line, the dash-dot line and the dotted line are for $|(c\bar{c})_{\bf 1}[^1S_0]\rangle$, $|(c\bar{c})_{\bf 1}[^3S_1]\rangle$, $|(c\bar{c})_{\bf 1}[^1P_1]\rangle$, $|(c\bar{c})_{\bf 1}[^3P_0]\rangle$, $|(c\bar{c})_{\bf 1}[^3P_1]\rangle$ and $|(c\bar{c})_{\bf 1}[^3P_2]\rangle$ respectively.} \label{CCcsdiscos}
\end{figure}

\begin{figure}[!htb]
\includegraphics[width=0.38\textwidth]{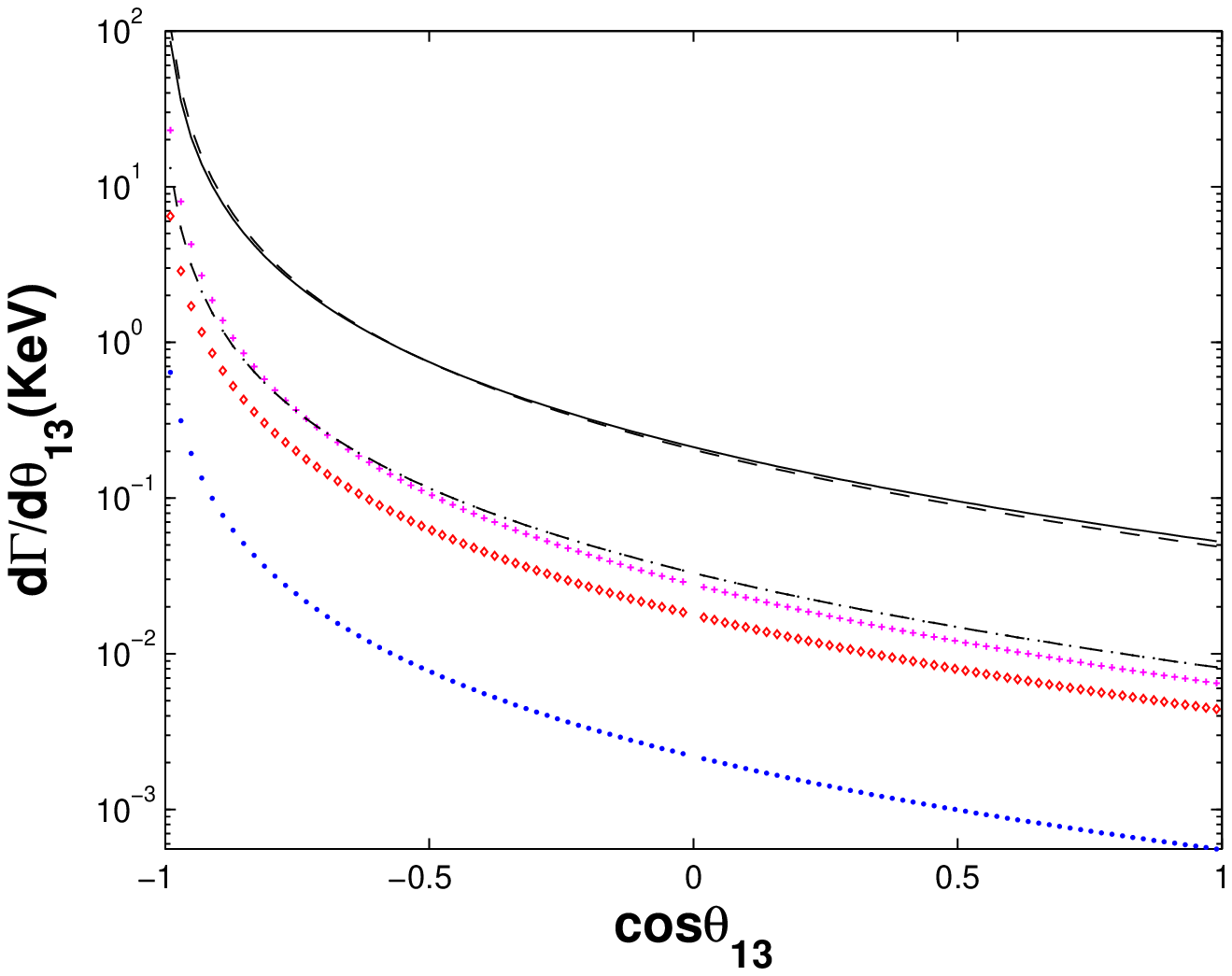}
\hspace{0.2cm}
\includegraphics[width=0.38\textwidth]{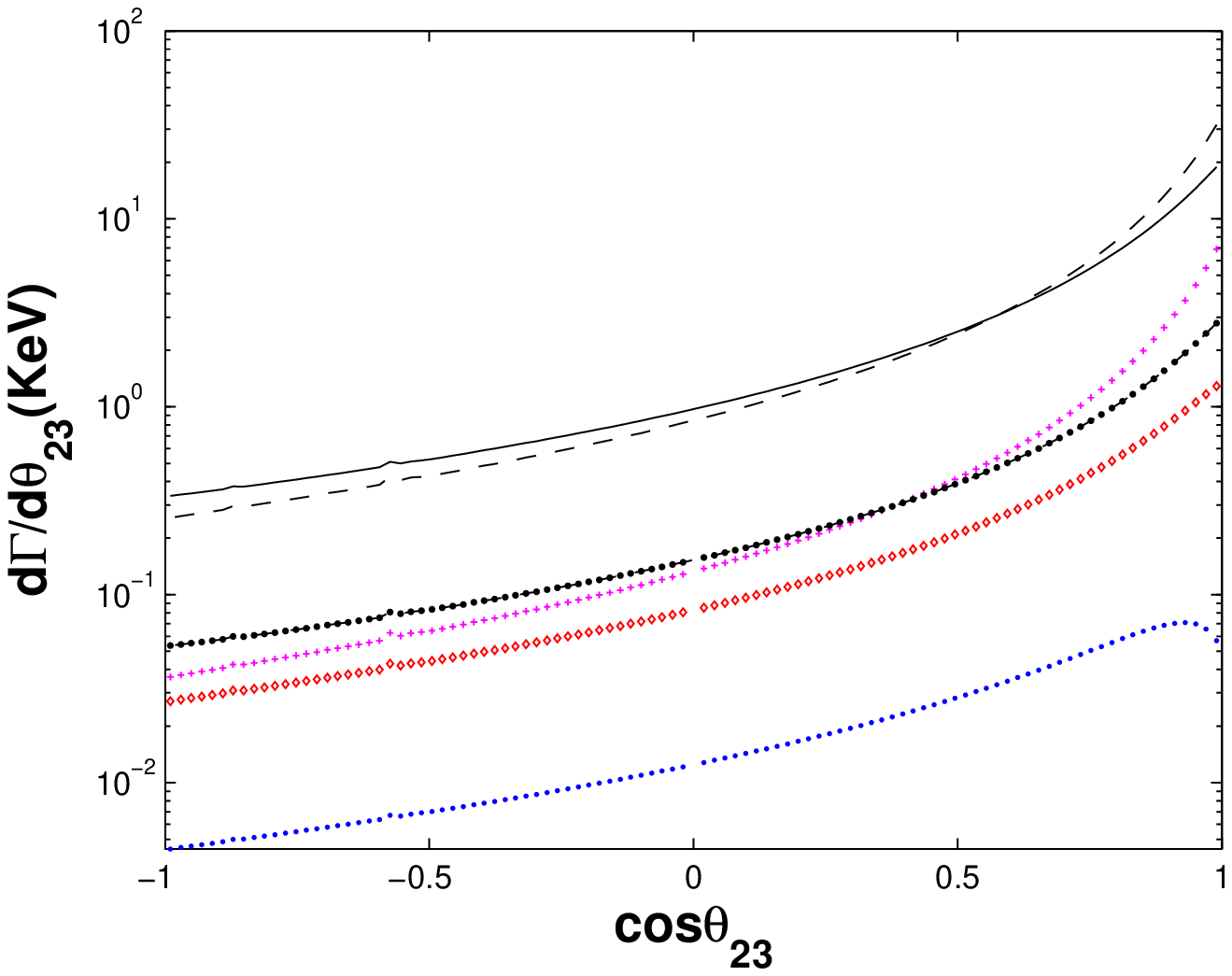}
\caption{Differential decay widths $d\Gamma/d\cos{\theta_{13}}$ (Up) and $d\Gamma/d\cos{\theta_{23}}$ (Down) for $W^+\rightarrow (c\bar{b})[n] +b\bar s$, where the dashed line, the solid line, the diamond line, the crossed line, the dash-dot line and the dotted line are for $|(c\bar{b})_{\bf 1}[^1S_0]\rangle$, $|(c\bar{b})_{\bf 1}[^3S_1]\rangle$, $|(c\bar{b})_{\bf 1}[^1P_1]\rangle$, $|(c\bar{b})_{\bf 1}[^3P_0]\rangle$, $|(c\bar{b})_{\bf 1}[^3P_1]\rangle$ and $|(c\bar{b})_{\bf 1}[^3P_2]\rangle$ respectively.} \label{W+Bcbsdcos}
\end{figure}

\begin{figure}[!htb]
\includegraphics[width=0.38\textwidth]{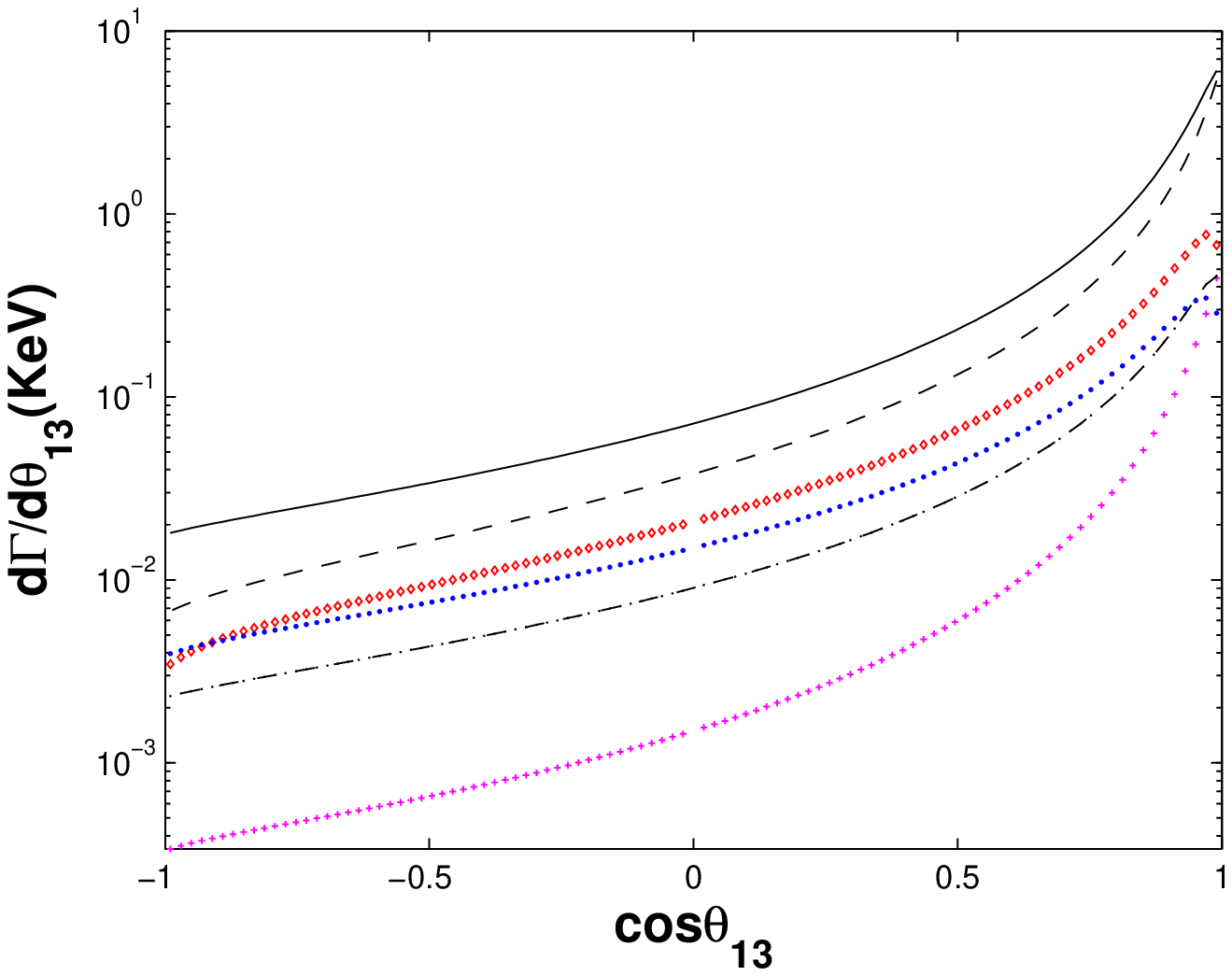}
\hspace{0.2cm}
\includegraphics[width=0.38\textwidth]{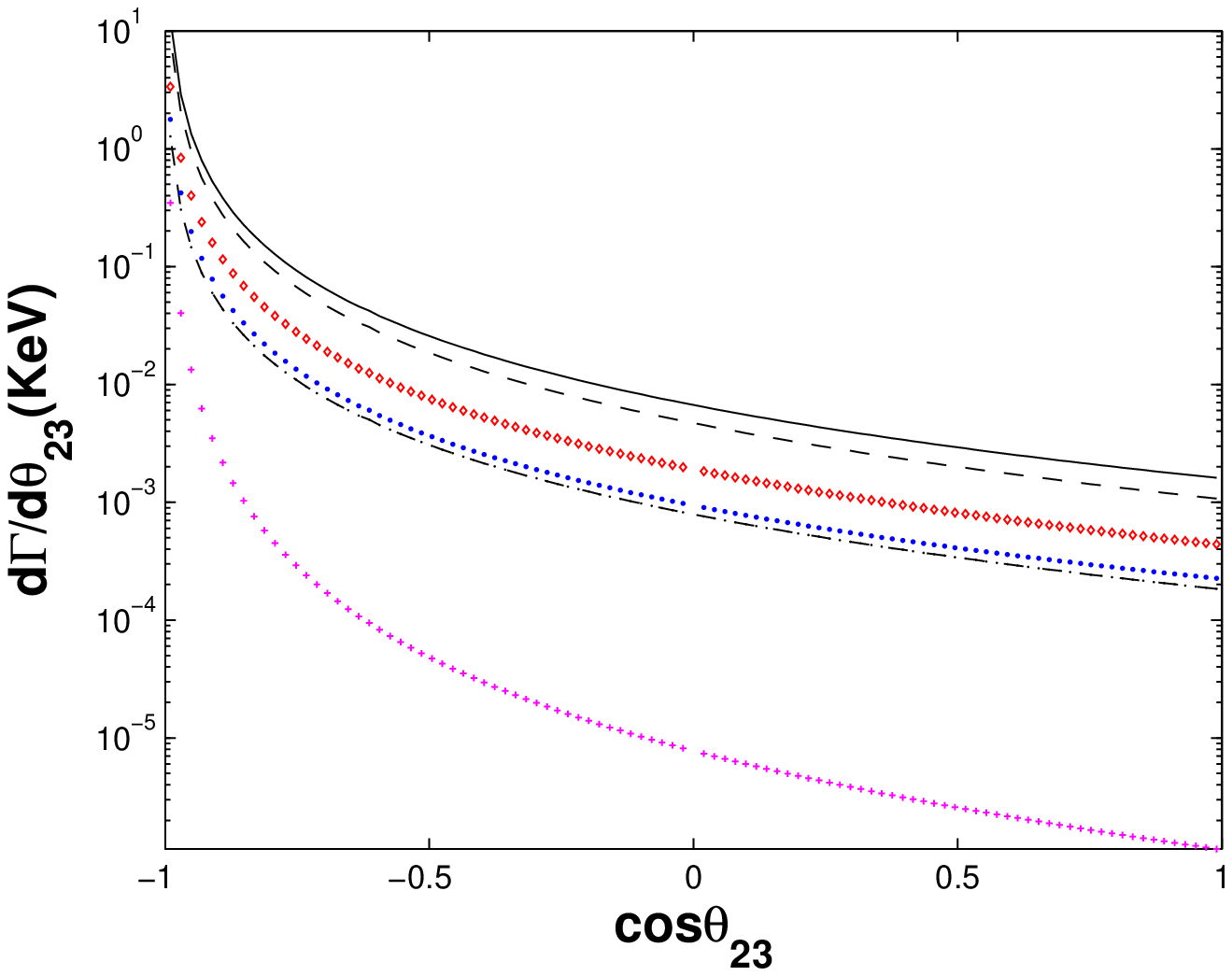}
\caption{Differential decay widths $d\Gamma/d\cos{\theta_{13}}$ (Up) and $d\Gamma/d\cos{\theta_{23}}$ (Down) for $W^+\rightarrow (c\bar{b})[n] +c\bar{c}$, where the dashed line, the solid line, the diamond line, the crossed line, the dash-dot line and the dotted line are for $|(c\bar{b})_{\bf 1}[^1S_0]\rangle$, $|(c\bar{b})_{\bf 1}[^3S_1]\rangle$, $|(c\bar{b})_{\bf 1}[^1P_1]\rangle$, $|(c\bar{b})_{\bf 1}[^3P_0]\rangle$, $|(c\bar{b})_{\bf 1}[^3P_1]\rangle$ and $|(c\bar{b})_{\bf 1}[^3P_2]\rangle$ respectively.} \label{W+Bcccdcos}
\end{figure}

\begin{figure}[!htb]
\includegraphics[width=0.38\textwidth]{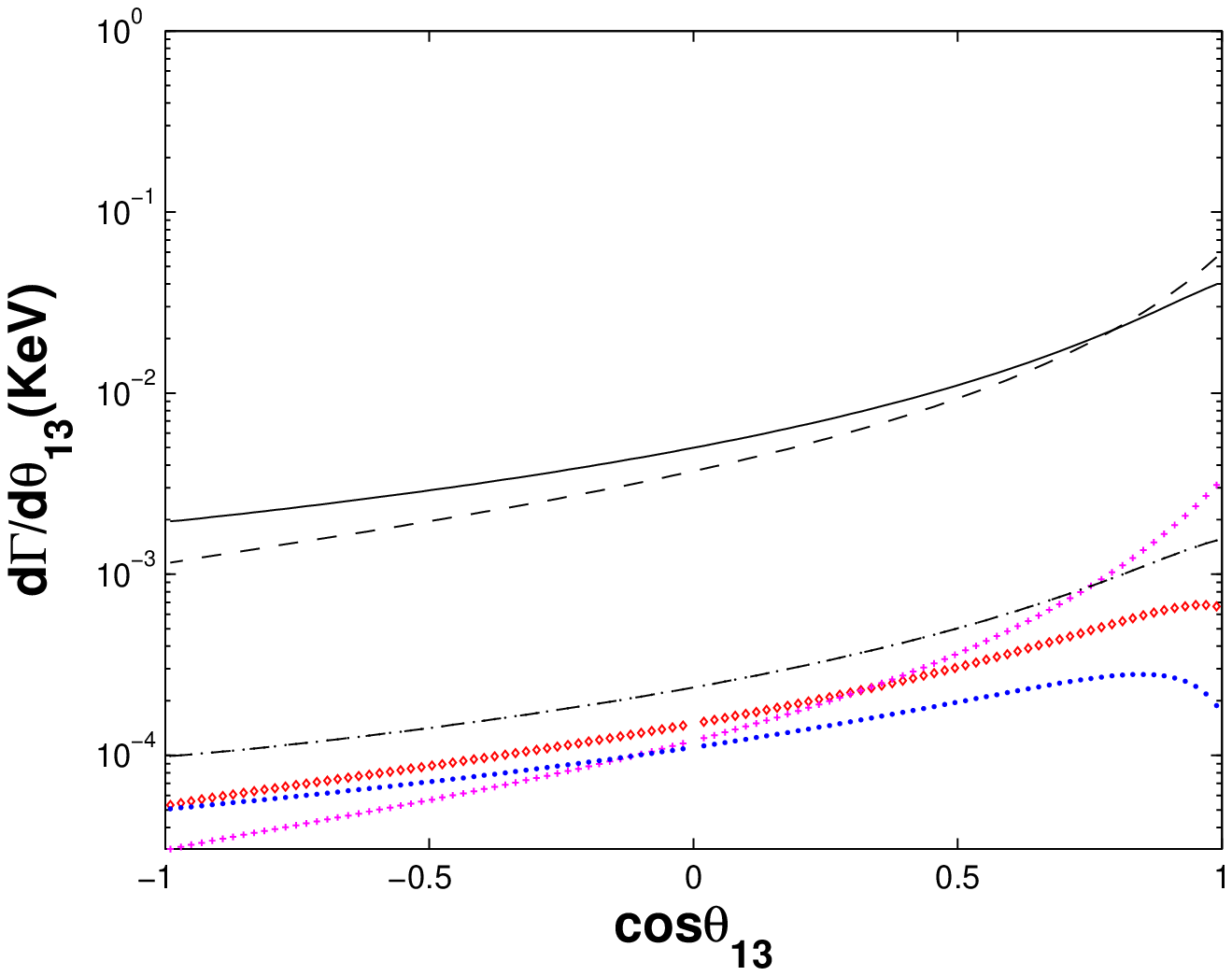}
\hspace{0.2cm}
\includegraphics[width=0.38\textwidth]{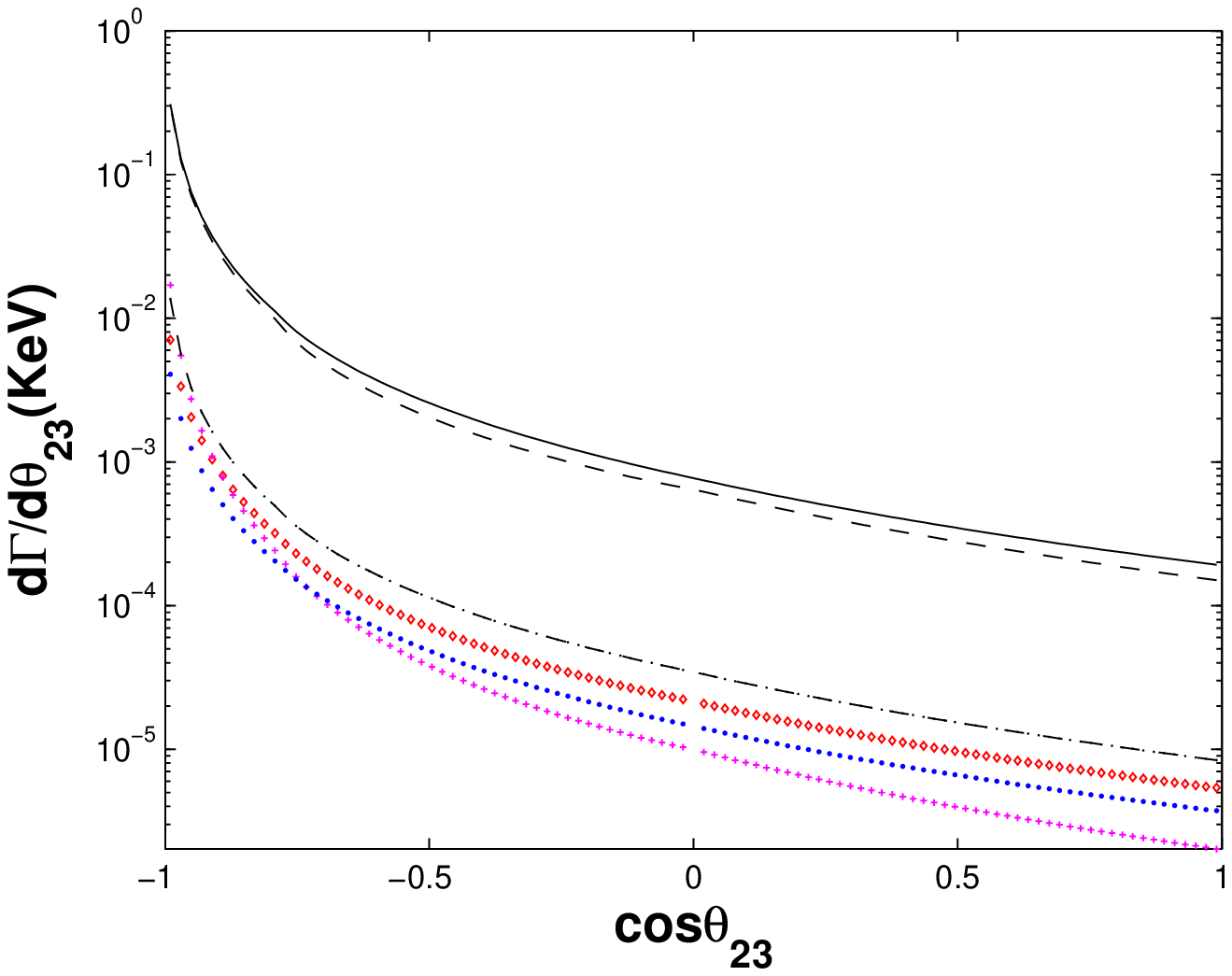}
\caption{Differential decay widths $d\Gamma/d\cos{\theta_{13}}$ (Up) and $d\Gamma/d\cos{\theta_{23}}$ (Down) for $W^+\rightarrow (b\bar{b})[n] +c\bar{b}$, where the dashed line, the solid line, the diamond line, the crossed line, the dash-dot line and the dotted line are for $|(b\bar{b})_{\bf 1}[^1S_0]\rangle$, $|(b\bar{b})_{\bf 1}[^3S_1]\rangle$, $|(b\bar{b})_{\bf 1}[^1P_1]\rangle$, $|(b\bar{b})_{\bf 1}[^3P_0]\rangle$, $|(b\bar{b})_{\bf 1}[^3P_1]\rangle$ and $|(b\bar{b})_{\bf 1}[^3P_2]\rangle$ respectively.} \label{W+rbcdiscos}
\end{figure}

To show the relative importance among different Fock states more clearly, we present the differential distributions $d\Gamma/ds_1$ and $d\Gamma/ds_2$ in Figs.(\ref{CCcsdiss1s2},\ref{W+Bcbsds},\ref{W+Bcccds},\ref{rbcdiss1s2}), and the differential distributions $d\Gamma/d\cos\theta_{13}$ and $d\Gamma/d\cos\theta_{23}$ in Figs.(\ref{CCcsdiscos},\ref{W+Bcbsdcos},\ref{W+Bcccdcos},\ref{W+rbcdiscos}). Here two invariant variables, $s_1=(q_1+q_3)^2$ and $s_2=(q_1+q_2)^2$, $\theta_{13}$ stands for the angle between $\vec{q}_1$ and $\vec{q}_3$, and $\theta_{23}$ stands for the angle between $\vec{q}_2$ and $\vec{q}_3$ in the $W^+$-rest frame. The curves for $|(Q\bar{Q}')_{\bf 1}[^1S_0]\rangle$, $|(Q\bar{Q}')_{\bf 1}[^3S_1]\rangle$, $|(Q\bar{Q}')_{\bf 1}[^1P_1]\rangle$ and $|(Q\bar{Q}')_{\bf 1}[^3P_J]\rangle$ are presented. Since the difference between the color-singlet $S$-wave states and the color-octet $S$-wave states is an overall color factor, the shapes of their curves are the same, so we do not present the curves for the color-octet ones in these figures.

As for the decay channels $W^{+}\rightarrow (c\bar{c})[n] + c \bar{s}$ and $W^{+}\rightarrow (c\bar{b})[n] + b \bar{s}$, because $\bar{s}$ is much lighter than the heavy quarks, the largest $d\Gamma/d\cos\theta_{13}$ is achieved when the quarkonium and $\bar{s}$ move back to back ($\theta_{13}=180^{\circ}$) or the quarkonium and the outgoing heavy quark move in the same direction ($\theta_{23}=0^{\circ}$), which is shown in Figs.(\ref{CCcsdiscos},\ref{W+Bcbsdcos}). While for the decay channels $W^{+}\rightarrow (c\bar{b})[n] +c\bar{c}$ and $W^{+}\rightarrow (b\bar{b}) +c \bar{b}$, as shown in Figs.(\ref{W+Bcccdcos},\ref{W+rbcdiscos}), the largest $d\Gamma/d\cos\theta_{13}$ is achieved when $\theta_{13}=0^{\circ}$ and the largest $d\Gamma/d\cos\theta_{23}$ is achieved when $\theta_{23}=180^{\circ}$. This shows that the maximum differential decay width is obtained when the quarkonium and the outgoing antiquark move in the same direction or the quarkonium and the outgoing quark move back to back in the $W^+$ rest frame.

Considering that the LHC runs at the center-of-mass energy $\sqrt{S}=14$ TeV with luminosity $10^{34} cm^{-2} s^{-1}$, one expects that about $3.07 \times 10^{10}$ $W^+$-events per year can be generated \cite{w}. Then we can estimate the heavy-quarkonium events generated through $W^+$ decays, i.e. $2.57\times10^6$ $\eta_c$, $2.65\times10^6$ $J/\Psi$ and $2.40\times10^6$ $P$-wave charmonium events per year can be generated; $1.01\times10^5$ $B_c$, $9.11\times10^4$ $B^*_c$ and $3.16\times10^4$ $P$-wave $(c\bar{b})$-quarkonium events per year can be generated; $2.74\times10^2$ $\eta_b$, $2.59\times10^2$ $\Upsilon$ and $35$ $P$-wave bottomonium events per year can be generated. A small event numbers for the bottomonium shows that it is hard to find the bottomonium through $W^+$ decays even at LHC. However, it might be possible to measure $J/\psi$ and $B_c$ events through $W^+$ decays, since $W^+$ is charged particle and one may identify these particles through their cascade decay channels as $J/\psi\to \mu^+\mu^-$ and $B_c \to J/\psi+\pi$ or $B_c\to J/\psi +e\nu_e$ with clear signal. Bearing the situation pointed out here and the possible upgrade for the LHC (SLHC, DLHC, etc. \cite{slhc}) in mind, the possibility to study the charmonium and the $(c\bar{b})$-quarkonium via the $W$ boson decays is worth thinking seriously about.

\subsection{Uncertainty analysis}

In this subsection, we discuss the uncertainties for the charmonium and the $(c\bar{b})$-quarkonium production through $W^+$ decays.

For the present leading-order calculation, their main uncertainty sources include the non-perturbative bound-state matrix elements, CKM matrix elements, the renormalization scale $\mu_R$ and the quark masses $m_b$, $m_c$ and $m_s$. In the literature, the Brodsky-Lepage-Mackenzie (BLM) method \cite{blm} or the principle of maximum conformality (PMC) \cite{pmc} provides a feasible way to derive a precise QCD predictions. The main idea of BLM/PMC is to sum all the non-conformal $\{\beta_i\}$ terms in the perturbative expansion into the running coupling, and then the remaining terms are identical to that of a conformal theory and are renormalization-scheme independent. At the present, the bound state and CKM matrix elements and $\alpha_s$ emerge as overall factors and their uncertainties can be conveniently discussed when we know their values well, so we shall not discuss their uncertainties in the present paper. In the following, we shall concentrate our attention on the uncertainties caused by $m_b$, $m_c$ and $m_s$, whose values are taken as $m_b=4.90\pm0.40$ GeV, $m_c=1.35\pm0.25$ GeV and $m_s=0.105\pm0.025$ GeV. And for clarity, when discussing the uncertainty caused by one parameter, the other parameters are fixed to be their center values.

\begin{table}
\begin{tabular}{|c||c|c|c|}
\hline
~~$m_c$({\rm GeV})~~        & ~~1.10~~   & ~~1.35~~   & ~~1.60~~  \\
\hline\hline
$\Gamma_{|(c\bar{c})_{\bf 1}[^1S_0]\rangle}({\rm KeV})$ & 326.0  & 174.8   & 104.0  \\
\hline
$\Gamma_{|(c\bar{c})_{\bf 1}[^3S_1]\rangle}({\rm KeV})$ & 336.6  & 180.6  & 107.5  \\
\hline
$\Gamma_{|(c\bar{c})_{\bf 1}[P-wave]\rangle}({\rm KeV})$ & 246.9    & 165.4     &  54.26   \\
\hline\hline
$\Gamma_{|(c\bar{b})_{\bf 1}[^1S_0]\rangle}({\rm KeV})$ & 6.33  & 6.32   & 6.30  \\
\hline
$\Gamma_{|(c\bar{b})_{\bf 1}[^3S_1]\rangle}({\rm KeV})$ & 5.30  & 5.38  & 5.46  \\
\hline
$\Gamma_{|(c\bar{b})_{\bf 1}[P-wave]\rangle}({\rm KeV})$ & 2.48   &  1.77    &  1.35   \\
\hline
\end{tabular}
\caption{Uncertainties for the decay width of the processes $W^+\rightarrow (c\bar{c})[n]$ and $W^+\rightarrow (c\bar{b})[n]$, where $|(c\bar{c})_{\bf 1}[P-wave]\rangle$ and $|(c\bar{b})_{\bf 1}[P-wave]\rangle$ stands for the sum of the four color-singlet $P$-wave states for the $(c\bar{c})$- and $(c\bar{b})$- quarkonium accordingly. }
\label{tabuncernmc}
\end{table}

\begin{table}
\begin{tabular}{|c||c|c|c|}
\hline ~~$m_b$ ({\rm GeV})~~    & ~~4.50~~   & ~~4.90~~   & ~~5.30~~  \\
\hline \hline
$\Gamma_{|(c\bar{b})_{\bf 1}[^1S_0]\rangle}({\rm KeV})$ & 8.31   & 6.32   & 4.89  \\
\hline
$\Gamma_{|(c\bar{b})_{\bf 1}[^3S_1]\rangle}({\rm KeV})$ & 7.14  & 5.38  & 4.14  \\
\hline
$\Gamma_{|(c\bar{b})_{\bf 1}[P-wave]\rangle}({\rm KeV})$ & 2.41   &  1.77    &  1.33  \\
\hline
\end{tabular}
\caption{Uncertainties for the decay width of the process $W^+\rightarrow (c\bar{b})[n]$ with varying $m_b$, where $|(c\bar{b})_{\bf 1}[P-wave]\rangle$ stands for the sum of the four color-singlet $P$-wave states. }
\label{tabuncernmb}
\end{table}

Typical uncertainties for $m_c$ and $m_b$ are presented in TABs.(\ref{tabuncernmc},\ref{tabuncernmb}), where $W^+\rightarrow (c\bar{c})[n]$ stands for the process $W^+\rightarrow (c\bar{c})[n]+c\bar{s}$, $W^+\rightarrow (c\bar{b})[n]$ stands for the processes $W^+\rightarrow (c\bar{b})[n]+b\bar{s}$ and $W^+\rightarrow (c\bar{b})[n]+c\bar{c}$ respectively. TABs.(\ref{tabuncernmc},\ref{tabuncernmb}) show that sizable uncertainties can be found for varying $m_b$ and $m_c$. The decay width will decrease with the increment of $m_b$ and $m_c$, and such tendency slow down with a heavier quark mass. Taking the process $W^+(k) \rightarrow (c\bar{c})[n](q_3)+c (q_2)\bar{s}(q_1)$ as an explicit example. One may observe that even though its phase-space is slightly affected by $m_c$, i.e. the maximum value of $s_1=(q_1 +q_3)^2$ ($s_{1}^{max}=m^2_W (1-{m_c}/{m_W})^2 \sim m^2_W$) remains almost unchanged, the total decay width shall be decreased by about $3-5$ times for various $(c\bar{c})$-quarkonium states by varying $m_c$ from $1.10$ GeV to $1.60$ GeV. Such a big uncertainty is mainly caused by the fact that it is harder for an intermediate hard gluon to generate a heavier $(c\bar{c})$-pair. More explicitly, at the specific momentum region with $q_g^2 \simeq 4 m_c^2$ (where $q_g$ stands for the intermediate gluon momentum), which gives the main contribution to the decay width, there is a strong suppression factor of $(1.60^4/1.10^4)\sim 4.5$ for the $S$-wave production by varying the $c$-quark masses from $1.10$ to $1.60$ GeV \footnote{As for the $P$-wave cases, because of the derivation of the amplitude over the bound-state relative momentum, this suppression factor shall become even bigger.}. Moreover, one may observe that the decay widths for $P$-wave states are more sensitive to the quark masses than the case of $S$-wave states. Varying $m_s\in[0.080,0.130]$, one may observe that the decay width of $W^+\rightarrow (c\bar{c})[n]+c\bar{s}$ is almost unchanged for the $S$-wave states, but shall cause sizable changes for the $P$-wave states, i.e. $\Gamma_{|(c\bar{c})_{\bf 1}[P-wave]\rangle}|_{m_s=0.080 {\rm GeV}}=139.2$ KeV, $\Gamma_{|(c\bar{c})_{\bf 1}[P-wave]\rangle}|_{m_s=0.105 {\rm GeV}}=165.4$ KeV and $\Gamma_{|(c\bar{c})_{\bf 1}[P-wave]\rangle}|_{m_s=0.130 {\rm GeV}}=185.3$ KeV.

Adding all the uncertainties caused by the constituent quark masses in quadrature, for $W^+\rightarrow (c\bar{c})[n]+c\bar{s}$, we obtain
\begin{eqnarray}
\Gamma_{|(c\bar{c})_{\bf 1}[^1S_0]\rangle}&=&174.8^{+151.2}_{-70.8} \;{\rm KeV},\nonumber\\
\Gamma_{|(c\bar{c})_{\bf 1}[^3S_1]\rangle}&=&180.6^{+156.0}_{-73.1} \;{\rm KeV},\nonumber\\
\Gamma_{|(c\bar{c})_{\bf 1}[P-wave]\rangle}&=&165.4^{+85.6}_{-112.9} \;{\rm KeV},\nonumber\\
\Gamma_{|(c\bar{c})_{\bf 8}[S-wave]g\rangle}&=&44.5^{+38.4}_{-18.0} \times{v^4}  \;{\rm KeV}. \nonumber
\end{eqnarray}
For $W^+\rightarrow (c\bar{b})[n] + b\bar {s}$, we obtain
\begin{eqnarray}
\Gamma_{|(c\bar{b})_{\bf 1}[^1S_0]\rangle}&=&6.32^{+1.99}_{-1.43} \;{\rm KeV},\nonumber\\
\Gamma_{|(c\bar{b})_{\bf 1}[^3S_1]\rangle}&=&5.38^{+1.76}_{-1.24} \;{\rm KeV},\nonumber\\
\Gamma_{|(c\bar{b})_{\bf 1}[P-wave]\rangle}&=&1.77^{+0.96}_{-0.61} \;{\rm KeV},\nonumber\\
\Gamma_{|(c\bar{b})_{\bf 8}[S-wave]g\rangle}&=&1.46^{+0.47}_{-0.33} \times{v^4}  \;{\rm KeV}. \nonumber
\end{eqnarray}
And for $W^+\rightarrow (c\bar{b})[n] + c\bar {c}$, we obtain
\begin{eqnarray}
\Gamma_{|(c\bar{b})_{\bf 1}[^1S_0]\rangle}&=&0.546^{+0.478}_{-0.223} \;{\rm KeV},\nonumber\\
\Gamma_{|(c\bar{b})_{\bf 1}[^3S_1]\rangle}&=&0.810^{+0.800}_{-0.354} \;{\rm KeV},\nonumber\\
\Gamma_{|(c\bar{b})_{\bf 1}[P-wave]\rangle}&=&0.379^{+0.699}_{-0.156} \;{\rm KeV},\nonumber\\
\Gamma_{|(c\bar{b})_{\bf 8}[S-wave]g\rangle}&=&0.170^{+0.160}_{-0.072} \times{v^4}  \;{\rm KeV}. \nonumber
\end{eqnarray}

If assuming the higher excited heavy-quarkonium states decay to the ground color-singlet and spin-singlet state with $100\%$ efficiency via electromagnetic or hadronic interactions, then we obtain the total decay width of $W^+$ decay channels,
\begin{eqnarray}
\Gamma_{(W^+\to (c\bar{c})_{\bf 1}[^1S_0] +c\bar{s})} &=& 524.8^{+396.3}_{-258.4} \;{\rm KeV} \label{cct} \\
\Gamma_{(W^+\to (c\bar{b})_{\bf 1}[^1S_0] +c\bar{s})} &=& 13.5^{+4.73}_{-3.29} \;{\rm KeV} \label{cbt1} \\
\Gamma_{(W^+\to (c\bar{b})_{\bf 1}[^1S_0] +c\bar{c})} &=& 1.74^{+1.98}_{-0.73} \;{\rm KeV} \label{cbt2} \\
\Gamma_{(W^+\to (b\bar{b})_{\bf 1}[^1S_0] +c\bar{b})} &=& 38.6^{+13.4}_{-9.69} \;{\rm eV} \label{bbt}
\end{eqnarray}
where $v^2 \in [0.10,0.30]$ is adopted.

\begin{figure}[!htb]
\includegraphics[width=0.38\textwidth]{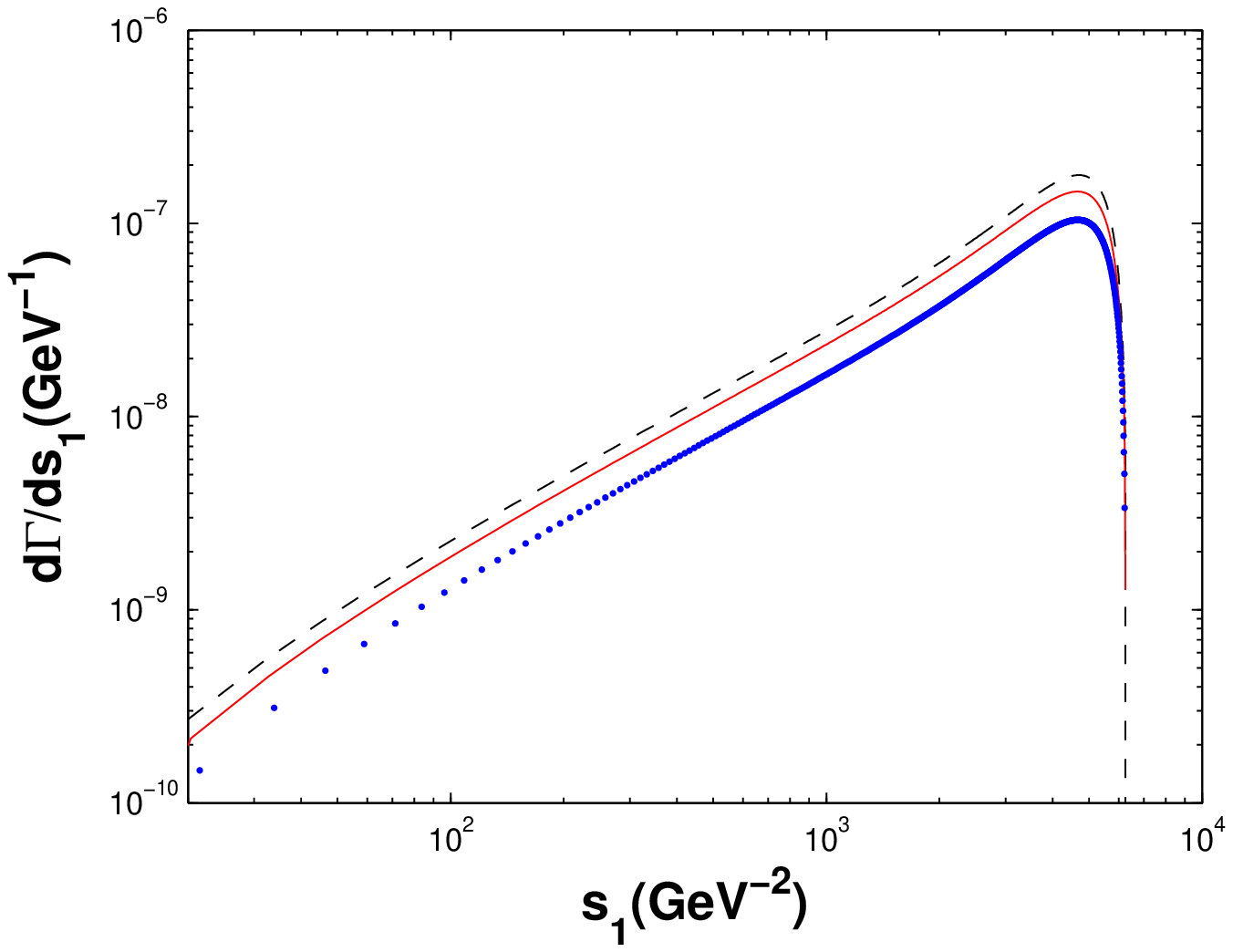}
\hspace{0.20cm}
\includegraphics[width=0.38\textwidth]{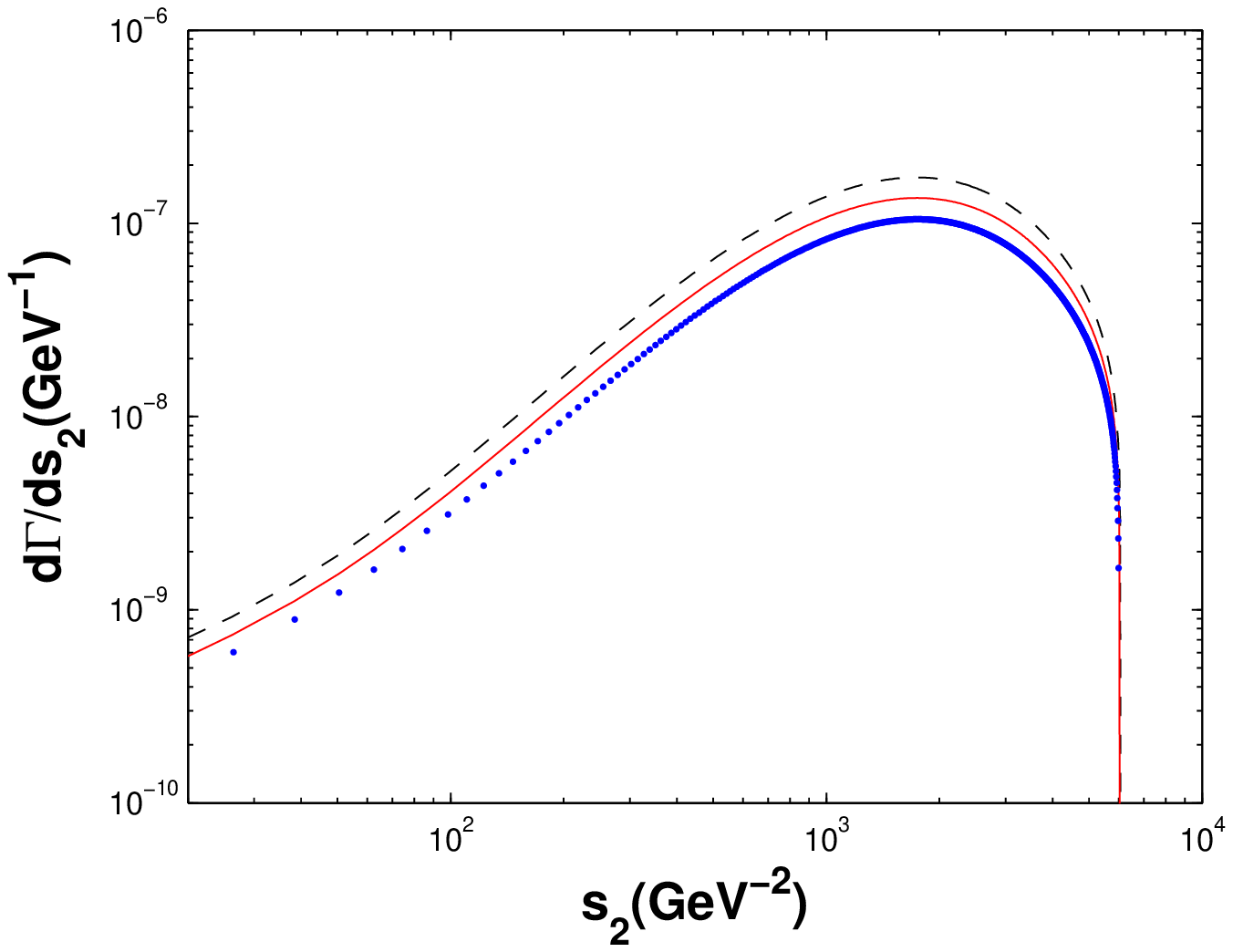}
\caption{Uncertainties of $d\Gamma/ds_1$ (Up) and $d\Gamma/ds_2$ (Down) for $W^+\rightarrow (c\bar{c})[n] +c\bar{s}$, where contributions from the color-singlet $S$-wave and $P$- wave states have been summed up. The dashed line, the solid line, the dotted line are for $m_c=1.25$GeV, $1.35$GeV and $1.45$GeV respectively. } \label{CCcssum}
\end{figure}

\begin{figure}[!htb]
\includegraphics[width=0.38\textwidth]{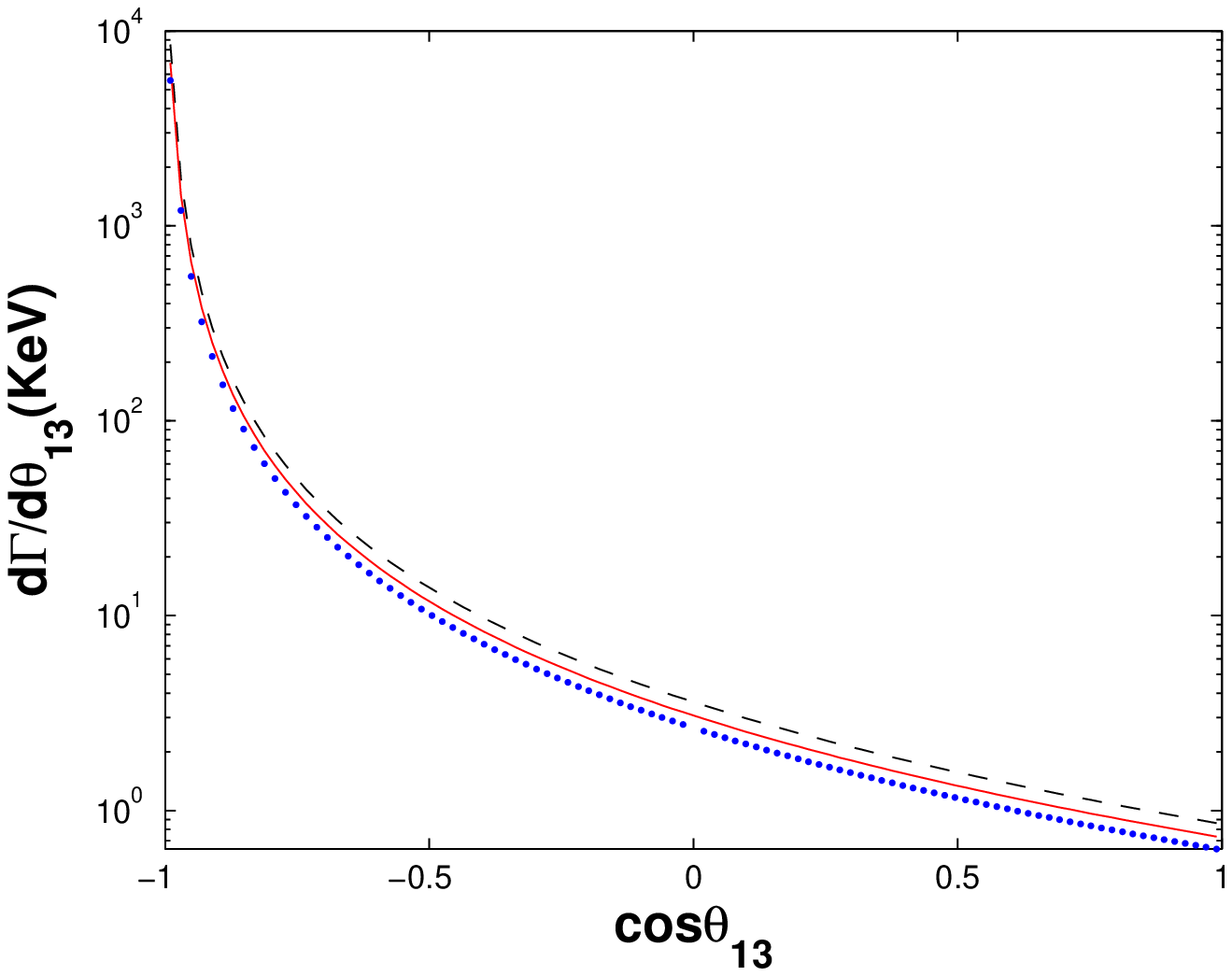}
\hspace{0.20cm}
\includegraphics[width=0.38\textwidth]{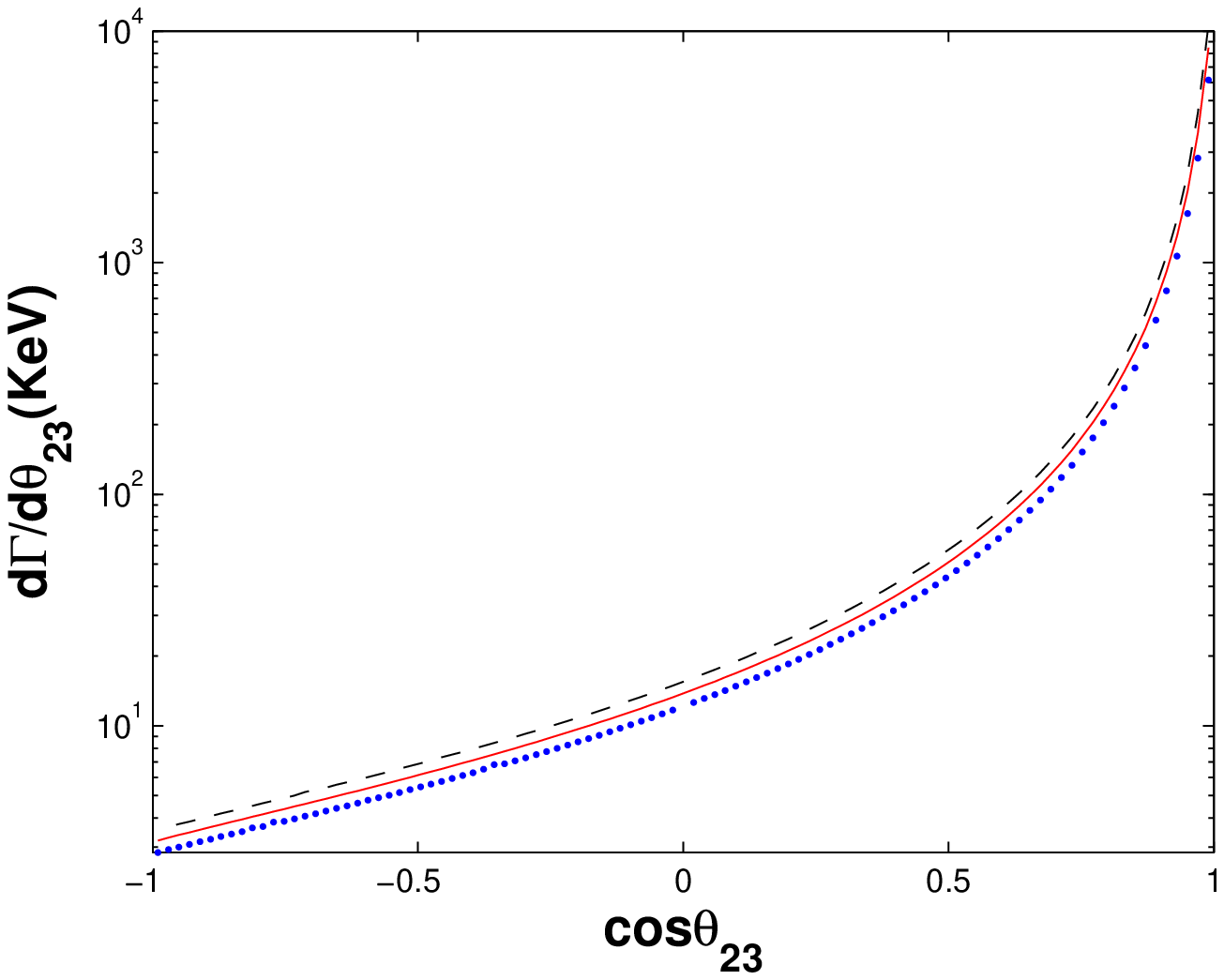}
\caption{Uncertainties of $d\Gamma/d\cos{\theta_{13}}$ (Up) and $d\Gamma/d\cos{\theta_{23}}$ (Down) for $W^+\rightarrow (c\bar{c})[n] +c\bar{s} $, where contributions from the color-singlet $S$-wave and $P$- wave states have been summed up. The dashed line, the solid line, the dotted line are for $m_c=1.25$GeV, $1.35$GeV and $1.45$GeV respectively. } \label{CCcscossum}
\end{figure}

Finally, to show how the decay width depends on the corresponding quark mass more clearly, we take the channel $W^+\rightarrow (c\bar{c})[n]+c\bar{s}$ as an explicit example, whose differential decay widths with several typical $m_c$ are drawn in Figs.(\ref{CCcssum},\ref{CCcscossum}). The contributions from the color-singlet $S$-wave and $P$- wave states have been summed up for convenience. And in Figs.(\ref{CCcssum},\ref{CCcscossum}), the center solid lines are for $m_c=1.35 {\rm GeV}$, the dashed lines are for $m_c=1.25 {\rm GeV}$ and the dotted lines are for $m_c=1.45 {\rm GeV}$ respectively. Similar to the observation from the Tab.\ref{tabuncernmc}, large uncertainties are caused for $m_c$ varying within the region of $[1.25,1.45]$ GeV. At the specific momentum regions with $q_g^2 \simeq 4 m_c^2$, we have $s_1 \to m^2_W$, which explains why there is a peak near $s_1 \sim m^2_W$ for the differential decay width $d\Gamma/ds_1$ as shown by Fig.(\ref{CCcssum}).

\section{Conclusions}

In the present paper, we have made a detailed study on the heavy-quarkonium production through $W$ boson semi-inclusive decays, $W^+\to (c\bar{c})[n] +c\bar{s}$, $W^+\to (c\bar{b})[n] +c\bar{s}$, $W^+ \to (c\bar{b})[n] +c\bar{c}$ and $W^+\to (b\bar{b})[n] +c\bar{b}$ within the NRQCD framework. Results for eight quarkonium Fock states, i.e. $|(Q\bar{Q'})_{\bf 1,8}[^1S_0]\rangle$, $|(Q\bar{Q'})_{\bf 1,8}[^3S_1]\rangle$, $|(Q\bar{Q'})_{\bf 1}[^1P_1]\rangle$ and $|(Q\bar{Q'})_{\bf 1}[^3P_J]\rangle$ have been presented. And to provide the analytical expressions as simply and compactly as possible, we have adopted the improved trace technology to derive the Lorentz invariant expressions for the $W$ boson decay processes at the amplitude level. Such a calculation technology shall be very helpful to derive simpler results for the processes with massive spinors.

Numerical results show that the $P$-wave states in addition to the $S$-wave states can also provide sizable contributions to the heavy-quarkonium production, so one needs to take the $P$-wave states into consideration for a sound estimation. More explicitly, for the charmonium production channel $W^{+}\rightarrow (c\bar{c})[n] + c \bar{s}$, the total decay width for all the $P$-wave states is $165$ KeV, which is comparable to that of $\eta_c$ or $J/\psi$. For the $(c\bar{b})$-quarkonium production, the total decay width for all the $P$-wave states is about $28\%$ ($33\%$) of that of $B_c$ ($B^*_c$) for $W^{+}\rightarrow (c\bar{b})[n] + b\bar{s}$; and is about $69\%$ ($47\%$) of that of $B_c$ ($B^*_c$) for $W^{+}\rightarrow (c\bar{b})[n] + c\bar{c}$. If all the low-lying excited states decay to the ground state $|(Q\bar{Q'})_{\bf 1}[^1S_0]\rangle$ with $100\%$ efficiency, we can obtain the total decay width for the $(Q\bar{Q'})$ production through $W^+$ decays as shown by Eqs.(\ref{cct},\ref{cbt1},\ref{cbt2},\ref{bbt}). At the LHC, due to its high collision energy and high luminosity, sizable heavy-quarkonium events can be produced through $W^+$ boson decays, i.e. $2.57\times10^6$ $\eta_c$, $2.65\times10^6$ $J/\Psi$ and $2.40\times10^6$ $P$-wave charmonium events per year can be obtained; and $1.01\times10^5$ $B_c$, $9.11\times10^4$ $B^*_c$ and $3.16\times10^4$ $P$-wave $(c\bar{b})$-quarkonium events per year can be obtained. So these channels may be an important supplement for other measurements at the LHC. And large event numbers for the higher excited states may cause themselves to be deserving of careful study.

\hspace{2cm}

{\bf Acknowledgements}: This work was supported in part by the Fundamental Research Funds for the Central Universities under Grant No.CDJXS1102209, the Program for New Century Excellent Talents in University under Grant No.NCET-10-0882, and the Natural Science Foundation of China under Grant No.10805082 and No.11075225.

\appendix

\section*{Appendix: Amplitudes for $W^+(k)\rightarrow (c\bar{Q'})[n](q_3) + Q'(q_2)\bar{s}(q_1)$}

We shall only list the results for the first type of semi-inclusive process $W^+(k)\rightarrow (c\bar{Q'})[n](q_3) + Q'(q_2)\bar{s}(q_1)$. The results for the second type of process $W^+(k)\rightarrow (Q\bar{b})[n](q_3) + c(q_2)\bar{Q}(q_1)$ is similar.

\noindent{\it $\diamond$ Short notations.}

To shorten the notation, we first define some dimensionless parameters
\begin{eqnarray}
r_1&=& \frac{m_{Q'}} {m_W}, \ r_2=\frac{m_s} {m_W}, \ r_3=\frac{M} {m_W}, \nonumber\\
x&=& \frac{{m_W}^2+{M}^2-s_2} {2 {m_W}^2}  ,
y= \frac{{m_W}^2+{m_{Q'}}^2-s_1} {2 {m_W}^2} ,\nonumber\\
z&=& \frac{{m_W}^2+{m_s}^2-s_3} {2 {m_W}^2} ,
u=\frac{s_3-{m_{Q'}}^2-{M}^2} {2 {m_W}^2},\nonumber\\
v&=& \frac{s_1-{m_s}^2-{M}^2} {2 {m_W}^2},
w= \frac{s_2-{m_{Q'}}^2-{m_s}^2} {2 {m_W}^2} ,
\end{eqnarray}
where $s_1 =(q_1+q_3)^2$, $s_2=(q_1+q_2)^2$ and ${s_3}=(q_2+q_3)^2$, which satisfy the relation
$s_1+s_2+s_3={m_W}^2+{m_s}^2+{m_{Q'}}^2+{M}^2$. As for the normalization constants, we have
$L_{1,2}={1}/(2{m_W} \sqrt{w \mp r_1 r_2})$ and
\begin{displaymath}
N_{0}=\frac{1} {{m_W}^3\sqrt{{r_1}^2 {r_2}^2 -w^2-{r_2}^2 y^2-{r_1}^2 z^2+2 w y z}} .
\end{displaymath}
The short notations for the denominators are
\begin{eqnarray}
d_1&=&\frac{1} {s^2_3 -{m_{Q'}}^2} \frac{1} {(q_2+{q^0_{32}})^2},\nonumber\\
d_{11}&=&\frac{1} {s^2_3 -{m_{Q'}}^2} \frac{{m_W}^2} {({q_2}+{q^0_{32}})^4},\nonumber\\
d_2&=&\frac{1} {(q_1+q_2+q^0_{32})^2-{m_s}^2} \frac{1} {({q_2}+{q^0_{32}})^2},\nonumber\\
d_{21}&=&\frac{{m_W}^2} {((q_1+q_2+q^0_{32})^2-{m_s}^2)^2} \frac{1} {(q_2+{q^0_{32}})^2},\nonumber\\
d_{22}&=&\frac{1} {((q_1+q_2+q^0_{32})^2-{m_s}^2)^2} \frac{{m_W}^2} {({q_2}+q^0_{32})^4},
\end{eqnarray}
where $q^0_{31} = \frac{m_c}{M}{q_3}$ and $q^0_{32} = \frac{m_{Q'}}{M}{q_3}$. Finally, the following relations are useful for further simplification,
\begin{eqnarray}
&& x+y+z=1,\; u+v+w \simeq 1 , \nonumber\\
&& u+v+r^2_3 =x,\; w+u+r_1^2 =y,\;w+v+r_2^2 =z.
\end{eqnarray}

\noindent{\it $\diamond$ Basic Lorentz-invariant structures.}

For $|(c\bar{Q'})_{\bf 1}[^{1}S_{0}] \rangle$, there are three Lorentz structures,
\begin{eqnarray}
B_1 &=& \frac{q_{3}\cdot\epsilon(k)} {m_W},\
B_2= \frac{q_{2}\cdot\epsilon(k)} {m_W}, \nonumber\\
B_3 &=& \frac{i} {{m_W}^3} \varepsilon(k,q_3,q_2,\epsilon(k)),
\end{eqnarray}
where $\varepsilon(k,q_3,q_2,\epsilon(k))=\varepsilon(\alpha,\beta,\rho,\sigma) k_\alpha q_{3\beta} q_{2\rho} \epsilon_\sigma(k)$.

For $|(c\bar Q')_{\bf 1}[^{3}S_{1}] \rangle$ and $|(c\bar Q')_{\bf 1}[^{1}P_{1}] \rangle$, there are eleven Lorentz structures,
\begin{eqnarray}
B_1 &=&{\epsilon(q_3)\cdot\epsilon(k)},\;
B_2  = \frac{i}{m_W^2}\varepsilon(k,q_{3},\epsilon(k),\epsilon(q_{3})),\nonumber\\
B_3 &=& \frac{i}{m_W^2}\varepsilon(k,q_2,\epsilon(k),\epsilon(q_3)),\nonumber\\
B_4 &=& \frac{i}{m_W^2}\varepsilon(q_3,q_2,\epsilon(k),\epsilon(q_3)),\nonumber\\
B_5 &=& \frac{{k\cdot\epsilon(q_3)}{q_3\cdot\epsilon(k)}}{m_W^2},\;
B_6= \frac{{k\cdot\epsilon(q_3)}{q_2\cdot\epsilon(k)}}{m_W^2},\nonumber\\
B_7 &=& \frac{{q_2\cdot\epsilon(q_3)}{q_3\cdot\epsilon(k)}}{m_W^2},\;
B_8  = \frac{{q_2\cdot\epsilon(k)}{q_2\cdot\epsilon(q_3)}}{m_W^2},\nonumber\\
B_9 &=& \frac{i}{m_W^4}{\varepsilon(k,q_3,q_2,\epsilon(k))}{k\cdot\epsilon(q_3)},\nonumber\\
B_{10} &=& \frac{i}{m_W^4}{\varepsilon(k,q_3,q_2,\epsilon(q_3))}{q_3\cdot\epsilon(k)},\nonumber\\
B_{11} &=& \frac{i}{m_W^4}{\varepsilon(k,q_3,q_2,\epsilon(k))}{q_2\cdot\epsilon(q_3)} ,
\end{eqnarray}
where $\varepsilon(q_3)$ stands for the polarization vector relating to the spin or the orbit angular momentum of $|(c\bar Q')_{\bf 1}[^{3}S_{1}] \rangle$ and $|(c\bar Q')_{\bf 1}[^{1}P_{1}] \rangle$ respectively.

For $|(c\bar Q')_{\bf 1}[^{3}P_{J}] \rangle$ with $J = 0,1,2$, there are thirty-four Lorentz structures,
\begin{widetext}
\begin{eqnarray}
B_1&=& \frac{1}{m_W}q_{2}\cdot\epsilon(k)\varepsilon^J_{\alpha\alpha},\;
B_2 =\frac{1}{m_W}q_{3}\cdot\epsilon(k)\varepsilon^J_{\alpha\alpha},\;
B_3 = \frac{1}{m_W}q_{2\alpha}\epsilon_\beta(k)\varepsilon^J_{\alpha\beta},\;
B_4 = \frac{1}{m_W}k_{\alpha}\epsilon_\beta(k)\varepsilon^J_{\alpha\beta},\nonumber \\
B_5&=&\frac{i\varepsilon^J_{\alpha\alpha}}{m_W^3}\varepsilon(k,q_2,q_3,\epsilon(k)),\;
B_6 =\frac{i\varepsilon^J_{\alpha\beta}}{m_W^3}\varepsilon(k,q_2,q_3,\alpha)\epsilon_\beta(k),\;
B_7 = \frac{i\varepsilon^J_{\alpha\beta}}{m_W^3}\varepsilon(k,q_3,\epsilon(k),\alpha)q_{2\beta} ,\nonumber\\
B_8&=&\frac{i\varepsilon^J_{\alpha\beta}}{m_W^3}\varepsilon(k,q_2,\alpha,\beta)q_{2}\cdot\epsilon(k),\;
B_9 = \frac{i\varepsilon^J_{\alpha\beta}}{m_W^3}\varepsilon(k,q_3,\epsilon(k),\alpha)k_{\beta},\;
B_{10}=\frac{i\varepsilon^J_{\alpha\beta}}{m_W^3}\varepsilon(k,q_2,\epsilon(k),\alpha)k_{\beta} ,\nonumber\\
B_{11}&=&\frac{i\varepsilon^J_{\alpha\beta}}{m_W^3}\varepsilon(k,q_3,\alpha,\beta)q_{3}\cdot\epsilon(k),\;
B_{12}=\frac{i\varepsilon^J_{\alpha\beta}}{m_W^3}\varepsilon(k,q_2,\alpha,\beta)q_{3}\cdot\epsilon(k),\;
B_{13} = \frac{i\varepsilon^J_{\alpha\beta}}{m_W^3}\varepsilon(k,q_3,\alpha,\beta)q_{2}\cdot\epsilon(k),\nonumber\\
B_{14} &=&\frac{i\varepsilon^J_{\alpha\beta}}{m_W^3}\varepsilon(k,q_2,\epsilon(k),\alpha)q_{2\beta},\;
B_{15} = \frac{i\varepsilon^J_{\alpha\beta}}{m_W^3}\varepsilon(q_2,q_3,\epsilon(k),\alpha)q_{2\beta},\;
B_{16}=\frac{i\varepsilon^J_{\alpha\beta}}{m_W^3}\varepsilon(q_2,q_3,\alpha,\beta)q_{2}\cdot\epsilon(k),\nonumber\\
B_{17} &=& \frac{i\varepsilon^J_{\alpha\beta}}{m_W^3}\varepsilon(q_2,q_3,\alpha,\beta)q_{3}\cdot\epsilon(k),\;
B_{18} = \frac{i\varepsilon^J_{\alpha\beta}}{m_W^3}\varepsilon(q_2,q_3,\epsilon(k),\alpha)k_{\beta},\;
B_{19} =\frac{i\varepsilon^J_{\alpha\beta}}{m_W}\varepsilon(k,\epsilon(k),\alpha,\beta),\nonumber\\
B_{20} &=&\frac{i\varepsilon^J_{\alpha\beta}}{m_W}\varepsilon(q_2,\epsilon(k),\alpha,\beta),\;
B_{21} =\frac{i\varepsilon^J_{\alpha\beta}}{m_W}\varepsilon(q_3,\epsilon(k),\alpha,\beta),\;
B_{22}  = \frac{i\varepsilon^J_{\alpha\beta}}{m_W^5}\varepsilon(k,q_2,q_3,\alpha)k_{\beta} q_3\cdot\epsilon(k),\nonumber\\
B_{23} &=&\frac{i\varepsilon^J_{\alpha\beta}}{m_W^5}\varepsilon(k,q_2,q_3,\alpha)q_{2\beta} q_2\cdot\epsilon(k),\;
B_{24}=\frac{i\varepsilon^J_{\alpha\beta}}{m_W^5}\varepsilon(k,q_2,q_3,\alpha)k_{\beta} q_2\cdot\epsilon(k),\;
B_{25} =\frac{i\varepsilon^J_{\alpha\beta}}{m_W^5}\varepsilon(k,q_2,q_3,\alpha)q_{2\beta} q_3\cdot\epsilon(k),\nonumber\\
B_{26}&=& \frac{i\varepsilon^J_{\alpha\beta}}{m_W^5}\varepsilon(k,q_2,q_3,\epsilon(k))q_{2\alpha} q_{2\beta},\;
B_{27} =\frac{i\varepsilon^J_{\alpha\beta}}{m_W^5}\varepsilon(k,q_2,q_3,\epsilon(k))k_{\alpha} k_{\beta},\;
B_{28} = \frac{i\varepsilon^J_{\alpha\beta}}{m_W^5}\varepsilon(k,q_2,q_3,\epsilon(k))k_{\alpha} q_{2\beta},\nonumber\\
B_{29} &=&\frac{\varepsilon^J_{\alpha\beta}}{m_W^3}k_{\alpha}k_{\beta}q_{3}\cdot\epsilon(k),\;
B_{30} = \frac{\varepsilon^J_{\alpha\beta}}{m_W^3}k_{\alpha}q_{2\beta}q_{3}\cdot\epsilon(k),\;
B_{31} =\frac{\varepsilon^J_{\alpha\beta}}{m_W^3}k_{\alpha}k_{\beta}q_{2}\cdot\epsilon(k),\nonumber\\
B_{32} &=&\frac{\varepsilon^J_{\alpha\beta}}{m_W^3}k_{\alpha}q_{2\beta}q_{2}\cdot\epsilon(k),\;
B_{33} = \frac{\varepsilon^J_{\alpha\beta}}{m_W^3}q_{2\alpha}q_{2\beta}q_{3}\cdot\epsilon(k),\;
B_{34}=\frac{\varepsilon^J_{\alpha\beta}}{m_W^3}q_{2\alpha}q_{2\beta}q_{2}\cdot\epsilon(k).
\end{eqnarray}
\end{widetext}

\noindent{\it $\diamond$ Non-zero coefficients for $|(c\bar{Q'})_{\bf 1}[^1S_0]\rangle$.}

Non-zero coefficients $A^1_j$ and $A^{3'}_j$ are
\begin{widetext}
\begin{eqnarray}
A^1_1 &=& -\frac{2 {L_1} {m_W}^{7/2}}{\sqrt{r_3}}({d_2} ((2 {r_3}-4 {r_2}) {r_1}^2+({r_3}^2-4 {r_2}^2+
2 u+2 z) {r_1} -2({r_2}-{r_3}) (u-y))\nonumber\\
&&+2 {d_1} (-2 {r_1}^3-2 ({r_2}-{r_3}) {r_1}^2+({r_3}^2+2 {r_2} {r_3}-2 u + 2 y) {r_1} + 4 {r_3} u - 2 {r_3} y)) ,\\
A^1_2 &=&\frac{2 {L_1} {m_W}^{7/2}}{\sqrt{r_3}} ({d_2} ({r_1}+{r_2}) ({r_3}^2+2 {r_2} {r_3}-2 x)-2 {d_1} (-2 {r_3}^3+{r_2} {r_3}^2 +2 x {r_3}+2 {r_2} u  +{r_1} ({r_3}^2-2 v))),\\
A^1_3 &=&-\frac{4 {L_1} {m_W}^{7/2} }{\sqrt{r_3}}({d_2} ({r_1}+{r_2})-2 {d_1} ({r_1}-{r_3})),\\
A^{3'}_1&=&-\frac{{m_W}^{9/2} {N_0}}{2 {L_2}\sqrt{r_3}}(-4 {d_2} y^2+(2 {d_1} (-2 {r_1}^2+2 ({r_2}+{r_3}) {r_1}+{r_3}^2-2 {r_2} {r_3} +2 u)+{d_2} ((2 {r_1}-2 {r_2}+{r_3}) (2{r_2}+{r_3})\nonumber\\
&&+4 u+2)) y+2 (2 {d_1} {r_1} ({r_1}-{r_3}) (1-2 x)-{d_2} (u+{r_1} ({r_1}-{r_2}-{r_3}) (x-1)))),\\
A^{3'}_2&=& \frac{{N_0} {m_W}^{9/2}} {2 {L_2} \sqrt{r_3}} ({d_2} ((x+2 y-3) {r_3}^2-2 {r_2} (x+2 y-1) {r_3}+2 x-4 x y)\nonumber\\
&&-2 {d_1} ((x+2 y-1) {r_3}^2+2 ({r_1}+{r_2}) x {r_3}-2 u-2 {r_1} ({r_1}+{r_2}) x+2 u (x+2 y))) ,\\
A^{3'}_3 & = &\frac{{m_W}^{9/2} {N_0}} {2 {L_2}\sqrt{r_3}}(4 {d_1} ({r_1} + {r_2}) ({r_1} - {r_3})+2{d_1}({r_3}^2 + 2 u)- {d_2}({r_3}^2 + 2{r_2} {r_3} + 4 y -2)).
\end{eqnarray}
\end{widetext}

Non-zero coefficients $A^2_j$ and $A^{4'}_j$ are
\begin{widetext}
\begin{eqnarray}
A^2_1 &=&-\frac{2 {L_2} {m_W}^{7/2}}{\sqrt{r_3}} ({d_2} (2 (2 {r_2}+{r_3}) {r_1}^2+({r_3}^2-4 {r_2}^2+
2 u+2 z) {r_1}+2({r_2}+{r_3}) (u-y))\nonumber\\
&&+2 {d_1} (-2 {r_1}^3+2 ({r_2}+{r_3}) {r_1}^2+({r_3}^2-2 {r_2} {r_3}-2 u+2 y) {r_1}+4{r_3} u-2 {r_3} y)),\\
A^2_2 &=&\frac{2 {L_2} {m_W}^{7/2}} {\sqrt{r_3}}(2 {d_1} (2 {r_3}^3+{r_2} {r_3}^2-2 x {r_3}+2 {r_2} u-{r_1} ({r_3}^2 -2 v))-{d_2} ({r_1}-{r_2}) (-{r_3}^2+2 {r_2} {r_3}+2 x)),\\
A^2_3 &=&\frac{4 {L_2} {m_W}^{7/2}}{\sqrt{r_3}} ({d_2} ({r_2}-{r_1})+2 {d_1} ({r_1}-{r_3})),\\
A^{4'}_1 &=&\frac{{m_W}^{9/2} {N_0}} {2 {L_1} \sqrt{r_3}}(-4 {d_1} {r_1} ({r_1}-{r_3}) (2 x-1)-2 {d_1} (-{r_3}^2-2 ({r_1} +{r_2}) {r_3}+2 {r_1} ({r_1}+{r_2})-2 u) y\nonumber\\
&&+{d_2} (-4 y^2+({r_3}^2+2 {r_1} {r_3}-4 {r_2} ({r_1}+{r_2})+4 u) y+2 y-2 u-2 {r_1}({r_1}+{r_2}-{r_3}) (x-1))),\\
A^{4'}_2 &=&-\frac{{m_W}^{9/2} {N_0}}{2 {L_1} \sqrt{r_3}} ({d_2} ((x+2 y-3) {r_3}^2+2 {r_2} ( y-z) {r_3}+2 x-4 x y)\nonumber\\
&&-2 {d_1} ((y-z) {r_3}^2+2 ({r_1}-{r_2}) x {r_3}+2 ({r_1} ({r_2}-{r_1}) x+u (y-z)))),\\
A^{4'}_3 &=&\frac{{m_W}^{9/2} {N_0}} {2 {L_1} \sqrt{r_3}}({d_2} ({r_3}^2-2 {r_2} {r_3}+4 y-2)-2 {d_1} (2 {r_1}^2-2 ({r_2}+{r_3}) {r_1}+{r_3}^2+2 {r_2} {r_3}+2 u)).
\end{eqnarray}
\end{widetext}

\noindent{\it $\diamond$ Non-zero coefficients for $|(c\bar{Q'})_{\bf 1}[^3S_1]\rangle$.}

Non-zero coefficients $A^1_j$ and $A^{3'}_j$ are
\begin{widetext}
\begin{eqnarray}
A^1_1 &=&\frac{2 {L_1} {m_W}^{7/2}} {\sqrt{r_3}} ({d_2} (2 ({r_3}^2-2 {r_2} {r_3}-2 x) {r_1}^2+({r_3}^3+(u-4 {r_2}^2 - x+2 z) {r_3}-2 {r_2} (u+x)) {r_1}-2 {r_2}^2 u\nonumber\\&&
+2({r_3}^2-2 x) (u-y)+{r_2} {r_3} (2 y-3 u))-2 {d_1} {r_3} ({r_2}u -{r_1} v)),\\
A^1_2 &=&\frac{2 {L_1} {m_W}^{7/2}} {\sqrt{r_3}} ({d_2}({r_1} (4 {r_1}+2 {r_2}+{r_3})+8 (u-y))-2 {d_1} {r_1} {r_3}),\\
A^1_3 &=&\frac{4 {L_1} {d_2} {m_W}^{7/2}} {\sqrt{r_3}} (({r_1}+{r_2}-2 {r_3}) {r_3}+2 x),\\
A^1_4 &=&-\frac{2 {L_1} {m_W}^{7/2}} {\sqrt{r_3}} (2 {d_1} ({r_1}+{r_2}) {r_3}+{d_2} (({r_1}+{r_2}) (2 {r_2}+{r_3})-4 x+4)),\\
A^1_5 &=&\frac{2 {L_1} {m_W}^{7/2}} {\sqrt{r_3}} ({d_2} ({r_1} (4 {r_1}+2 {r_2}+{r_3})+4 (u-y))-2 {d_1} {r_1} {r_3}),\\
A^1_6 &=&\left(\frac{d_2}{2d_1}\right) A^1_8={4 {L_1}  {d_2} {m_W}^{7/2}} {\sqrt{r_3}} (r_1+r_2),\\
A^1_7 &=&\frac{2 {L_1} {m_W}^{7/2}} {\sqrt{r_3}} (2 {d_1} (3 {r_1}+{r_2}) {r_3}+{d_2} ({r_1}+{r_2}) (2 {r_2}+{r_3})),\\
A^1_9 &=& A^1_{10} = -\frac{8 {L_1} {d_2} {m_W}^{7/2}} {\sqrt{r_3}},\\
A^{3'}_1 &=&-\frac{{N_0} {m_W}^{9/2}} {2 {L_2} \sqrt{r_3}} (-2 {d_1} {r_3} (u+{r_1}({r_1}+{r_2}) x)+2 {d_1} {r_3}({r_3}^2+2 u) y +{d_2}(({r_3} (x-2)+2 {r_2} x){r_1}^2+(2 x{r_2}^2+{r_3} (-3 x\nonumber\\
&&-4 y+2){r_2}+2 ({r_3}^2-2 x) z) {r_1}+4 {r_2}^2{r_3} y+2 {r_2} (u-2 (u+x)y)-{r_3} (2 y u+u+({r_3}^2+2) y-4 y (x+y)))),\\
A^{3'}_2 &=&-\frac{{N_0} {m_W}^{9/2}} {2 {L_2} \sqrt{r_3}} (2 {d_1} {r_3} ({r_1} ({r_1}+{r_2})+2 u)+{d_2} (({r_3}-2 {r_2}) {r_1}^2 +({r_2} ({r_3}-2 {r_2})-4 x+4) {r_1}-2 (2 {r_2}-{r_3}) (u-2y))),\\
A^{3'}_3 &=&\frac{{N_0} {m_W}^{9/2}} {2 {L_2} \sqrt{r_3}} (2 {d_1} {r_3}^3+{d_2} ({r_3}^3-2 {r_2}{r_3}^2+(4 y-2) {r_3}+4 ({r_1}+{r_2}) x)),\\
A^{3'}_4 &=&-\frac{{N_0} {m_W}^{9/2}} {2 {L_2} \sqrt{r_3}}(-2 {d_1} {r_3}+4 {d_1} (x+y) {r_3}+{d_2} (4{r_1}+(2 {r_2}-{r_3}) (1+2 z))),\\
A^{3'}_5 &=&-\frac{{N_0} {m_W}^{9/2}} {2 {L_2} \sqrt{r_3}} ((2 {d_1} {r_3} ({r_1} ({r_1}+{r_2})+2 u))+({d_2} (({r_3}-2 {r_2}) {r_1}^2+({r_2} ({r_3}-2 {r_2})+4 z) {r_1}+2 {r_3} u+4 {r_2} y-4 {r_3} y))),\\
A^{3'}_6 &=&\frac{{N_0} {m_W}^{9/2} \sqrt{r_3}} {2 L_2} (2 {d_1} {r_3}^2+{d_2} ({r_3}^2+2 {r_2} {r_3}+4 y-2)),\\
A^{3'}_7 &=&\frac{{N_0} {m_W}^{9/2}} {2 L_2 \sqrt{r_3}} (2 {d_1} {r_3} (2 x+4 y-1)+{d_2} ({r_2} (2-4 y)-{r_3} (1+2 z))),\\
A^{3'}_8 &=&\frac{2 {d_1} {N_0} {m_W}^{9/2} \sqrt{r_3}} {L_2} ( y-z),\;
A^{3'}_9 =-\frac{2 {d_2} {N_0} {m_W}^{9/2}} {{L_2} \sqrt{r_3}} ({r_1}+{r_2}),\;
A^{3'}_{10} = \left(\frac{d_2 r_2}{d_1 r_3}\right) A^{3'}_{12}= -\frac{2 {d_2} {N_0} {m_W}^{9/2} {r_2}} {{L_2} \sqrt{r_3}} .
\end{eqnarray}
\end{widetext}

Non-zero coefficients $A^2_j$ and $A^4_j$ are
\begin{widetext}
\begin{eqnarray}
A^2_1 &=&\frac{2 {L_2} {m_W}^{7/2}} {\sqrt{r_3}} (2 {d_1} {r_3} ({r_2} u+{r_1} v)+{d_2} (2({r_3}^2+2 {r_2} {r_3} -2 x){r_1}^2+({r_3}^3+(u-4 {r_2}^2-x+2 z) {r_3}\nonumber\\
&&+2 {r_2} (u+x)) {r_1}-u(({r_2}-2 {r_3}) (2 {r_2}+{r_3})+4 x)-2 ({r_3}({r_2}+{r_3})-2 x) y)),\\
A^2_2 &=&\frac{2 {L_2} {m_W}^{7/2}} {\sqrt{r_3}} {({d_2} ({r_1} (4 {r_1}-2 {r_2}+{r_3})+8 (u-y))-2 {d_1} {r_1} {r_3})},\\
A^2_3 &=&\frac{4 {d_2} {L_2} {m_W}^{7/2}} {\sqrt{r_3}} {({r_1} {r_3}-({r_2}+2 {r_3}) {r_3}+2 x)},\\
A^2_4 &=&\frac{2 {L_2} {m_W}^{7/2}} {\sqrt{r_3}} {(2 {d_1} ({r_2}-{r_1}) {r_3}+{d_2} (({r_1}-{r_2}) (2 {r_2}-{r_3})+4 (x-1)))},\\
A^2_5 &=&\frac{2 {L_2} {m_W}^{7/2}} {\sqrt{r_3}} {({d_2} ({r_1} (4 {r_1}-2 {r_2}+{r_3})+4 (u-y))-2 {d_1} {r_1} {r_3})},\\
A^2_6 &=& \left(\frac{d_2}{2d_1}\right) A^2_8 =4 {d_2} {L_2} {m_W}^{7/2} (r_1-r_2) \sqrt{r_3} ,\\
A^2_7 &=&\frac{2 {L_2} {m_W}^{7/2}} {\sqrt{r_3}} {(2 {d_1} (3 {r_1}-{r_2}) {r_3}-{d_2} ({r_1}-{r_2}) (2 {r_2}-{r_3}))}, \\
A^2_9 &=&A^2_{10}= -\frac{8 {d_2} {L_2} {m_W}^{7/2}} {\sqrt{r_3}},\\
A^{4'}_1 &=&\frac{{N_0} {m_W}^{9/2}} {2 {L_1} \sqrt{r_3}} (2 {d_1} {r_3} (y {r_3}^2+{r_1} ({r_2}-{r_1}) x
+u (2 y-1))+{d_2} (({r_3} (x-2)-2 {r_2} x) {r_1}^2+(2 x {r_2}^2+{r_3} (3 x+4 y-2) {r_2}\nonumber\\
&&+2 ({r_3}^2-2 x) z) {r_1}-2 {r_2} u+4 {r_2}^2 {r_3} y+4 {r_2} (u+x) y-{r_3} (2 y u+u+({r_3}^2+2) y-4 y (x+y)))),\\
A^{4'}_2 &=&\frac{{N_0} {m_W}^{9/2}} {2 {L_1} \sqrt{r_3}} {(2 {d_1} {r_3} ({r_1}^2-{r_2} {r_1}+2 u)+{d_2} ((2 {r_2}+{r_3}) {r_1}^2-(2 {r_2}^2+{r_3} {r_2}+4 x-4) {r_1}+2 (2 {r_2}+{r_3}) (u-2 y)))},\\
A^{4'}_3 &=&-\frac{{N_0} {m_W}^{9/2}} {2 {L_1} \sqrt{r_3}} {(2 {d_1} {r_3}^3+{d_2} ({r_3}^3+2
{r_2} {r_3}^2+(4 y-2) {r_3}+4({r_1}-{r_2}) x))},\\
A^{4'}_4 &=&\frac{{N_0} {m_W}^{9/2}} {2 {L_1} \sqrt{r_3}} {(-2 {d_1} {r_3}+4 {d_1} (x+y) {r_3}+{d_2} (4 {r_1}-(2 {r_2}+{r_3}) (1+2 z)))},\\
A^{4'}_5 &=&\frac{{N_0} {m_W}^{9/2}} {2 {L_1} \sqrt{r_3}} (2 {d_1} {r_3} ({r_1}^2-{r_2} {r_1}+2 u)+{d_2} ((2 {r_2}+{r_3}) {r_1}^2 -({r_2} (2 {r_2}+{r_3})-4 z) {r_1}+2 {r_3} u-4 ({r_2}+{r_3}) y)),\\
A^{4'}_6 &=&-\frac{{N_0} {m_W}^{9/2} \sqrt{r_3}} {2 {L_1} } (2 {d_1} {r_3}^2+{d_2} ({r_3}^2-2 {r_2} {r_3}+4 y-2)),\\
A^{4'}_7 &=&-\frac{{N_0} {m_W}^{9/2} } {2 {L_1} \sqrt{r_3}} (2 {d_1} {r_3} (2 x+4 y-1)-{d_2} ({r_3} (1+2 z)-{r_2} (4 y-2))) ,\\
A^{4'}_8 &=&-\frac{2 {d_1} {N_0} {m_W}^{9/2} \sqrt{r_3}} {L_1} ( y-z),\;
A^{4'}_9 =\frac{2 {d_2} {N_0} {m_W}^{9/2} } {{L_1} \sqrt{r_3}} {(r_1-r_2)} ,\;
A^{4'}_{10} = -\left(\frac{d_2 r_2}{d_1 r_3} \right) A^{4'}_{12}= -\frac{2 {d_2} {r_2} {N_0} {m_W}^{9/2}} {{L_1} \sqrt{r_3}} .
\end{eqnarray}
\end{widetext}

\noindent{\it $\diamond$ Non-zero coefficients for $|(c\bar{Q'})_{\bf 1}[^1P_1]\rangle$.}

Non-zero coefficients $A^1_j$ and $A^{3'}_j$ are
\begin{widetext}
\begin{eqnarray}
A^1_1 &=&\frac{2 {L_1} {m_W}^{5/2}} {{r_1}^2 {r_3}^{3/2}} ({d_2} {r_3}^2 (-2 {r_3} {r_1}^3+(2 x-2 {r_2} {r_3}) {r_1}^2+({r_2} (2 x-v)+2 {r_3} (y-u)) {r_1}+{r_2}^2 u+2 u x-2 x y)\nonumber\\
&&-2 {d_{21}} {r_1}^2 (({r_1}^2-{r_2}^2+1) {r_3}-2 {r_1} x) (2 {r_3} {r_1}^2+(2 {r_2} {r_3}-v) {r_1}+{r_2} u+2 {r_3} u-2 {r_3} y)),\\
A^1_2 &=&-\frac{2 {L_1} {m_W}^{5/2}} {{r_1}^2 {r_3}^{3/2}} (2 {d_{21}} (({r_1}^2-{r_2}^2+1) {r_3}-2 {r_1} x) {r_1}^3+{d_2} {r_3}^2 (2 {r_1}^2+{r_2} {r_1}+4 u-4 y)),\\
A^1_3 &=&\frac{4 {d_2} {L_1} {m_W}^{5/2} \sqrt{r_3}} {{r_1}^2} ({r_3}^2-x),\\
A^1_4 &=&\frac{2 {L_1} {m_W}^{5/2}} {{r_1}^2 {r_3}^{3/2}} ({d_2} {r_3}^2 ({r_2} ({r_1}+{r_2})-2 x+2)-2 {d_{21}} {r_1}^2 ({r_1}+{r_2}) (({r_1}^2-{r_2}^2+1) {r_3}-2 {r_1} x)),\\
A^1_5 &=&-\frac{2 {L_1} {m_W}^{5/2}} {{r_1}^2 {r_3}^{3/2}} (2 {d_{21}} (3 {r_3} {r_1}^3+2 ({r_3}^2+2 u-x) {r_1}^2+(-3 {r_3} {r_2}^2-2  {r_3}^2 {r_2}+4 u {r_2}\nonumber\\
 &&+{r_3}+4 {r_3} u-6 {r_3} y) {r_1}+2 {r_2} {r_3} (y-2 u)) {r_1}^2+{d_2} {r_3}^2 ({r_2} {r_1}-2 w)),\\
A^1_6 &=&\frac{8 {d_{21}} {L_1} {m_W}^{5/2}} {\sqrt{r_3}} (r_1+r_2) ((r_1+r_2) {r_3}-x),\\
A^1_7 &=&-\frac{2 {L_1} {m_W}^{5/2}} {{r_1}^2 {r_3}^{3/2}} (2 {d_{21}} ({r_1}+{r_2}) (({r_1}^2-{r_2}^2+1) {r_3}-2 {r_1} x) {r_1}^2+4 {d_{22}} (2 {r_3} {r_1}^3-({r_3}^2+2 v) {r_1}^2\nonumber\\
&&+(-2 {r_3} {r_2}^2-{r_3}^2 {r_2}+2 u {r_2}+{r_3}+2 {r_3} u-3 {r_3} y) {r_1}+{r_2} {r_3} (y-2 u)) {r_1}^2 +{r_3} (4 {d_{11}} (-2 {r_1}^3 -2 ({r_2}-{r_3}) {r_1}^2\nonumber\\
&&+({r_3}^2+2 {r_2} {r_3}-2 u+2 y) {r_1}+4 {r_3} u-2 {r_3} y) {r_1}^2+{d_2} {r_2}({r_1}+{r_2}) {r_3})),\\
A^1_8 &=&\frac{8 {L_1} {m_W}^{5/2}} {\sqrt{r_3}} ({d_{22}} ({r_1}+{r_2}) (({r_1}+{r_2}) {r_3}-x)-{d_{11}} (-2 {r_3}^3+{r_2} {r_3}^2+2 x {r_3}+2 {r_2} u+{r_1} ( {r_3}^2-2 v))),\\
A^1_9 &=&\frac{4 {L_1} {m_W}^{5/2}} {{r_1}^2 \sqrt{r_3}} ({d_2} {r_3}-2 {d_{21}} {r_1}^2
({r_1}+{r_2})),\\
A^1_{10} &=&\frac{4 {d_2} {L_1} {m_W}^{5/2} \sqrt{r_3}} {{r_1}^2},\\
A^1_{11} &=&-\frac{8 {L_1} {m_W}^{5/2}} {\sqrt{r_3}} ({d_{22}} ({r_1}+{r_2})-2 {d_{11}}({r_1}-{r_3})),\\
A^{3'}_1 &=&\frac{{N_0} {m_W}^{7/2}} {2 {L_2} {r_1}^2 {r_3}^{3/2}} (2 {d_{21}} (({r_1}^2-{r_2}^2+1) {r_3}-2 {r_1} x) (2 {r_1} {r_3}+u+{r_1} ({r_1}+{r_2}-2 {r_3}) x-({r_3} (2 {r_1}-2 {r_2}+{r_3})+2 u) y) {r_1}^2\nonumber\\
&&+{d_2} {r_3}^2 (({r_2} x+2 {r_3} z) {r_1}^2+x({r_2}^2+2 x-2) {r_1}+2 ({r_2} {r_3}+x) y {r_1}+{r_2} (u-({r_3}^2+2 (u+x)) y))),\\
A^{3'}_2 &=&-\frac{{N_0} {m_W}^{7/2}} {2 {L_2} {r_1}^2 {r_3}^{3/2}} (2 {d_{21}} ({r_1} ({r_1}+{r_2})+2 u) (({r_1}^2-{r_2}^2+1) {r_3}-2 {r_1} x) {r_1}^2\nonumber\\
 &&+{d_2} {r_3}^2 ({r_2} {r_1}^2+({r_2}^2+2 x-2) {r_1}+2 {r_2} (u-2 y))),\\
A^{3'}_3 &=&\frac{{N_0} {m_W}^{7/2} \sqrt{r_3}} {2 {L_2} {r_1}^2} (2 {d_{21}} (({r_1}^2-{r_2}^2+1) {r_3}-2 {r_1} x) {r_1}^2+{d_2} {r_2} {r_3}^2-2 {d_2} ({r_1}+{r_2}) x),\\
A^{3'}_4 &=&-\frac{{N_0} {m_W}^{7/2}} {2 {L_2} {r_1}^2 {r_3}^{3/2}} (2 {d_{21}} (({r_1}^2-{r_2}^2+1) {r_3}-2 {r_1} x) (1-2 z) {r_1}^2 -{d_2} {r_3}^2 ({r_2} (1+2 z)+2 {r_1})),\\
A^{3'}_5 &=&\frac{{N_0} {m_W}^{7/2}} {2 {L_2} {r_1}^2 {r_3}^{3/2}} (2 {d_{21}} {r_1}^2 (4 {r_3} y^2-2 ({r_3} (2 {r_1}^2+{r_3} {r_1}+{r_2} ({r_3}-2 {r_2})+1)+4 {r_1} u) y\nonumber\\
&&+{r_1} (4 u+({r_1}+{r_2}) ({r_3} ({r_1}^2-{r_2}^2-4 x+3)+2 {r_1} x)))-{d_2} {r_3}^2 ({r_2} {r_1}^2+({r_2}^2-2 z) {r_1}-2 {r_2} y)),\\
A^{3'}_6 &=&\frac{{N_0} {m_W}^{7/2}} {2 {L_2} {r_1}^2 \sqrt{r_3}} (2 {d_{21}} {r_1}^2 ({r_1}^2 {r_3}^2-({r_3}+{r_2} ({r_2} {r_3}+2 x-2)) {r_3}+2 {r_1} (2 y-1) {r_3}+2 x-4 ({r_2} {r_3}+x) y)-{d_2} {r_2} {r_3}^3),\\
A^{3'}_7 &=&-\frac{{N_0} {m_W}^{7/2}} {2 {L_2} {r_1}^2 {r_3}^{3/2}} (-8 {d_{22}} x {r_1}^5-4 {d_{22}} {r_3} {r_1}^4-8 {d_{22}} {r_2} x {r_1}^4+8 {d_{22}} {r_3} x {r_1}^4-4 {d_{22}} {r_2} {r_3} {r_1}^3-8 {d_{22}} u {r_1}^3\nonumber\\
&&+8 {d_{22}} {r_2} {r_3} x {r_1}^3-8 {d_{22}} {r_3} y^2 {r_1}^2+2 {d_{21}} (({r_1}^2-{r_2}^2+1) {r_3}-2 {r_1} x) (2 y-1) {r_1}^2+4 {d_{11}} {r_3} ((-2 {r_1}^2+2 ({r_2}\nonumber\\
&&+{r_3}) {r_1}+{r_3}^2-2 {r_2} {r_3}+2 u) y-2 {r_1} ({r_1}-{r_3}) (2 x-1)) {r_1}^2+{d_2} {r_2} {r_3}^2+2 (2 {d_{22}} {r_1}^2 ({r_3} (2 {r_1}^2-2 {r_2}^2\nonumber\\
&&+({r_1}+{r_2}) {r_3}+1)+4 {r_1} u)-{d_2} {r_2} {r_3}^2) y),\\
A^{3'}_8 &=&-\frac{2 {N_0} {m_W}^{7/2}} {{L_2} \sqrt{r_3}} ({d_{22}} ({r_3}^2-{r_1} ( y-z) {r_3}+{r_2} (y-z) {r_3}-x+2 x y)\nonumber\\
&&+{d_{11}} ((y-z) {r_3}^2+2 ({r_1}+{r_2}) x {r_3}-2 u-2 {r_1} ({r_1}+{r_2}) x+2 u (x+2 y))),\\
A^{3'}_9 &=&-\frac{{N_0} {m_W}^{7/2}} {{L_2} {r_1}^2 \sqrt{r_3}} (2 {d_{21}} {r_1}^2 (({r_1}+{r_2}) {r_3}+2 y-1) -{d_2} ({r_1}+{r_2}) {r_3}),\\
A^{3'}_{10} &=&\frac{{N_0} {m_W}^{7/2}} {{L_2} {r_1}^2 {r_3}^{3/2}} (2 {d_{21}} (({r_1}^2-{r_2}^2+1) {r_3}-2 {r_1} x) {r_1}^2+{d_2} {r_2} {r_3}^2),\\
A^{3'}_{11} &=&\frac{2 {N_0} {m_W}^{7/2}} {{L_2} \sqrt{r_3}} ({d_{11}} ({r_3}^2-2 ({r_1}+{r_2}) {r_3}+2 {r_1} ({r_1}+{r_2})+2 u)-{d_{22}} (({r_1}+{r_2}) {r_3}+2 y-1)).
\end{eqnarray}
\end{widetext}

Non-zero coefficients $A^2_j$ and $A^4_j$ are
\begin{widetext}
\begin{eqnarray}
A^2_1 &=&-\frac{2 {L_2} {m_W}^{5/2}} {{r_1}^2 {r_3}^{3/2}} (2 {d_{21}} (({r_1}^2-{r_2}^2+1) {r_3}-2 {r_1} x) (2 {r_3} {r_1}^2-(v+2 {r_2}  {r_3}) {r_1}-{r_2} u+2 {r_3} u-2 {r_3} y) {r_1}^2\nonumber\\
 &&+{d_2} {r_3}^2 (2 {r_3} {r_1}^3-2 ({r_2} {r_3}+x) {r_1}^2+({r_2} ({r_3}^2+u+x)+2 {r_3} (u-y)) {r_1}-u ({r_2}^2+2 x)+2 x y)),\\
A^2_2 &=&-\frac{2 {L_2} {m_W}^{5/2}} {{r_1}^2 {r_3}^{3/2}} (2 {d_{21}} (({r_1}^2-{r_2}^2+1) {r_3}-2 {r_1} x) {r_1}^3+{d_2} {r_3}^2 (2  {r_1}^2-{r_2} {r_1}+4 u-4 y)),\\
A^2_3 &=&\frac{4 {d_2} {L_2} {m_W}^{5/2} \sqrt{r_3}} {{r_1}^2} ({r_3}^2-x),\\
A^2_4 &=&\frac{2 {L_2} {m_W}^{5/2}} {{r_1}^2 {r_3}^{3/2}} ({d_2} {r_3}^2 ({r_2}^2-{r_1} {r_2}-2 x+2)-
2 {d_{21}} {r_1}^2 ({r_1}-{r_2}) (({r_1}^2-{r_2}^2+1) {r_3}-2 {r_1} x)),\\
A^2_5 &=&-\frac{2 {L_2} {m_W}^{5/2}} {{r_1}^2 {r_3}^{3/2}} (2 {d_{21}} (3 {r_3} {r_1}^3+2 ({r_3}^2+2 u-x) {r_1}^2 +(-3 {r_3} {r_2}^2+2 {r_3}^2 {r_2}-4 u {r_2}+{r_3}+4 {r_3} u\nonumber\\
&&-6 {r_3} y) {r_1}+2 {r_2} {r_3} (2 u-y)) {r_1}^2+{d_2} {r_3}^2 (2 {r_1}^2-{r_2} {r_1}+2 u-2 y)),\\
A^2_6 &=&-\frac{8 {d_{21}} {L_2} {m_W}^{5/2}} {\sqrt{r_3}} ({r_1}-{r_2}) (-{r_1} {r_3}+{r_2} {r_3}+x),\\
A^2_7 &=&-\frac{2 {L_2} {m_W}^{5/2}} {{r_1}^2 {r_3}^{3/2}} (2 {d_{21}} ({r_1}-{r_2}) (({r_1}^2-{r_2}^2+1) {r_3}-2 {r_1} x)  {r_1}^2+4 {d_{22}} (2 {r_3} {r_1}^3+({r_3}^2+2 u-2 x) {r_1}^2+(-2 {r_3} {r_2}^2\nonumber\\
&&+{r_3}^2 {r_2}-2 u {r_2}+{r_3}+2 {r_3} u-3 {r_3} y) {r_1}+{r_2} {r_3} (2 u-y)) {r_1}^2+{r_3} (4 {d_{11}} (-2 {r_1}^3+2 ({r_2}\nonumber\\
&&+{r_3}){r_1}^2+({r_3}^2-2 {r_2} {r_3}-2 u+2 y) {r_1}+4 {r_3} u-2 {r_3} y) {r_1}^2+{d_2} {r_2} ({r_2}-{r_1}) {r_3})),\\
A^2_8 &=&\frac{8 {L_2} {m_W}^{5/2}} {\sqrt{r_3}} ({d_{22}} ({r_1}-{r_2}) ({r_1} {r_3}-{r_2} {r_3}-x)+{d_{11}} (2 {r_3}^3 +{r_2} {r_3}^2-2 x {r_3}+2 {r_2} u+{r_1} (-3 {r_3}^2-2 u+2 x))),\\
A^2_9 &=&\frac{4 {L_2} {m_W}^{5/2}} {{r_1}^2 \sqrt{r_3}} (2 {d_{21}} ({r_2}-{r_1}){r_1}^2+{d_2} {r_3}),\\
A^2_{10} &=&\frac{4 {d_2} {L_2} {m_W}^{5/2} \sqrt{r_3}} {{r_1}^2} ,\\
A^2_{11} &=&\frac{8 {L_2} {m_W}^{5/2}} {\sqrt{r_3}} ({d_{22}} ({r_2}-{r_1})+2 {d_{11}}({r_1}-{r_3})),\\
A^{4'}_1 &=&\frac{{N_0} {m_W}^{7/2}} {2 {L_1} {r_1}^2 {r_3}^{3/2}} (2 {d_{21}} (({r_1}^2-{r_2}^2+1) {r_3}-2 {r_1} x) (-x {r_1}^2+({r_2} x-2 {r_3} z) {r_1}+{r_3} (2{r_2}+{r_3}) y+u (2 y-1)) {r_1}^2\nonumber\\
&&+{d_2} {r_3}^2 (({r_2} x-2 {r_3} z) {r_1}^2-x ({r_2}^2+2 x-2) {r_1}+2 ({r_2} {r_3}-x) y {r_1}+{r_2} (u-({r_3}^2+2 (u+x)) y))),\\
A^{4'}_2 &=&\frac{{N_0} {m_W}^{7/2}} {2 {L_1} {r_1}^2 {r_3}^{3/2}} (2 {d_{21}} ({r_1}^2-{r_2} {r_1}+2 u) (({r_1}^2-{r_2}^2+1) {r_3}-2 {r_1}  x) {r_1}^2\nonumber\\
 &&+{d_2} {r_3}^2 (-{r_2} {r_1}^2+({r_2}^2+2 x-2) {r_1}-2 {r_2} (u-2 y))),\\
A^{4'}_3 &=&\frac{{N_0} {m_W}^{7/2} \sqrt{r_3}} {2 {L_1} {r_1}^2} ({d_2} ({r_2} ({r_3}^2-2 x)+2 {r_1} x)
-2 {d_{21}} {r_1}^2 (({r_1}^2-{r_2}^2+1) {r_3}-2 {r_1} x)),\\
A^{4'}_4 &=&\frac{{N_0} {m_W}^{7/2}} {2 {L_1} {r_1}^2 {r_3}^{3/2}} (2 {d_{21}} {r_1}^2 (({r_1}^2
-{r_2}^2+1) {r_3}-2 {r_1} x) (1-2 z)-{d_2} {r_3}^2 (2 {r_1}-{r_2} (1+2 z))),\\
A^{4'}_5 &=&-\frac{{N_0} {m_W}^{7/2}} {2 {L_1} {r_1}^2 {r_3}^{3/2}} (2 {d_{21}} (4 {r_3} y^2
-2 ({r_3} (2 {r_1}^2+{r_3} {r_1}-{r_2} (2 {r_2}+{r_3})+1)+4 {r_1} u) y+{r_1} (4 u+({r_1}\nonumber\\
&&-{r_2}) ({r_3} ({r_1}^2-{r_2}^2-4 x+3)+2 {r_1} x))) {r_1}^2+{d_2} {r_3}^2 ({r_1} (({r_1}-{r_2}) {r_2}-2 x+2) -2 ({r_1}+{r_2}) y)),\\
A^{4'}_6 &=&-\frac{{N_0} {m_W}^{7/2}} {2 {L_1} {r_1}^2 \sqrt{r_3}} ({d_2} {r_2} {r_3}^3+2 {d_{21}} {r_1}^2 ({r_3} ({r_3} {r_1}^2-2 {r_1} -{r_3}-{r_2} ({r_2} {r_3}-2 x+2))+2 x+4 (({r_1}+{r_2}){r_3}-x) y)),\\
A^{4'}_7 &=&\frac{{N_0} {m_W}^{7/2}} {2 {L_1} {r_1}^2 {r_3}^{3/2}} (-8 {d_{22}} x {r_1}^5-4 {d_{22}} {r_3} {r_1}^4 +8 {d_{22}} {r_2} x {r_1}^4+8 {d_{22}} {r_3} x {r_1}^4+4 {d_{22}} {r_2} {r_3} {r_1}^3\nonumber\\
&&-8 {d_{22}} u {r_1}^3-8 {d_{22}} {r_2} {r_3} x {r_1}^3-8 {d_{22}} {r_3} y^2 {r_1}^2+2 {d_{21}} (({r_1}^2-{r_2}^2+1) {r_3}-2 {r_1} x) (2 y-1) {r_1}^2\nonumber\\
&&+4 {d_{11}} {r_3} (({r_3}^2+2 ({r_1}+{r_2}) {r_3}-2 {r_1} ({r_1}+{r_2})+2 u) y-2 {r_1} ({r_1}-{r_3}) (2 x-1)) {r_1}^2\nonumber\\
&&-{d_2} {r_2} {r_3}^2+2 (2 {d_{22}} ({r_3} (2 {r_1}^2+{r_3} {r_1}-{r_2} (2 {r_2}+{r_3})+1)+4 {r_1} u) {r_1}^2+{d_2} {r_2} {r_3}^2) y),\\
A^{4'}_8 &=&\frac{2 {N_0} {m_W}^{7/2}} {{L_1} \sqrt{r_3}} ({d_{22}} ({r_3}^2-{r_1} (y-z) {r_3}-{r_2} (y-z) {r_3}-x+2 x y)+{d_{11}} ((y-z) {r_3}^2\nonumber\\
&&+2 ({r_1}-{r_2}) x {r_3}+2 ({r_1} ({r_2}-{r_1}) x+u (y-z)))),\\
A^{4'}_9 &=&\frac{{N_0} {m_W}^{7/2}} {{L_1} {r_1}^2 \sqrt{r_3}} (2 {d_{21}} ({r_1} {r_3}-{r_2} {r_3}+2 y-1) {r_1}^2+{d_2} ({r_2}-{r_1}) {r_3}),\\
A^{4'}_{10} &=&-\frac{{N_0} {m_W}^{7/2}} {{L_1} {r_1}^2 {r_3}^{3/2}} (2 {d_{21}} {r_1}^2 (({r_1}^2-{r_2}^2+1) {r_3}-2 {r_1} x)-{d_2} {r_2} {r_3}^2),\\
A^{4'}_{11} &=&-\frac{2 {N_0} {m_W}^{7/2}} {{L_1} \sqrt{r_3}} ({d_{11}} (2 {r_1}^2-2 ({r_2}+{r_3}) {r_1}+{r_3}^2+2 {r_2} {r_3}+2 u)+{d_{22}} (-{r_1} {r_3}+{r_2} {r_3}-2 y+1)).
\end{eqnarray}
\end{widetext}

\noindent{\it $\diamond$ Non-zero coefficients for $|(c\bar{Q'})_{\bf 1}[^3P_J]\rangle (J=0,1,2)$.}

It is noted that $\varepsilon^{0,2}_{\alpha\beta}$ is the symmetric tensor and $\varepsilon^{1}_{\alpha\beta}$ is the anti-symmetric tensor, and the fact that $\varepsilon^{1}_{\alpha\alpha}=\varepsilon^{2}_{\alpha\alpha}=0$. so the terms involving the following coefficients have no contributions to the square of the amplitude, and practically, we can safely set the coefficients before them to be zero
\begin{equation}
\begin{array}{c}
A^i_j(|(c\bar{Q'})_{\bf 1}[^3P_0]\rangle)=0 \;\;\;\;\;\;\;\;{\rm for}\;i=(1-4),\\
j=(8,9,11,12,13,15,16,17,19,20,21,22,23,24,25)\\
\end{array}
\end{equation}
\begin{equation}
\begin{array}{c}
A^i_j(|(c\bar{Q'})_{\bf 1}[^3P_1]\rangle)=0 \;\;\;\;\;\;\;\;{\rm for}\;i=(1-4), \\
j=(1,2,5,26,27,29,31,33,34)\\
\end{array}
\end{equation}
\begin{equation}
\begin{array}{c}
A^i_j(|(c\bar{Q'})_{\bf 1}[^3P_2]\rangle)=0 \;\;\;\;\;\;\;\;{\rm for}\;i=(1-4),\\
j=(1,2,5,8,11,12,13,16,17,19,20,21) .
\end{array}
\end{equation}

Non-zero coefficients $A^1_j$ and $A^{3'}_j$ for $|(c\bar{Q'})_{\bf 1}[^3P_J]\rangle$ are
\begin{widetext}
\begin{eqnarray}
A^1_1 &=&-\frac{2 {L_1} {m_W}^{5/2}} {{r_1}^2 \sqrt{r_3}} (({r_1}+{r_2}) (2 {d_{21}} ({r_3} {r_1}^2-2 x {r_1}-{r_2}^2 {r_3}+{r_3}) {r_1}^2+{d_2} ({r_3}^3 -3 {r_1} {r_3}^2+2 {r_1}^2 {r_3}-2 {r_1} {r_2} {r_3}-x {r_3}+2 {r_1} x))\nonumber\\
&&+{d_1} (8 {r_3} {r_1}^3+2 ({r_3}^2+4 {r_2} {r_3}+2 u) {r_1}^2+(-5 {r_3}^3+2 {r_2} {r_3}^2-2 u {r_3}+4 x {r_3}+4 {r_2} u){r_1}-{r_2} {r_3} ({r_3}^2+2 u))),\\
A^1_2 &=&\frac{2 {L_1} {m_W}^{5/2}} {{r_1}^2 {r_3}^{3/2}} ({d_2} (4 {r_3} {r_1}^4+(8 {r_2} {r_3}-6 {r_3}^2-4 v) {r_1}^3-(3 {r_3}^3+4 {r_2} {r_3}^2+2 (-2 {r_2}^2+u-3 x+y+1) {r_3}-4 {r_2} u) {r_1}^2\nonumber\\
&&+{r_3} ({r_3}^3+(-2 {r_2}^2-2 u-2 x+3 y+1) {r_3}-2 {r_2} (u+y)) {r_1}+{r_2} {r_3}^2 y)-{r_1} (2 {d_{21}} {r_1} (({r_1}^2-{r_2}^2+1) {r_3}\nonumber\\
&&-2 {r_1} x) ({r_1} (2 ({r_1}+{r_2})+{r_3})+2 (u-y))+{d_1} {r_3} (12 {r_3} {r_1}^2+2 ({r_3}^2+2 {r_2} {r_3}+2 u) {r_1}-{r_3} ({r_3}^2-6 u+4 y)))),\\
A^1_3 &=&-\frac{8{L1}{m_W}^{5/2}}{\sqrt{r_3}}({d_{11}}{r_3}({r_2} u - {r_1}v) - {d_{21}} ({r_1} + {r_2})(({r_1}^2 - {r_2}^2 + 1){r_3} - 2{r_1} x) \nonumber\\
&&+ {d_{22}}(2{r_3}{r_1}^3 + (-{r_3}^2 + 4{r_2}{r_3} + u +x){r_1}^2 - ({r_3}^3 + {r_2}{r_3}^2 + (-2{r_2}^2 - u - 2 x +y + 1){r_3}\nonumber\\
 &&- {r_2} (2 u +x)){r_1} + {r_2}^2 u + {r_2}{r_3} (u - y) -2({r_3}^2 - x) (u - y))),\\
A^1_4 &=&-\frac{8 d_{21} L_1 {m_W}^{5/ 2}}{\sqrt{r_3}}(3{r_3}{r_1}^3 + (-{r_3}^2 + 4{r_2}{r_3} + u -x){r_1}^2 + ({r_3}{r_2}^2 + (-{r_3}^2 + 2 u +  x){r_2}\nonumber\\
 && - {r_3}({r_3}^2 - u - 2 x + y)){r_1} + {r_2}^2 u + {r_2}{r_3} (u - y) - 2({r_3}^2 - x) (u - y)),\\
A^1_5 &=&\frac{2 {L_1} {m_W}^{5/2}} {{r_1}^2 \sqrt{r_3}} {({d_2} ({r_1}+{r_2}) ({r_3}-2 {r_1})-4 {d_1} {r_1} {r_3})},\\
A^1_{7}&=&\frac{8{L_1}{m_W}^{5/ 2}}{\sqrt{r_3}}({d_{22}}(3{r_1}^2 + {r_2}{r_1} + 4 u -4 y) - {d_{11}}{r_1}{r_3}),\\
A^1_{9}&=&\frac{8 d_{21}  {L_1}{m_W}^{5/2}} {\sqrt{r_3}}(3{r_1}^2 + {r_2}{r_1} + 4 u - 4 y),\\
A^1_{10}&=& \left(\frac{d_{21}}{d_{22}}\right) A^1_{14}=\frac{8 d_{21} {L_1}{m_W}^{5/2} }{\sqrt{r_3}}(({r_1} + {r_2} - 2{r_3}){r_3} + 2 x),\\
A^1_{11} &=&\frac{4 {d_2} {L_1} {m_W}^{5/2}} {\sqrt{r_3}},\\
A^1_{12} &=&\frac{2 {L_1} {d_2} {m_W}^{5/2} \sqrt{r_3}} {{r_1}^2} {(r_1+r_2)},\\
A^1_{13} &=&\frac{2 {L_1} {m_W}^{5/2}} {{r_1}^2 {r_3}^{3/2}} {(4 {d_{21}} (({r_1}^2-{r_2}^2+1) {r_3}-2 {r_1} x) {r_1}^2+{d_2} ({r_1}+3 {r_2}) {r_3}^2)},\\
A^1_{15} &=&-\frac{8{L_1}{m_W}^{5/ 2}}{\sqrt{r_3}}({d_{11}} ({r_1} + {r_2}){r_3} + {d_{22}}(-{r_1}^2 + {r_3}{r_1} + {r_2} ({r_2} + {r_3}) - 2 x + 2)),\\
A^1_{17} &=&\frac{4 {L_1} {m_W}^{5/2}} {{r_1}^2 {r_3}^{3/2}} {({d_2} {r_3} ({r_1} ({r_1}+{r_2})-({r_1}+2 {r_2}) {r_3})-2 {d_{21}} {r_1}^2 (({r_1}^2-{r_2}^2+1) {r_3}-2 {r_1} x))},\\
A^1_{18} &=&-\frac{8 d_{21} {L_1} {m_W}^{5/2}}{\sqrt{r_3}}(-{r_1}^2 + {r_3}{r_1} + {r_2} ({r_2} + {r_3}) - 2 x +2),\\
A^1_{19} &=&-\frac{2 {L_1} {m_W}^{5/2}} {{r_1}^2 {r_3}^{3/2}} (2 {d_{21}} ({r_1} {r_3}+2 u) (({r_1}^2-{r_2}^2+1) {r_3}-2 {r_1} x) {r_1}^2\nonumber\\
&&+{r_3}^2 ({r_1} {r_3} ({d_2} {r_1}+{d_1} {r_3})+(2 {d_1} {r_1}+{d_2} ({r_1}+3 {r_2})) u)),\\
A^1_{20} &=&\frac{2 {L_1} {m_W}^{5/2}} {{r_1}^2 {r_3}^{3/2}} (2 {d_{21}} (({r_1}+{r_2}-2 {r_3}) {r_3}+2 x) (({r_1}^2-{r_2}^2+1) {r_3}-2 {r_1} x) {r_1}^2+{r_3}^2 ({d_2} {r_3} ({r_1} ({r_1}+{r_2})\nonumber\\
&&-({r_1}+3 {r_2}) {r_3})+{d_1} ({r_1}+{r_2}) ({r_3}^2+2 u)+{d_2} ({r_1}+3 {r_2}) x)),\\
A^1_{21} &=&\frac{2 {L_1} {m_W}^{5/2}} {{r_1}^2 {r_3}^{3/2}} (2 {d_{21}} (({r_1}^2-{r_2}^2+1) {r_3}-2 {r_1} x) ({r_1} (2 ({r_1}+{r_2})+{r_3})+4 u-2 y) {r_1}^2\nonumber\\
 &&+{r_3}^2 ({d_1} {r_1} ({r_3}^2+2 u)+{d_2} ((2 {r_2}+{r_3}) {r_1}^2+(2 {r_2}^2-{r_3}^2+ x-z) {r_1}+{r_2} (4 u-y)))),\\
\frac {A^1_{22}} {d_{21}} &=&\frac {A^1_{25}} {d_{22}}=\frac {A^1_{27}} {d_{21}}=\frac {A^1_{28}} {d_{22}}= -\frac{16 {L_1} {m_W}^{5/2}} {\sqrt{r_3}},\\
A^1_{29} &=&\frac{8 {d_{21}} {L_1} {m_W}^{5/2}} {\sqrt{r_3}} {({r_1}^2+{r_2} {r_1}-2 w)},\\
A^1_{30} &=&-\frac{8 L_1 {m_W}^{5/ 2}} {\sqrt{r_3}}(d_{21} ({r_1} + {r_2}) ({r_1} - {r_2} - {r_3}) + {d_{11}}{r_1}{r_3} - {d_{22}}({r_1}^2 + {r_2}{r_1} - 2 w)),\\
\frac{{}A^1_{31}}{d_{21}} &=&\frac{{}A^1_{32}}{d_{22}}=\frac{{}A^1_{34}}{2 d_{11}}=8 {L_1} {m_W}^{5/2} \sqrt{r_3} {(r_1+r_2)},\\
A^1_{33} &=&\frac{8 {L_1} {m_W}^{5/2}} {\sqrt{r_3}} {({d_{11}} (3 {r_1}+{r_2}) {r_3}+{d_{22}} (-{r_1}^2+{r_3} {r_1}+{r_2} ({r_2}+{r_3})))},\\
A^{3'}_1 &=&-\frac{{N_0} {m_W}^{7/2}} {2 {L_2} {r_1}^2 \sqrt{r_3}}  (2 {d_{21}} (({r_1}^2-{r_2}^2+1) {r_3}-2{r_1} x) (y-z) {r_1}^2+{d_2} (2 {r_3} (y-z) {r_1}^2\nonumber\\
&&+((-3 x-6 y+5) {r_3}^2+2 {r_2} (y-z) {r_3}-2 x+4 x y) {r_1}+{r_3} ((x+2 y-2) {r_3}^2+x-2 x y))+{d_1} (4 {r_3} (3 x\nonumber\\
&&+4 y-2) {r_1}^2+2 ((y-z) {r_3}^2+2 {r_2} x {r_3}+2 u (y-z)) {r_1}-{r_3} ({r_3}^2+2 u) (y-z))),\\
A^{3'}_2 &=&-\frac{{N_0} {m_W}^{7/2}} {2 {L_2} {r_1}^2 \sqrt{r_3}} (2 {d_{21}} (({r_1}^2-{r_2}^2+1) {r_3}-2 {r_1} x) (({r_3}-2 {r_2}) y -2 {r_1} z) {r_1}^2+{d_1} {r_3} (4 {r_3} (2 x+3 y-1) {r_1}^2\nonumber\\
&&+2 ({r_3}^2-2 {r_2} {r_3}+2 u) y {r_1}-{r_3} ({r_3}^2+2 u) y)+{d_2} (4 x {r_1}^4+(4 {r_2} x-2 {r_3} (5 x+2 y-3)) {r_1}^3\nonumber\\
&&+((6 x+2 y-5) {r_3}^2-2 {r_2} (x-4 y+1) {r_3}+4 u-8 u y) {r_1}^2+{r_3} (-4 y^2+(-4 {r_2}^2-4 {r_3} {r_2}+3 {r_3}^2+12 u+2) y\nonumber\\
&&+{r_2} {r_3}-6 u) {r_1}-{r_3}^2 ((-2 {r_2}^2+{r_3}^2-2 y+1) y+u (4 y-2)))),\\
A^{3'}_3 &=&\frac{2{m_W}^{7/2}{N_0}} {{L_2}\sqrt{r_3}}({d_{21}}(({r_1}^2 - {r_2}^2 + 1){r_3} - 2{r_1} x) (y-z) + {d_{11}}{r_3}(-y{r_3}^2 + u + {r_1} ({r_1} + {r_2}) x -2 u y) \nonumber\\
&&+ {d_{22}}(x{r_1}^3 + {r_3} (y +3 z){r_1}^2 + (-x{r_2}^2 + {r_3} (3 y - z){r_2} + u - 2 (x - 1) x - 2 (u + x) y\nonumber\\
 &&- {r_3}^2 ( y + 2 z)){r_1}-2{r_2}^2{r_3} y + {r_3}({r_3}^2 + 2 u + 2 z- 1) y + {r_2}(-y{r_3}^2 - u + 2 (u + x) y))),\\
A^{3'}_4 &=&\frac{2{d_{21}}{m_W}^{7/2}{N_0}} {{L_2} \sqrt{r_3}} (-{r_3}{r_1}^4 + (3 x - {r_2}{r_3}){r_1}^3 + (2{r_2} x + {r_3}({r_2}^2 - u -  x + 2 z)){r_1}^2 + ({r_3}{r_2}^3 -x{r_2}^2\nonumber\\
&& + {r_3} (x + 4 y - 2){r_2}+ u + 2 (u - x) x + 2 x -2 (u + x) y - {r_3}^2 (x + 2 z)){r_1} + {r_2}^2{r_3} (u - 2 y)  \nonumber\\
 &&+ {r_2}(-y{r_3}^2 - u +2 (u + x) y)+ {r_3}(({r_3}^2 + 1) y - 2 (x + y) y +u (2 y - 1))),\\
A^{3'}_5 &=&\frac{{N_0} {m_W}^{7/2}} {2 {L_2} {r_1}^2 \sqrt{r_3}} (2 {d_{21}} (({r_1}^2-{r_2}^2+1) {r_3}-2 {r_1} x) {r_1}^2+{d_1} (4 {r_3} {r_1}^2+2 ({r_3}^2-2 {r_2} {r_3}+2 u) {r_1}-{r_3} ({r_3}^2+2 u))\nonumber\\
&&+{d_2} (2 {r_3} {r_1}^2-(3 {r_3}^2+2 {r_2} {r_3}+4 y-2){r_1}+{r_3} ({r_3}^2+2 y-1))),\\
A^{3'}_6 &=&-\frac{2 d_{21} {m_W}^{7/2} {N_0}} {L_2 \sqrt{r_3}} (({r_1}^2 - {r_2}^2 + 1){r_3} - 2{r_1} x)\\
A^{3'}_7 &=&- \frac{2{m_W}^{7/2}{N_0}}{{L_2}\sqrt{r_3}}({d_{11}}{r_3} ({r_1} ({r_1} + {r_2}) +  2 u) - {d_{22}}({r_1}^3 + (2{r_2} - {r_3}){r_1}^2\nonumber\\
 &&+ ({r_2} ({r_2} -{r_3}) + 2 (u + x - 1)){r_1} +2 ({r_2} u - {r_3} u - 2{r_2} y + {r_3} y))),\\
A^{3'}_{9} &=&\frac{2{d_{21}}{m_W}^{7/2}{N_0}}{{L_2}\sqrt{r_3}}({r_1}^3 + (2{r_2} - {r_3}){r_1}^2 + ({r_2}^2 - {r_3}{r_2} +  2 (u + x - 1)){r_1} + 2 ({r_2} (u - 2 y) + {r_3} (y -u)))\nonumber,\\
A^{3'}_{10} &=&\frac{2{d_{21}}{m_W}^{7/2}{N_0}}{{L_2}\sqrt{r_3}}({r_3}^3 - ({r_1} + {r_2}){r_3}^2 + (2 y - 1){r_3} + 2 ({r_1} + {r_2}) x) ,\\
A^{3'}_{11} &=&\frac{{N_0} {m_W}^{7/2}} {{L_2} {r_1}^2 {r_3}^{3/2}} {(2{d_{21}} (({r_1}^2-{r_2}^2+1) {r_3}-2 {r_1} x) {r_1}^3+{d_2} {r_3}  ({r_1}^3+({r_2}-{r_3}) {r_1}^2+(2 u-2 {r_2} {r_3}) {r_1}+2 {r_3} (y-u)))} ,\\
A^{3'}_{12} &=&\frac{{N_0} {m_W}^{7/2}} {2 {L_2} {r_1}^2 \sqrt{r_3}} (2 {d_{21}} ({r_3} {r_1}^2-2 x {r_1}-{r_2}^2 {r_3}+{r_3}) {r_1}^2+{d_1}  {r_3} ({r_3}^2+2 u)+{d_2} {r_3} ({r_3}^2-{r_1} {r_3}+2 y-1)) ,\\
A^{3'}_{13} &=&\frac{{N_0} {m_W}^{7/2}} {2 {L_2} {r_1}^2 {r_3}^{3/2}} (2 {d_{21}} (2 ({r_1}+{r_2})-{r_3}) (({r_1}^2-{r_2}^2+1) {r_3}-2 {r_1} x) {r_1}^2\nonumber\\
&&+{r_3}^2 ({d_2} ({r_3}^2-{r_1} {r_3}-2 {r_2} ({r_1}+{r_2})+2 y-1)-{d_1} ({r_3}^2+2 u))),\\
A^{3'}_{14} &=&\frac{2{m_W}^{7/ 2}{N_0}}{{L_2}\sqrt{r_3}}({d_{11}}{r_3}^3 + {d_{22}}({r_3}^3 - ({r_1} + {r_2}){r_3}^2 + (2 y - 1){r_3} + 2 ({r_1} + {r_2}) x)),\\
A^{3'}_{15} &=&\frac{2{m_W}^{7/2}{N_0} }{{L_2}\sqrt{r_3}}({d_{11}}{r_3} - 2{d_{11}} (x + y){r_3} + {d_{22}} (-3{r_2} + 2{r_3} +  2 ({r_2} - {r_3}) (x + y) - {r_1} (2z + 1))),\\
A^{3'}_{17} &=&\frac{{N_0} {d_2} {m_W}^{7/2}} {{L_2} {r_1}^2 \sqrt{r_3}} {({r_1}(1-2 z)+2 {r_3} z)},\\
A^{3'}_{18} &=&\frac{2{d_{21}}{m_W}^{7/2}{N_0}}{{L_2}\sqrt{r_3}} (-3{r_2} + 2{r_3} +2 ({r_2} - {r_3}) (x + y) - {r_1} (1+2 z)),\\
A^{3'}_{19} &=&-\frac{{N_0} {m_W}^{7/2}} {2 {L_2} {r_1}^2 {r_3}^{3/2}} (2 {d_{21}} ({r_1} {r_3} ({r_1}+{r_2}+2 {r_3})+(2 ({r_1}+{r_2})+{r_3}) u) (({r_1}^2-{r_2}^2+1) {r_3}-2 {r_1} x) {r_1}^2\nonumber\\
&&+{r_3}^2 ({d_1} ({r_1} ({r_1}+{r_2})+u) ({r_3}^2+2 u)+{d_2} ({r_3} {r_1}^3+({r_2}-{r_3}) {r_3} {r_1}^2\nonumber\\
&&+(-3 {r_2} {r_3}^2+u {r_3}-2 {r_2} u) {r_1}-(2 {r_2}^2+{r_3}^2+1) u+2 ({r_3}^2+u) y))),\\
A^{3'}_{20} &=&\frac{{N_0} {m_W}^{7/2}} {2 {L_2} {r_1}^2 {r_3}^{3/2}} (2 {d_{21}} {r_1}^2 (({r_1}^2-{r_2}^2+1) {r_3}-2 {r_1} x) (2 ({r_1}+{r_2}) x+{r_3} (y-z))-{r_3}^2 ({d_2} (2 x {r_2}^2\nonumber\\
&&+x-2 x y-{r_3}^2 (x+2 z)+{r_1} (2 {r_2} x-{r_3} (y-z)))-{d_1} ({r_3}^2+2 u) (y-z))),\\
A^{3'}_{21} &=&\frac{{N_0} {m_W}^{7/2}} {2 {L_2} {r_1}^2 {r_3}^{3/2}} (2 {d_{21}} (({r_1}^2-{r_2}^2+1) {r_3}-2 {r_1} x) (({r_3}-2 {r_2}) y+2 {r_1} (x-z)) {r_1}^2+{r_3}^2 ({d_1} ({r_3}^2+2 u) y\nonumber\\
&&-{d_2} (-2 y^2+({r_3}^2-{r_1} {r_3}+2 ({r_1}-{r_2}) {r_2}-4 x) y+y+2 u+{r_1} ({r_1}-{r_2}+4 {r_2} x)))),\\
\frac{A^{3'}_{22}}{d_{21}} &=&\frac{A^{3'}_{25}}{d_{22}}=-\frac{4  {N_0} {m_W}^{7/2} {r_2}} {{L_2} \sqrt{r_3}},\\
A^{3'}_{26} &=&-\frac{4 {d_{11}} {N_0} {m_W}^{7/2}  \sqrt{r_3}} {{L_2}},\\
\frac{A^{3'}_{27}}{d_{21}} &=&\frac{A^{3'}_{28}}{d_{22}}= -\frac{4  {N_0} {m_W}^{7/2}} {{L_2} \sqrt{r_3}} {(r_1+r_2)},\\
A^{3'}_{29} &=&\frac{2 {d_{21}} {N_0} {m_W}^{7/2}} {{L_2} \sqrt{r_3}} ({r_1}^3+(2 {r_2}-{r_3}) {r_1}^2+({r_2}^2-{r_3} {r_2}+2 (u-z))  {r_1}-2 ({r_3} (u-y)+{r_2} y)),\\
A^{3'}_{30} &=&- \frac{2{m_W}^{7/ 2}{N_0}}{{L_2}\sqrt{r_3}}({d_{11}}{r_3} ({r_1} ({r_1} + {r_2}) +  2 u) + {d_{21}} (2{r_3} z + {r_2} (2 y - 1) + {r_1} (1-2 z))\nonumber\\
 && - {d_{22}}({r_1}^3 + (2{r_2} - {r_3}){r_1}^2 + ({r_2}^2 - {r_3}{r_2} + 2 (u -v)){r_1} - 2 ({r_3} (u - y) + {r_2} y))),\\
A^{3'}_{31} &=&\frac{2 {N_0} {d_{21}} {m_W}^{7/2} \sqrt{r_3}} {L_2} ({r_3} (-r_1+r_2+r_3)+2 y-1),\\
A^{3'}_{32} &=&\frac{2{m_W}^{7/ 2}{N_0}\sqrt{r_3}} {{L_2}}  ({d_{11}}{r_3}^2 + {d_{22}} ({r_3} (-{r_1} + {r_2} + {r_3}) + 2 y - 1)),\\
A^{3'}_{33} &=&\frac{2 {N_0} {m_W}^{7/2}} {{L_2} \sqrt{r_3}} ({d_{11}} {r_3} (2 x+4 y-1)+{d_{22}} (-2 x {r_1}+{r_1}+{r_2}-2 {r_3}+2 {r_3} x-2 ({r_1}+{r_2}-{r_3}) y)),\\
A^{3'}_{34} &=&\frac{4 {N_0} {d_{11}} {m_W}^{7/2} \sqrt{r_3}} {L_2} (y-z).
\end{eqnarray}
\end{widetext}

Non-zero coefficients $A^2_j$ and $A^4_j$ for $|(c\bar{Q'})_{\bf 1}[^3P_J]\rangle$ are
\begin{widetext}
\begin{eqnarray}
A^2_1 &=&-\frac{2 {L_2} {m_W}^{5/2}} {{r_1}^2 \sqrt{r_3}} (({r_1}-{r_2}) ({r_3} (2 {d_{21}} ({r_1}^2-{r_2}^2+1) {r_1}^2+{d_2} ({r_3}^2-3 {r_1} {r_3} +2 {r_1}({r_1}+{r_2})))\nonumber\\
&&+(-4 {d_{21}} {r_1}^3+2 {d_2} {r_1}-{d_2} {r_3}) x)+{d_1} (8 {r_3} {r_1}^3+2({r_3}^2-4 {r_2} {r_3}+2 u) {r_1}^2\nonumber\\
&&-(5 {r_3}^3+2 {r_2} {r_3}^2+2 u {r_3}-4 x {r_3}+4 {r_2} u) {r_1}+{r_2} {r_3} ({r_3}^2+2 u))),\\
A^2_2&=&\frac{2 {L_2} {m_W}^{5/2}} {{r_1}^2 {r_3}^{3/2}} ({r_1} ({d_1} {r_3} ({r_3}^3-2 {r_1} {r_3}^2-12 {r_1}^2 {r_3}+4 {r_1} {r_2} {r_3}-6 u {r_3} +4 y{r_3}-4 {r_1} u)\nonumber\\
&&-2 {d_{21}} {r_1} (({r_1}^2-{r_2}^2+1) {r_3}-2 {r_1} x) ({r_1} (2 {r_1} -2 {r_2}+{r_3})+2 (u-y)))+{d_2} (4 {r_3} {r_1}^4\nonumber\\
&&-2 ({r_3}^2+4 {r_2} {r_3}-2 u+2 x) {r_1}^3-(3 {r_3}^3-4 {r_2} {r_3}^2 +2 (-2 {r_2}^2+u-3 x+y+1) {r_3}\nonumber\\
&&+4 {r_2} u) {r_1}^2+{r_3} ({r_3}^3+(-2 {r_2}^2-2 u-2 x+3 y+1) {r_3}+2 {r_2} (u+y)) {r_1}-{r_2} {r_3}^2 y)),\\
A^2_3&=&\frac{8{L_2}{m_W}^{5/2}}{\sqrt{r_3}}({d_{11}}{r_3}({r_2} u + {r_1}v) + {d_{21}} ({r_1} - {r_2})(({r_1}^2 - {r_2}^2 + 1){r_3} -  2{r_1} x) + d_{22}(-2{r_3}{r_1}^3 \nonumber\\
&&+ ({r_3}^2 + 4{r_2}{r_3} - u - x){r_1}^2 + ({r_3}^3 - {r_2}{r_3}^2 + (-2{r_2}^2 - u - 2 x + y + 1){r_3} + {r_2} (2 u + x)){r1}\nonumber\\
&&- {r2}^2 u + {r_2}{r_3} (u - y) + 2({r_3}^2 - x) (u - y))),\\
A^2_4&=&\frac{8{d_{21}}{L_2}{m_W}^{5/ 2}}{\sqrt{r_3}}(-3{r_3}{r_1}^3 + ({r_3}^2 + 4{r_2}{r_3} - u + x){r_1}^2  + (-{r_3}{r_2}^2 + (-{r_3}^2 + 2 u + x){r_2} \nonumber\\
 &&+ {r_3}({r_3}^2 - u - 2 x + y)){r_1} - {r_2}^2 u + {r_2}{r_3} (u - y) +2({r_3}^2 - x) (u - y)),\\
A^2_5&=&-\frac{2 {L_2} {m_W}^{5/2}} {{r_1}^2 \sqrt{r_3}} ({d_2} ({r_1}-{r_2}) (2 {r_1}-{r_3})+4 {d_1} {r_1} {r_3}),\\
A^2_{7}&=&-\frac{8 L_2 {m_W}^{5/2} }{ \sqrt{r_3}}({d_{11}}{r_1}{r_3} + {d_{22}} ({r_1} ({r_2} - 3{r_1}) - 4 u +4 y)),\\
A^2_{9}&=&\frac{8{d_{21}} {L_2} {m_W}^{5/2}} {\sqrt{r_3}}(3{r_1}^2 - {r_2}{r_1} + 4 u - 4 y),\\
A^2_{10}&=&\frac{8 d_{21} {L_2}{m_W}^{5/2} } {\sqrt{r_3}}({r_1}{r_3} - ({r_2} + 2{r_3}){r_3} + 2 x),\\
A^2_{11}&=&\frac{4 {d_2} {L_2} {m_W}^{5/2}} {\sqrt{r_3}},\\
A^2_{12}&=&\frac{2 {d_2} {L_2} {m_W}^{5/2} \sqrt{r_3}} {{r_1}^2} {(r_1-r_2)},\\
A^2_{13}&=&\frac{2 {L_2} {m_W}^{5/2}} {{r_1}^2 {r_3}^{3/2}} {(4 {d_{21}} (({r_1}^2-{r_2}^2+1) {r_3}-2 {r_1} x) {r_1}^2+{d_2} ({r_1}-3 {r_2}){r_3}^2)},\\
A^2_{14}&=&\frac{8 d_{22} {L_2} {m_W}^{5/2}}{\sqrt{r_3}}({r_1}{r_3} - ({r_2} + 2{r_3}){r_3} + 2 x),\\
A^2_{15}&=&\frac{8 {L_2} {m_W}^{5/ 2}} {\sqrt{r_3}}({d_{11}} ({r_2} - {r_1}){r_3} + {d_{22}}({r_1}^2 - {r_3}{r_1} - {r_2}^2 + {r_2}{r_3} + 2 x - 2)),\\
A^2_{17}&=&\frac{4 {L_2} {m_W}^{5/2}} {{r_1}^2 {r_3}^{3/2}} {({d_2} {r_3} ({r_1}^2-({r_2}+{r_3}) {r_1}+2 {r_2} {r_3})-2 {d_{21}} {r_1}^2 (({r_1}^2-{r_2}^2+1) {r_3}-2 {r_1} x))},\\
A^2_{18}&=&\frac{8 d_{21} L_2 {m_W}^{5/2}} {\sqrt{r3}} ({r_1}^2 - {r_3}{r_1} - {r_2}^2 + {r_2}{r_3} + 2 x - 2),\\
A^2_{19}&=&-\frac{2 {L_2} {m_W}^{5/2}} {{r_1}^2 {r_3}^{3/2}} (2 {d_{21}} ({r_1} {r_3}+2 u) (({r_1}^2-{r_2}^2+1) {r_3}-2 {r_1} x) {r_1}^2 +{r_3}^2 ({r_1} {r_3} ({d_2} {r_1}+{d_1} {r_3})\nonumber\\
&&+(2 {d_1} {r_1}+{d_2} ({r_1}-3 {r_2})) u)),\\
A^2_{20}&=&\frac{2 {L_2} {m_W}^{5/2}} {{r_1}^{2} {r_3}^{3/2}} (2 {d_{21}} ({r_1} {r_3}-({r_2}+2 {r_3}) {r_3}+2 x) (({r_1}^2-{{r_2}^2}+1) {r_3}-2 {r_1} x) {r_1}^2\nonumber\\
&&+{r_3}^2 ({d_2} {r_3} ({r_1}^2-({r_2}+{r_3}) {r_1}+3 {r_2} {r_3})+{d_1} ({r_1}-{r_2}) ({r_3}^2+2 u)+{d_2} ({r_1}-3 {r_2}) x)),\\
A^2_{21}&=&\frac{2 {L_2} {m_W}^{5/2}} {{r_1}^2 {r_3}^{3/2}} (2 {d_{21}} (({r_1}^2-{r_2}^2+1) {r_3}-2 {r_1} x) ({r_1} (2 {r_1}-2 {r_2}+{r_3})+4 u-2 y) {r_1}^2\nonumber\\
&&+{r_3}^2 ({d_1} {r_1} ({r_3}^2+2 u)+{d_2} (({r_3}-2 {r_2}) {r_1}^2+(2 {r_2}^2-{r_3}^2+ x-z){r_1}+{r_2} (y-4 u)))),\\
\frac{A^2_{22}}{d_{21}}&=&\frac{A^2_{25}}{d_{22}}=\frac{A^2_{27}}{d_{21}}=\frac{A^2_{28}}{d_{22}}= -\frac{16  {L_2} {m_W}^{5/2}} {\sqrt{r_3}},\\
A^2_{29}&=&\frac{8 {d_{21}} {L_2} {m_W}^{5/2}} {\sqrt{r_3}} {( {r_1}^2-{r_2} {r_1}-2 w)},\\
A^2_{30}&=&-\frac{8{L_2}{m_W}^{5/ 2}} {\sqrt{r_3}}({d_{21}} ({r_1} - {r_2}) ({r_1} + {r_2} - {r_3}) + {d_{11}}{r_1}{r_3} + {d_{22}} ({r_1} ({r_2} - 3{r_1}) - 2 u + 2 y)),\\
\frac{{}A^2_{31}}{d_{21}}&=&\frac{{}A^2_{32}}{d_{22}}=\frac{{}A^2_{34}}{2 d_{11}}=8  {L_2} {m_W}^{5/2} \sqrt{r_3} {(r_1-r_2)},\\
A^2_{33}&=&-\frac{8 {L_2} {m_W}^{5/2}}{\sqrt{r_3}} {({d_{22}} ({r_1}-{r_2}) ({r_1}+{r_2}-{r_3})+{d_{11}} ({r_2}-3 {r_1}) {r_3})},\\
A^{4'}_1&=&\frac{{N_0} {m_W}^{7/2}} {2 {L_1} {r_1}^2 \sqrt{r_3}} (2 {d_{21}} (({r_1}^2-{r_2}^2+1) {r_3}-2 {r_1} x) (y-z) {r_1}^2+{d_1} (4 {r_3} (3 x+4 y-2) {r_1}^2\nonumber\\
&&+2 ((y-z) {r_3}^2-2 {r_2} x {r_3}+2 u ( y-z)) {r_1}-{r_3} ({r_3}^2+2 u) (y-z))+{d_2} (2 {r_3} ( y-z) {r_1}^2\nonumber\\
&&-((3 x+6 y-5) {r_3}^2+2 {r_2} (y-z) {r_3}+2 x-4 x y) {r_1}+{r_3} ((x+2 y-2) {r_3}^2+x-2 x y))),\\
A^{4'}_2&=&\frac{{N_0} {m_W}^{7/2}} {2 {L_1} {r_1}^2 {r_3}^{3/2}} (2 {d_{21}} (({r_1}^2-{r_2}^2+1) {r_3}-2 {r_1} x) ((2 {r_2}+{r_3}) y-2 {r_1} z) {r_1}^2+{d_1} {r_3} (4 {r_3} (2 x+3 y-1) {r_1}^2\nonumber\\
&&+2 ({r_3}^2+2 {r_2} {r_3}+2 u) y {r_1}-{r_3} ({r_3}^2+2 u) y)+{d_2} (4 x {r_1}^4-2 (2 {r_2} x+{r_3} (5 x+2 y-3)) {r_1}^3\nonumber\\
&&+((6 x+2 y-5) {r_3}^2+2 {r_2} (x-4 y+1) {r_3}+4 u-8 u y) {r_1}^2+{r_3} (-4 y {r_2}^2+{r_3} (4 y-1) {r_2}+(3 {r_3}^2-4 y+2) y\nonumber\\
&&+6 u (2 y-1)) {r_1}-{r_3}^2 ((-2 {r_2}^2+{r_3}^2-2 y+1) y+u (4 y-2)))),\\
A^{4'}_3&=&-\frac{2{N_0}{m_W}^{7/ 2}} {{L_1} \sqrt{r_3}} ({d_{21}}(({r_1}^2 - {r_2}^2 + 1){r_3} - 2{r_1} x) (y - z) + {d_{11}}{r_3}(-y{r_3}^2 + u + {r_1} ({r_1} - {r_2}) x - 2 u y) \nonumber\\
&&+ {d_{22}}(x{r_1}^3 + {r_3} ( y + 3 z){r_1}^2 + (-x{r_2}^2 - {r_3} (3 y - z){r_2} +u - 2 (x - 1) x - 2 (u + x) y - {r_3}^2 (y +2 z)){r_1}\nonumber\\
&& - 2{r_3} y^2 + {r_2} u + {r_3} y - (({r_2} - {r_3})({r_3}^2 + 2{r_2}{r_3} + 2 u) + 2 ({r_2} + {r_3}) x) y)),\\
A^{4'}_4&=&\frac{2{d_{21}}{m_W}^{7/ 2}{N_0}} {{L_1}\sqrt{r_3}} ({r_3}{r_1}^4 - ({r_2}{r_3} + 3 x){r_1}^3 + (2{r_2} x + {r_3}(-{r_2}^2 + u + x- 2 z)){r_1}^2 + ({r_3}{r_2}^3 + x{r_2}^2 \nonumber\\
 &&+ {r_3} (x + 4 y - 2){r_2} + 2 x^2 - u - 2 u x - 2 x + 2 (u + x) y + {r_3}^2 (y + 2 z)){r_1} +2{r_3} y^2 - ({r_2} + ({r_2}^2 - 1){r_3}) u\nonumber\\
  &&- ({r_3}^3 + {r_2}{r_3}^2 + (-2{r_2}^2 + 2 u - 2 x + 1){r_3} - 2{r_2} (u + x)) y),\\
A^{4'}_{5}&=&-\frac{{N_0} {m_W}^{7/2}} {2 {L_1} {r_1}^2 \sqrt{r_3}} (2 {d_{21}} (({r_1}^2-{r_2}^2+1) {r_3}-2 {r_1} x) {r_1}^2+{d_1} (4 {r_3} {r_1}^2+2
 ({r_3}^2+2 {r_2} {r_3}+2 u) {r_1}-{r_3} ({r_3}^2+2 u))\nonumber\\
&&+{d_2} (2 {r_3} {r_1}^2+(-3 {r_3}^2+2 {r_2} {r_3}-4 y+2) {r_1}+{r_3} ({r_3}^2+2 y-1))),\\
A^{4'}_{6}&=&\frac{2{d_{21}}{m_W}^{7/2}{N_0}} {{L_1}\sqrt{r_3}}(({r_1}^2 - {r_2}^2 + 1){r_3} - 2{r_1} x),\\
A^{4'}_{7}&=&- \frac{2{m_W}^{7/ 2}{N_0}} {{L_1}\sqrt{r_3}} ({d_{22}}({r_1}^3 - (2{r_2} + {r_3}){r_1}^2 + ({r_2} ({r_2} + {r_3}) + 2 (u + x - 1)){r_1} - 2 ({r_2} + {r_3}) u + 2 (2{r_2} + {r_3}) y) \nonumber\\
 &&- {d_{11}}{r_3}({r_1}^2 - {r_2}{r_1} +2 u)) ,\\
A^{4'}_{9}&=&- \frac{2{d_{21}}{m_W}^{7/ 2}{N_0}} {{L_1}\sqrt{r_3}}({r_1}^3 - (2{r_2} + {r_3}){r_1}^2 + ({r_2} ({r_2} + {r_3}) +  2 (u + x - 1)){r_1} - 2 ({r_2} + {r_3}) u +2 (2{r_2} + {r_3}) y),\\
A^{4'}_{10}&=&- \frac{2{d_{21}}{m_W}^{7/2}{N_0}} {{L_1}\sqrt{r_3}} ({r_3}^3 + ({r_2} - {r_1}){r_3}^2 + (2 y - 1){r_3} +2 ({r_1} - {r_2}) x),\\
A^{4'}_{11}&=&-\frac{{N_0} {m_W}^{7/2}} {{L_1} {r_1}^2 {r_3}^{3/2}} (2 {d_{21}} (({r_1}^2-{r_2}^2+1) {r_3}-2 {r_1} x) {r_1}^3 +{d_2} {r_3} ({r_1}^3-({r_2}+{r_3}) {r_1}^2+2 ({r_2} {r_3}+u) {r_1}+2 {r_3} (y-u))),\\
A^{4'}_{12}&=&-\frac{{N_0} {m_W}^{7/2}} {2 {L_1} {r_1}^2 \sqrt{r_3}} (2 {d_{21}} ({r_3} {r_1}^2-2 x {r_1}-{r_2}^2 {r_3}+{r_3}) {r_1}^2+{d_1} {r_3}
({r_3}^2+2 u)+{d_2} {r_3} ({r_3}^2-{r_1} {r_3}+2 y-1)),\\
A^{4'}_{13}&=&-\frac{{N_0} {m_W}^{7/2}} {2 {L_1} {r_1}^2 {r_3}^{3/2}} (2 {d_{21}} (2 {r_1}-2 {r_2}-{r_3}) (({r_1}^2-{r_2}^2+1) {r_3}-2 {r_1} x) {r_1}^2
+{r_3}^2 ({d_2} ({r_3}^2-{r_1} {r_3}\nonumber\\
&&+2 ({r_1}-{r_2}) {r_2}+2 y-1)-{d_1} ({r_3}^2+2 u))),\\
A^{4'}_{14}&=& - \frac{2{m_W}^{7/2}{N_0}} {{L_1}\sqrt{r_3}}({d_{11}}{r_3}^3 + {d_{22}}({r_3}^3 + ({r_2} - {r_1}){r_3}^2 + (2 y - 1){r_3} + 2 ({r_1} - {r_2}) x)),\\
A^{4'}_{15}&=&\frac{2{m_W}^{7/2}{N_0} } {{L_1}\sqrt{r_3}} ({d_{11}}{r_3} (1-2 z) + {d_{22}} (-3{r_2} - 2{r_3} + {r_1} (1+2 z) + 2 ({r_2} + {r_3}) (x + y))),\\
A^{4'}_{17}&=&-\frac{{N_0} {d_2}{m_W}^{7/2}} {{L_1} {r_1}^2 \sqrt{r_3}} {({r_1} (1-2 z)+2 {r_3} z)},\\
A^{4'}_{18}&=&\frac{2{d_{21}}{m_W}^{7/2}{N_0}} {{L_1}\sqrt{r_3}} (-3{r_2} - 2{r_3} + {r_1} (1+2 z) + 2 ({r_2} + {r_3}) (x +  y)),\\
A^{4'}_{19}&=&\frac{{N_0} {m_W}^{7/2}} {2 {L_1} {r_1}^2 {r_3}^{3/2}} (2 {d_{21}} ({r_1} {r_3} ({r_1}-{r_2}+2 {r_3})+(2 {r_1}-2 {r_2}+{r_3}) u)  (({r_1}^2-{r_2}^2+1) {r_3}-2 {r_1} x) {r_1}^2\nonumber\\
 &&+{r_3}^2 ({d_1} ({r_1}^2-{r_2} {r_1}+u) ({r_3}^2+2 u)+{d_2} ({r_3} {r_1}^3-{r_3} ({r_2}+{r_3}) {r_1}^2\nonumber\\
 &&+(3 {r_2} {r_3}^2+u {r_3}+2 {r_2} u) {r_1}-(2 {r_2}^2+{r_3}^2+1) u+2 ({r_3}^2+u) y))),\\
A^{4'}_{20}&=&\frac{{N_0} {m_W}^{7/2}} {2 {L_1} {r_1}^2 {r_3}^{3/2}} ({r_3}^2 ({d_2} (2 x {r_2}^2+x-2 x y-{r_3}^2 (x+2 z)-{r_1} (2 {r_2} x+{r_3} ( y-z)))\nonumber\\
&&-{d_1} ({r_3}^2+2 u) ( y-z))-2 {d_{21}} {r_1}^2 (({r_1}^2-{r_2}^2+1) {r_3}-2 {r_1} x) (2 ({r_1}-{r_2}) x+{r_3} ( y-z))),\\
A^{4'}_{21}&=&\frac{{N_0} {m_W}^{7/2}} {2 {L_1} {r_1}^2 {r_3}^{3/2}} ({r_3}^2 ({d_2} (-2 y^2-(-{r_3}^2+{r_1} {r_3}+2 {r_2} ({r_1}+{r_2})+4 x) y+y+2 u+{r_1} ({r_1}+{r_2}-4 {r_2} x)) \nonumber\\
 &&-{d_1} ({r_3}^2+2 u) y)-2 {d_{21}} {r_1}^2 (({r_1}^2-{r_2}^2+1) {r_3}-2 {r_1} x) ((2 {r_2}+{r_3}) y+2 {r_1} (x-z))),\\
\frac {A^{4'}_{22} } {d_{21}}&=& \frac {A^{4'}_{25}} {d_{22}}=-\frac{4  {N_0} {r_2} {m_W}^{7/2}} {{L_1} \sqrt{r_3}},\\
A^{4'}_{26}&=&\frac{4 {d_{11}} {N_0} {m_W}^{7/2} \sqrt{r_3}} {L_1},\\
\frac{A^{4'}_{27}}{d_{21}}&=&\frac{A^{4'}_{28}}{d_{22}}=\frac{4  {N_0} {m_W}^{7/2}}  {{L_1} \sqrt{r_3}}{(r_1-r_2)},\\
A^{4'}_{29}&=&-\frac{2 {d_{21}} {N_0} {m_W}^{7/2}} {{L_1} \sqrt{r_3}} ({r_1}^3-(2 {r_2}+{r_3}) {r_1}^2+({r_2} ({r_2}+{r_3})+2 (u-z)) {r_1}-2 {r_3} u+2 ({r_2}+{r_3}) y),\\
A^{4'}_{30}&=&\frac{2{m_W}^{7/2}{N_0}} {{L_1}\sqrt{r_3}} ({d_{11}}{r_3}({r_1}^2 - {r_2}{r_1} +2 u) - {d_{22}}({r_1}^3 - (2{r_2} + {r_3}){r_1}^2 + ({r_2} ({r_2} +{r_3}) + 2 (u + x - 1)){r_1} - 2{r_3} u) \nonumber\\
&&-  2{d_{22}} ({r_1} + {r_2} + {r_3}) y + {d_{21}} (-2 y{r_2} + {r_2} + 2{r_3} z + {r_1} (1-2 z))),\\
A^{4'}_{31}&=&\frac{2 {d_{21}} {N_0} {m_W}^{7/2}  \sqrt{r_3}} {L_1} ((r_1+r_2-r_3) {r_3}-2 y+1),\\
A^{4'}_{32}&=& - \frac{2{m_W}^{7/ 2}{N_0} \sqrt{r_3}} {{L_1}} ({d_{11}}{r_3}^2 + {d_{22}} ({r_3} (-{r_1} - {r_2} + {r_3}) + 2 y - 1)),\\
A^{4'}_{33}&=&\frac{2 {N_0} {m_W}^{7/2}} {{L_1} \sqrt{r_3}} ({d_{11}} {r_3} (-2 x-4 y+1)+{d_{22}} (-2 y {r_2}+{r_2}+2 {r_3} z+{r_1} (1-2 z ))),\\
A^{4'}_{34}&=&-\frac{4 {d_{11}} {N_0} {m_W}^{7/2} \sqrt{r_3}}  {L_1}{( y-z)}.
\end{eqnarray}
\end{widetext}


\begin{thebibliography}{99}

\bibitem{Teva} R.M. Thurman-Keup, A.V. Kotwal, M. Tecchio and A.B. Wagner, Rev.Mod.Phys. {\bf 73}, 267(2001).

\bibitem{Tevb} G. Weiglein {\it et al.}, LHC/LC Study Group Collaboration, Phys.Rept. {\bf 426}, 47(2006).

\bibitem{Tevc} J.R. Gaunt, C.H. Kom, A. Kulesza and W.J. Stirling, Eur.Phys.J. C{\bf 69}, 53(2010).

\bibitem{w} C.F. Qiao, L.P. Sun, D.S. Yang and R.L. Zhu, Eur.Phys.J. C{\bf 71}, 1766(2011).

\bibitem{w1} V.D. Barger, K. Cheung, W.Y. Keung, Phys.Rev. D{\bf 41}, 1541(1990).

\bibitem{nrqcd} G.T. Bodwin, E. Braaten and G.P. Lepage, Phys.Rev. D{\bf 51}, 1125 (1995); {\bf 55}, 5853 (E) (1997).

\bibitem{yellow1} N. Brambilla, {\it et al.}, Quarkonium Working Group, published as CERN Yellow Report, arXiv: 0412158[hep-ph].

\bibitem{yellow2} N. Brambilla, {\it et al.}, Quarkonium Working Group,  Eur.Phys.J. C{\bf 71}, 1534(2011).

\bibitem{cms} G.L. Bayatian, {\it et al.}, CMS technical design report volume II: Physics performance, J.Phys. G{\bf 34}, 995(2007).

\bibitem{cdf} F. Abe {\it et al.}, CDF Collaboration, Phys.Rev. D{\bf 58}, 112004(1998); A. Abulencia {\it et al.}, CDF Collaboration, Phys.Rev.Lett. {\bf 96}, 082002 (2006); A. Abulencia {\it et al.}, CDF Collaboration, Phys.Rev.Lett. {\bf 97}, 012002 (2006).

\bibitem{bc1} C.H. Chang and Y.Q. Chen, Phys.Rev. D{\bf 48}, 4086(1993); C.H. Chang, Y.Q. Chen, G.P. Han and H.T. Jiang, Phys.Lett. B{\bf 364}, 78(1995); C.H. Chang and X.G. Wu, Eur.Phys.J. C{\bf 38}, 267(2004).

\bibitem{bc2} A.V. Berezhnoi, A.K. Likhoded and M.V. Shevlyagin, Phys. Atom. Nucl. {\bf 58}, 672(1995); S.S. Gershtein, V.V. Kiselev, A.K. Likhoded, A.V. Tkabladze, Phys. Usp. {\bf 38}, 1(1995).

\bibitem{bc3}  C.H. Chang, J.X. Wang and X.G. Wu, Phys.Rev. D{\bf 70}, 114019(2004); C.H. Chang, C.F. Qiao, J.X. Wang and X.G. Wu, Phys.Rev. D{\bf 71}, 074012(2005); C.H. Chang, C.F. Qiao, J.X. Wang and X.G. Wu, Phys.Rev. D{\bf 72}, 114009(2005).

\bibitem{bcvegpy} C.H. Chang, C. Driouich, P. Eerola and X.G. Wu, Comput. Phys. Commun. {\bf 159}, 192(2004); C.H. Chang, J.X. Wang and X.G. Wu,  Comput. Phys. Commun. {\bf 174}, 241(2006); C.H. Chang, J.X. Wang and X.G. Wu, Comput. Phys. Commun. {\bf 175}, 624(2006); X.Y. Wang and X.G. Wu, Comput. Phys. Commun. {\bf 183}, 442(2012).

\bibitem{tbc1} C.F. Qiao, C.S. Li and K.T. Chao, Phys.Rev. D{\bf 54}, 5606(1996).

\bibitem{tbc2} C.H. Chang, J.X. Wang and X.G. Wu, Phys.Rev. D{\bf 77}, 014022(2008). X.G. Wu, Phys.Lett. B{\bf 671}, 318(2009).

\bibitem{zbc0} C.H. Chang and Y.Q. Chen, Phys.Rev. D{\bf 46}, 3845(1992).

\bibitem{zbc1} L.C. Deng, X.G. Wu, Z. Yang, Z.Y. Fang and Q.L. Liao, Eur.Phys.J. C{\bf 70}, 113(2010).

\bibitem{zbc2} Z. Yang, X.G. Wu, L.C. Deng, J.W. Zhang and G. Chen, Eur.Phys.J. C{\bf 71}, 1563(2011).

\bibitem{zbc3} C.F. Qiao, L.P. Sun and R.L. Zhu, JHEP {\bf 1108}, 131 (2011).

\bibitem{helicity} R. Kleiss and W.J. Stirling, Nucl.Phys. B{\bf 262}, 235(1985).

\bibitem{petrelli} A. Petrelli, M. Cacciari, M. Greco, F. Maltoni and M.L. Mangano, Nucl.Phys. B{\bf 514}, 245(1998).

\bibitem{pot1}  E. Eichten, K. Gottfried, T. Kinoshita, K.D. Lane and T.M. Yan, Phys.Rev. D{\bf 17}, 3090(1978); ibid. {\bf 21}, 313(E)(1980); ibid.{\bf 21}, 203(1980).

\bibitem{pot2} W. Buchm${\rm \ddot{u}}$ller and S.-H.H. Tye, Phys.Rev. D{\bf 24}, 132(1981).

\bibitem{pot3} A. Martin, Phys.Lett. B{\bf 93}, 338(1980).

\bibitem{pot4} C. Quigg and J.L. Rosner, Phys.Lett. B{\bf 71}, 153(1977).

\bibitem{pot5} Y.Q. Chen and Y.P. Kuang, Phys.Rev. D{\bf 46}, 1165(1992).

\bibitem{pot6} E.J. Eichten and C. Quigg, Phys.Rev. D{\bf 49}, 5845(1994).

\bibitem{pnrqcd1} N. Brambilla, A.Pineda, J. Soto and A. Vairo, Nucl. Phys. B {\bf 566}, 275 (2000);

\bibitem{pnrqcd2} N. Brambilla, A.Pineda, J. Soto and A. Vairo, Rev.Mod.Phys. {\bf 77}, 1423(2005).

\bibitem{lat1} G.T. Bodwin, D.K. Sinclair and S. Kim, Phys.Rev. Lett.{\bf 77}, 2376(1996).

\bibitem{bcdecay} X.G. Wu, C.H. Chang, Y.Q. Chen and Z.Y. Fang, Phys.Rev. D{\bf 67}, 094001(2003).

\bibitem{projector}  Y.Q. Chen, Phys.Rev. D{\bf 48}, 5181(1993).

\bibitem{wtd} J. Alcaraz, {\it etal.}, ariXiv:0911.2604[hep-ex].

\bibitem{pdg} K. Nakamura, {\it et al.}, Particle Data Group, J.Phys. G{\bf 37}, 075021(2010).

\bibitem{genxicc} C.H. Chang, J.X. Wang and X.G. Wu, Comput.Phys.Commun.{\bf 177}, 467(2007); C.H. Chang, J.X. Wang and X.G. Wu, Comput.Phys.Commun.{\bf 181}, 1144(2010).

\bibitem{xicc} J.P. Ma and Z.G. Si, Phys.Lett. B{\bf 568}, 135(2003); C.H. Chang, C.F. Qiao, J.X. Wang and X.G. Wu, Phys.Rev. D{\bf 73}, 094022(2006); C.H. Chang, J.P. Ma, C.F. Qiao, X.G. Wu, J.Phys. G{\bf 34}, 845(2007).

\bibitem{slhc} A. Blondel, {\it et al.}, arXiv: 0609102[hep-ph]; M.L. Mangano, arXiv:0910.0030[hep-ph].

\bibitem{blm} S.J. Brodsky, G.P. Lepage and P.B. Mackenzie, Phys.Rev. D{\bf 28}, 228(1983).

\bibitem{pmc} S.J. Brodsky and L.D. Giustino, arXiv: 1107.0338; S.J. Brodsky and X.G. Wu, arXiv: 1111.6175.

\end{thebibliography}
\end{document}